\documentclass{article}

\usepackage{arxiv}
\usepackage[T1]{fontenc}
\usepackage[utf8]{inputenc}
\usepackage{microtype}
\usepackage{amsmath,amssymb}
\usepackage{graphicx}
\usepackage{xcolor}
\usepackage{booktabs}
\usepackage{multirow}
\usepackage{longtable}
\usepackage{float}
\usepackage{placeins}
\usepackage{algorithm}
\usepackage{algpseudocode}
\usepackage[numbers,sort&compress]{natbib}
\usepackage{bibunits}
\usepackage[colorlinks=true,linkcolor=blue,citecolor=blue,urlcolor=blue]{hyperref}
\usepackage{doi}

\graphicspath{{figures/}}

\title{Subsystem Structure as an Inferential Resource for Coupled Engineered Systems}

\author{%
  Esmaeil Ghorbani\thanks{E.G.\ and J.H.\ contributed equally to this work.} \\
  Department of Civil and Environmental Engineering\\
  Princeton University, Princeton, NJ 08544\\
  \texttt{ghorbani@princeton.edu}
  \And
  J\"urgen Hackl\footnotemark[1]\ \thanks{Corresponding author.} \\
  Department of Civil and Environmental Engineering\\
  Princeton University, Princeton, NJ 08544\\
  \texttt{hackl@princeton.edu}
}
\date{\today}

\renewcommand{\headeright}{}
\renewcommand{\undertitle}{}
\renewcommand{\shorttitle}{Subsystem Structure as an Inferential Resource}

\hypersetup{
  pdftitle={Subsystem Structure as an Inferential Resource for Coupled Engineered Systems},
  pdfauthor={Esmaeil Ghorbani, Juergen Hackl},
  pdfkeywords={inverse problems, uncertainty quantification, complex engineered systems, message passing, digital twins},
}

\begin{document}
\maketitle

\begin{abstract}
Engineered infrastructure systems pose inverse problems in which hidden states, unknown parameters, and subsystem couplings must be inferred from sparse and noisy measurements. These problems are difficult because physical subsystems are heterogeneous, sensing is partial, uncertainty is distributed across subsystem interfaces, and computational cost grows rapidly with system size. We address this challenge with probabilistic compositional inference, a graph-based architecture that represents a coupled system as interacting subsystems, each retaining its own local model, estimator, and uncertainty representation, while coupling is handled through physically meaningful stochastic messages exchanged across subsystem interfaces. This formulation allows mechanistic, learned, and deterministic components to coexist within a single inference framework and propagates calibrated uncertainty without assembling a global augmented state or covariance. We validate the framework in three increasingly demanding settings: a sparse-sensing canonical inverse problem, where interface couplings can also be learned from data; infrastructure-scale power networks, where the method matches centralized joint state-and-parameter inference while reducing computational scaling from approximately cubic to approximately linear; and a multi-physics turbine embedded in a power-grid network, where heterogeneous subsystems compose hierarchically without degrading local inference or collapsing local posteriors into a global estimate. Together, these results show that subsystem structure can be exploited as the organizing principle for uncertainty-aware inverse inference in coupled engineered systems.
\end{abstract}


\begin{bibunit}[unsrtnat]

\noindent Engineered infrastructure systems give rise to inverse problems whose solution is essential for reliable operation, condition monitoring, and the construction of digital twins that evolve with their physical counterparts~\cite{Kapteyn2021probabilistica}.  The central task is to infer quantities that cannot be measured directly, including hidden internal states, unknown parameters, and latent couplings between subsystems, from sparse and noisy observations made at a limited set of accessible locations~\cite{Willcox2021imperative}.  What makes this problem difficult is not any one feature in isolation, but the simultaneous interaction of heterogeneity, sparse sensing, distributed uncertainty, and scale~\cite{Masison2021modular}.  The different parts of an infrastructure system, including hydraulic, mechanical, electrical, and control components, are governed by different equations, operate at different fidelities, and are sometimes represented by data-driven surrogates when first-principles descriptions are unavailable~\cite{Makke2024Datadriven}.  Most internal degrees of freedom remain unobserved during operation, so information must be propagated across subsystem boundaries to reach the quantities of interest~\cite{Champion2019Datadriven}. Uncertainty is itself distributed across the system: each subsystem carries its own measurement noise, model discrepancy, and parameter uncertainty, while coupling causes uncertainty in one location to affect inference elsewhere through interface variables that are themselves uncertain.  These difficulties are amplified further by size, because modern infrastructure networks may contain hundreds to thousands of interacting components and still require computationally tractable inference~\cite{Mohammadi2021Thinking}. Methods that address any one of these challenges in isolation are well developed; the central difficulty is to address them jointly within a single coupled architecture~\cite{Peherstorfer2018Survey}.

Existing approaches to inverse modeling under uncertainty address parts of this problem, but no single class of existing methods handles heterogeneity, sparse sensing, distributed uncertainty, and scale at the same time. Monolithic data-assimilation methods, including ensemble and variational approaches developed for geophysical~\cite{Arcucci2026convergence} and weather-forecasting applications~\cite{Leach2021Forecastbased}, treat the full system as a single coupled state-space model and infer global states and parameters within a unified probabilistic framework~\cite{Kalnay2023Review,Tokuda2026Selective}. These methods are highly effective when system components share a common mathematical structure~\cite{Karniadakis2021Physicsinformeda}, but their cost grows rapidly with the dimension of the joint state, and the requirement of a single forward model and observation operator becomes restrictive when subsystems span different physical domains, model classes, and fidelities~\cite{Carrassi2018Data}. 

Modular and co-simulation frameworks address the heterogeneity problem from the opposite direction by treating each subsystem as an independent simulator that exchanges inputs and outputs through standardized interfaces under a coordinating master algorithm~\cite{Gomes2018CoSimulation}. Such frameworks preserve modularity and are widely used for forward simulation across heterogeneous engineering domains~\cite{Tencer2023Network}, but they were not designed for joint inverse inference across subsystem boundaries: uncertainty is typically not propagated through the interfaces, and joint state-and-parameter estimation over the coupled system is rarely supported. Distributed and factor-graph approaches have addressed parts of this inverse problem within a more inferential setting, but they have typically remained confined to a single physical domain~\cite{Chavali2015Distributed}, to homogeneous estimator classes, or to settings in which interaction laws are fixed rather than learned~\cite{Sanchez-Gonzalez2020Learning}. More recently, graph-based learned dynamics models, including graph neural networks and physics-informed graph learning architectures, have emerged as flexible surrogates for forward simulation on networked systems~\cite{Sharma2026physicsinformed}. These approaches can absorb heterogeneous couplings and scale to large networks, but they generally replace underlying mechanistic models rather than compose with them~\cite{Yu2024Learning}, and they do not by themselves provide calibrated posterior uncertainty for the hidden states and unknown parameters of the physical system. What is therefore needed is a framework for inverse inference, rather than forward simulation alone, that combines mechanistic, data-driven, and learned components within a single coupled architecture, propagates calibrated uncertainty across subsystem interfaces, and remains tractable at infrastructure scale.

We recast inverse inference in coupled infrastructure systems as a graph of interacting subsystems rather than a single global state-space problem to be solved monolithically. Each node carries a local model of one subsystem together with an estimator that maintains a posterior over its hidden states and any unknown parameters. These local models may be mechanistic, data-driven, or hybrid, and different subsystems may employ different estimator classes according to the physics and data available.  Each edge carries an interface law that transforms outgoing interface variables from one subsystem into incoming coupling effects for another; when the coupling physics is known, this law can be specified analytically, and when it is not, it can be identified from data.  Inference then proceeds through local subsystem updates coordinated by message passing~\cite{Cantwell2019Message} on the interaction graph, with each subsystem combining its own measurements with incoming interface messages from neighboring subsystems over the next time interval. We call this framework probabilistic compositional inference: probabilistic because it propagates calibrated uncertainty across subsystem interfaces rather than only point estimates; compositional because heterogeneous subsystem models remain local and are coupled through the graph rather than absorbed into a single global state; and inference because the framework is designed to solve coupled inverse inference problems rather than only forward simulation. Its central architectural principle is that uncertainty is propagated across subsystems without collapsing all local structure into one global posterior, allowing heterogeneous model classes, estimators, and posterior representations to coexist within a single coupled architecture.

To validate this framework, we present three case studies that progressively isolate its capabilities under increasingly demanding conditions. On a canonical mechanical testbed, probabilistic compositional inference recovers hidden internal dynamics and unknown parameters from sparse boundary measurements, propagates calibrated uncertainty across subsystem interfaces, and accommodates a learned interaction law when the coupling physics is incomplete. On infrastructure-scale power networks ranging from 9 to 300 buses, the framework matches a centralized joint state-and-parameter estimator in accuracy and calibration while reducing computational scaling from approximately cubic to approximately linear, making uncertainty-aware inference tractable at the scale of real infrastructure. On a multi-physics turbine embedded within a power-grid network, the same architecture combines mechanistic estimators, deterministic controllers, and data-driven surrogates within a single coupled inference while preserving local inferential accuracy under hierarchical composition across scales, without collapsing subsystem posteriors into a global estimate. The contribution is therefore architectural rather than the introduction of a new filter, surrogate, or domain-specific method: the framework reorganizes how coupled inverse problems are formulated and solved, not which local components are used to solve them. This reorganization opens a path toward uncertainty-aware inference for engineered infrastructure systems whose complexity, heterogeneity, and scale have long challenged monolithic approaches.

\begin{figure}[tbp]
\includegraphics[width=\textwidth,height=140mm]{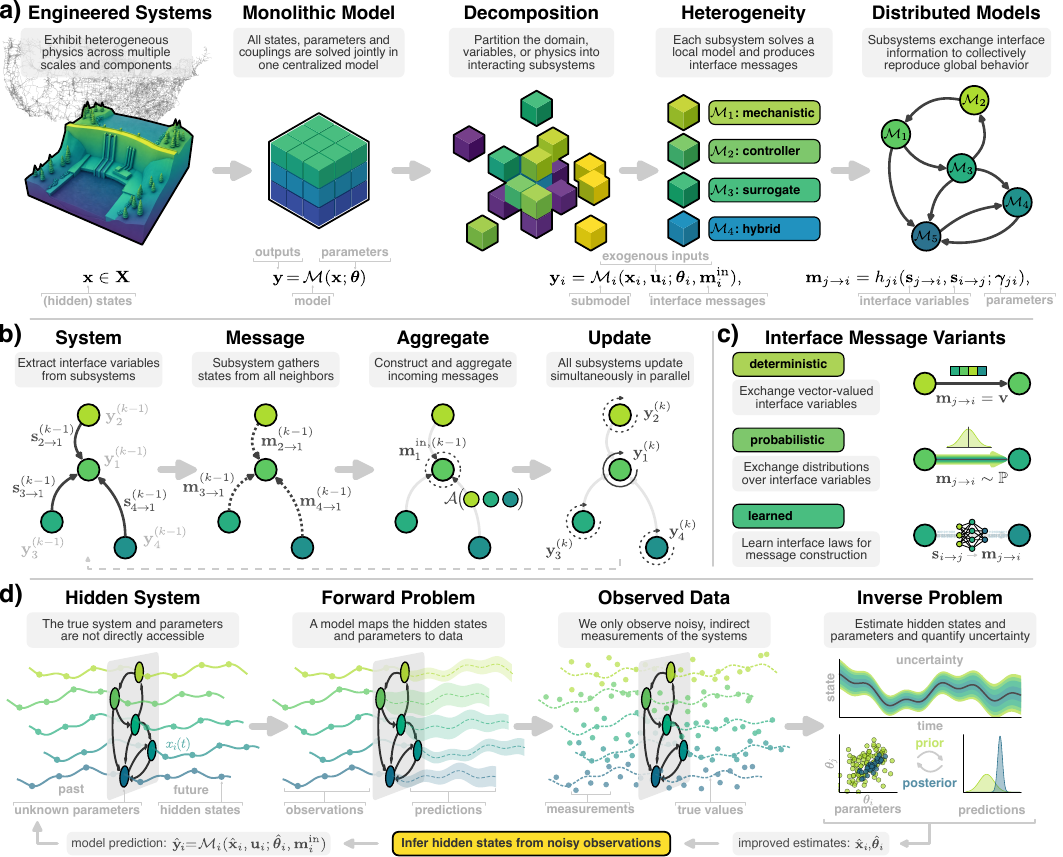}
\caption{\textbf{Probabilistic compositional inference represents coupled engineered systems as directed graphs of interacting subsystems.}
(\textit{a})~From system to graph: an engineered system can be treated monolithically, with all states, parameters, and couplings solved jointly in one centralized model $\mathcal{M}$, or decomposed into interacting subsystems, each carrying its own local model $\mathcal{M}_i$ of a possibly different class (mechanistic, controller, surrogate, or hybrid). The resulting interaction graph couples the local models through interface laws $h_{ji}$ that transform outgoing interface variables from the sender into incoming coupling effects for the receiver.
(\textit{b})~Message passing on the graph: at each iteration, every subsystem extracts its interface variables, gathers the states of its neighbors, constructs and aggregates the incoming messages, and updates in parallel with all other subsystems; no global augmented state is assembled and no system-wide covariance is propagated.
(\textit{c})~Interface message variants: deterministic messages exchange vector-valued interface variables, probabilistic messages exchange distributions over interface variables, and learned messages construct the interface law from data.
(\textit{d})~Inverse-problem setting: the true system and its parameters are not directly accessible; a forward model maps hidden states and parameters to observable outputs, only noisy and indirect measurements are available, and the inference task is to estimate hidden states and parameters with quantified uncertainty.}
\label{fig:framework}
\end{figure}

\section*{Results}
\subsection*{Framework overview}

We represent a coupled engineered system as a directed graph of interacting subsystems, in which each node carries a local model mapping subsystem states and parameters to observable outputs, and each edge carries an interface law that transforms outgoing interface variables from one subsystem into incoming coupling effects for another (Fig.~\ref{fig:framework}a).  Local models may be mechanistic, data-driven, or hybrid, and each can be paired with its own estimator maintaining a posterior over subsystem states and any unknown parameters. Interface variables are physically meaningful quantities such as forces, flows, currents, or torques exchanged across subsystem boundaries, and they form the primary channel through which coupling and uncertainty propagate across the graph.

This formulation differs from monolithic estimation in three structural respects (Fig.~\ref{fig:framework}a, b). First, no global augmented state is assembled; each subsystem updates only its own states and parameters, while coupling is represented entirely through interface messages exchanged along graph edges. Second, no system-wide covariance matrix is propagated; uncertainty remains local and crosses subsystem boundaries only through the interface messages. Third, the formulation does not require a common model or inference class across subsystems: mechanistic estimators, data-driven surrogates, deterministic controllers, and learned interaction laws can coexist within the same coupled inference architecture. Inference is therefore carried out as a collection of local subsystem updates coordinated by message passing on the graph.

This architecture provides, within one formulation, five capabilities that are central to the inverse problems considered here: it accommodates heterogeneous local models, propagates calibrated posterior uncertainty across subsystem interfaces, absorbs learned interaction laws when coupling physics is incomplete, scales to infrastructure-sized networks with approximately linear runtime growth, and composes hierarchically across scales of system organization (Fig.~\ref{fig:framework}b, c).
 The case studies that follow isolate these capabilities in progressively more demanding settings, showing how the same probabilistic compositional framework supports uncertainty-aware inference under sparse sensing, at network scale, and in heterogeneous multi-physics systems composed hierarchically across scales.

\begin{figure}[tbp]
\includegraphics[width=\textwidth]{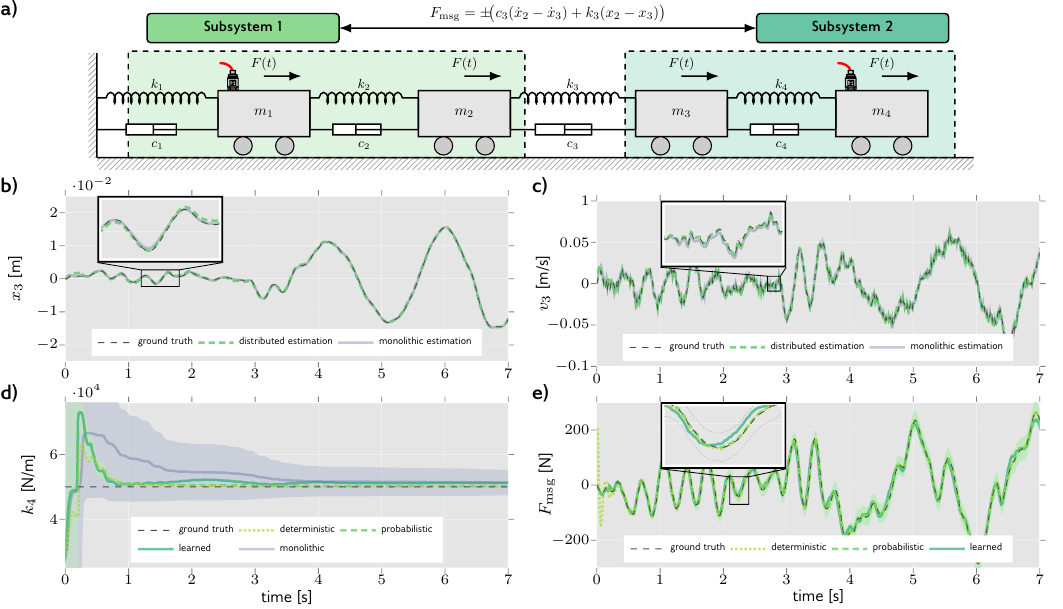}
\caption{\textbf{Probabilistic interface messages restore uncertainty calibration under sparse boundary sensing.}
(\textit{a})~Canonical testbed: a four-degree-of-freedom mass-spring-damper chain partitioned into two subsystems connected by an interface spring-damper pair ($k_3$, $c_3$), through which the interface force $F_{\mathrm{msg}}$ is exchanged. Each subsystem is observed only through a single boundary acceleration measurement; all interior states are latent and one stiffness parameter ($k_4$) is unknown.
(\textit{b},~\textit{c})~Hidden-state recovery at the unmeasured interior mass: displacement $x_3$ (\textit{b}) and velocity $v_3$ (\textit{c}) reconstructed by the distributed estimator (dashed) and the monolithic reference (solid band), both overlapping the ground truth under sparse boundary sensing.
(\textit{d})~Unknown stiffness parameter $k_4$ converging from biased initialization under four matched configurations: monolithic centralized unscented Kalman filter (UKF), deterministic Jacobi, probabilistic Jacobi, and the learned interaction law, with posterior uncertainty bounds where applicable.
(\textit{e})~Reconstructed interface force $F_{\mathrm{msg}}$ for the deterministic, probabilistic, and learned interface laws, overlaid on the ground truth; the probabilistic variant additionally carries the propagated force-uncertainty envelope used to restore calibration in the receiving subsystem.}
\label{fig:testbed}
\end{figure}

\subsection*{Probabilistic interface messages restore calibration under sparse sensing}

We isolate the central inference mechanism of the framework on a minimal coupled inverse problem: how uncertainty should be communicated across subsystem interfaces during joint state-and-parameter estimation. As a canonical testbed, we consider a four-degree-of-freedom mass-spring-damper chain partitioned into two interacting subsystems connected by a spring-damper interface (Fig.~\ref{fig:testbed}a). Each subsystem is observed only through a boundary acceleration measurement, leaving the interior states latent, and one stiffness parameter is treated as unknown and estimated online. Despite its simplicity, this system contains the essential ingredients of the general problem: subsystem decomposition, sparse sensing, coupling through an interface whose force depends on independently estimated interface states, and simultaneous estimation of hidden states and unknown parameters. To isolate the effect of inter-subsystem communication, we compare three otherwise matched estimators: a centralized unscented Kalman filter (UKF)~\cite{Julier2004Unscented} on the full augmented state, a distributed Jacobi~\cite{Ruozzi2013Messagepassing} scheme with deterministic mean-valued interface messages, and a distributed Jacobi scheme with probabilistic interface messages that transmit both mean and variance.  Observations, initialization, and filter hyperparameters are identical across all three cases, so performance differences reflect only how inter-subsystem information is represented and propagated.

With deterministic mean-valued messages, the distributed Jacobi estimator already recovers the latent dynamics and identifies the unknown stiffness under sparse boundary sensing (Fig.~\ref{fig:testbed}b--d). At each time step, the two subsystem filters update independently, exchange posterior means of the interface kinematics, and reconstruct from them a deterministic coupling force used as an external input in the subsequent prediction.  Deterministic messaging therefore transmits sufficient information to solve the coupled inverse problem at the level of point estimates. What it does not transmit is the uncertainty attached to the sender's estimate. The receiving subsystem conditions on a single interface value and therefore accounts only for local process and measurement uncertainty, not for uncertainty inherited through the coupling itself.  Deterministic messaging is thus uncertainty-blind at the interface.

To restore uncertainty awareness, we treat the interface message as a random variable. Each subsystem communicates the posterior mean of its interface-state estimate together with the associated covariance. For the linear spring-damper interface, the coupling force is a linear function of the relative interface displacement and velocity and, neglecting cross-covariances between subsystem interface states under the Jacobi exchange, its variance is
\begin{equation*}
  \mathrm{var}(F_b) = \mathbf{a}^\top (\mathbf{P}_{s_1} + \mathbf{P}_{s_2}) \mathbf{a},
\end{equation*}
where \(\mathbf{a}\) collects the interface stiffness and damping coefficients and \(\mathbf{P}_{s_1}\) and \(\mathbf{P}_{s_2}\) are the interface-state covariance submatrices of the two subsystems. The coupling-force uncertainty is therefore induced directly by the posterior uncertainty of the two interface-state estimates. This variance is then propagated into the receiving subsystem as an additional process-noise contribution on the interface-velocity component. Methods and the Supporting Information give the derivation and the incremental update rule that prevents repeated reinjection of previously accounted-for interface uncertainty. The key point is that uncertainty is not added heuristically; it is propagated through the interface law itself.

The three estimators perform similarly on point estimates but diverge in uncertainty calibration (Fig.~\ref{fig:testbed}b--d). State root mean squared error (RMSE) on the hidden internal states is $4.07\times10^{-5}$ for the centralized UKF, $2.02\times10^{-4}$ for deterministic Jacobi, and $1.05\times10^{-4}$ for probabilistic Jacobi, whereas parameter normalized RMSE (NRMSE) is $2.47\times10^{-2}$, $4.16\times10^{-3}$, and $4.65\times10^{-3}$, respectively. Probabilistic messaging is therefore not required to recover the latent dynamics or the unknown parameter; its effect appears in the statistical reliability of the posterior. At the nominal $95\%$ level, the centralized UKF and probabilistic Jacobi achieve empirical coverage of $1.00$ and $1.00$, both slightly conservative over this evaluation window, whereas deterministic Jacobi undercovers at $0.83$ because the receiver's posterior omits uncertainty transmitted through the interface. At the $68\%$ level, the centralized UKF and probabilistic Jacobi again remain conservative ($1.00$ and $1.00$), while deterministic Jacobi falls to $0.66$, below the nominal level. Time-averaged predictive negative log-likelihood, which penalizes both mean error and variance miscalibration, reinforces this distinction: it is $-4.29\times10^{1}$ for the centralized UKF, $-5.25\times10^{1}$ for probabilistic Jacobi, and $4.98\times10^{3}$ for deterministic Jacobi. The advantage of probabilistic messaging therefore lies not in improving point estimates, but in preserving calibrated uncertainty under subsystem decomposition.

The preceding comparison assumes that the interface law is known analytically. When the coupling physics is incomplete, the same architecture can accommodate a learned interaction law without changing the inference machinery. We therefore replace the analytical spring-damper relation by a sparse model~\cite{Brunton2016Discovering} identified from interface observations over a candidate library of linear, cubic, and dissipative terms, though any identification method producing an interface law would serve the same architectural role. The learned interaction recovers the dominant linear stiffness and damping components (\(\hat{k} = 5.61 \times 10^4\) N/m and \(\hat{c} = 3.31 \times 10^2\) Ns/m, within \(12.20\%\) and \(10.30\%\) of the true values), with only small higher-order corrections. Substituted into the probabilistic Jacobi estimator, the learned law leaves state RMSE, parameter NRMSE, and 95\% coverage at \(2.38 \times 10^{-3}\), \(2.39 \times 10^{-2}\), and \(1.00\), respectively, relative to the analytical-interface case (Fig.~\ref{fig:testbed}d, e). What changes is the interface model, not the message-passing schedule, the local estimators, or the mechanism by which interface uncertainty is propagated.

Together, these results show that distributed inference across subsystem interfaces is not merely a computational reformulation of a coupled inverse problem. Mean-only messages are sufficient to recover hidden states and unknown parameters under sparse sensing, but they cannot transmit the uncertainty attached to interface estimates and therefore yield miscalibrated posteriors. Probabilistic interface messages propagate that uncertainty through the coupling law and restore calibration while preserving the accuracy of distributed point estimates. The same architecture further accommodates learned interaction laws without altering the inferential scaffold. On this minimal testbed, probabilistic compositional inference thus integrates state estimation, parameter identification, uncertainty propagation, and interface learning within a single formulation.

\subsection*{Subsystem structure enables near-linear scaling without loss of calibration}

\begin{figure}[tbp]
\includegraphics[width=\textwidth]{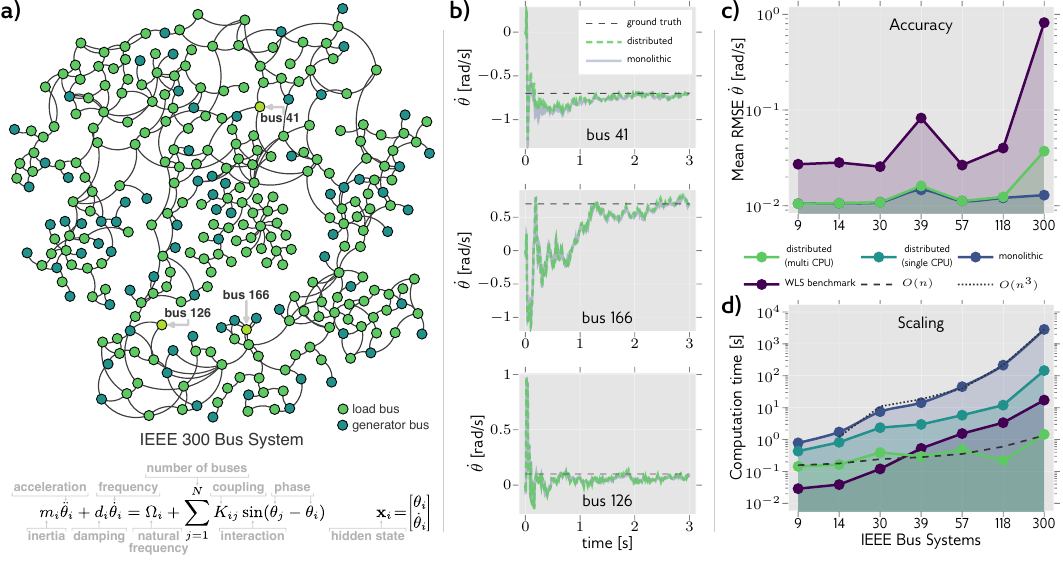}
\caption{\textbf{Subsystem structure enables near-linear computational scaling without loss of inferential accuracy or calibration.}
(\textit{a})~The IEEE 300-bus benchmark network with load and generator buses. Each bus is modeled as a second-order Kuramoto oscillator with inertia, damping, and admittance-weighted coupling; the phase and frequency at every bus are hidden states and the natural frequencies are unknown parameters. The three highlighted buses (41, 126, 166) are examined in (\textit{b}).
(\textit{b})~Representative frequency trajectories $\dot{\theta}$ at the three highlighted buses of the 300-bus network: the distributed estimate (dashed) overlaps the monolithic centralized estimate (solid band) and converges to the ground truth (dashed black), confirming at the level of individual trajectories the aggregate accuracy reported in (\textit{c}).
(\textit{c})~Accuracy across benchmark sizes (9 to 300 buses): mean state root mean squared error (RMSE) on the bus frequencies for the distributed estimator, the monolithic centralized UKF, and the weighted least squares (WLS) benchmark. The distributed estimator matches the centralized UKF across all network sizes; WLS addresses only static state estimation and does not solve the joint dynamic state-and-parameter inference problem.
(\textit{d})~Runtime scaling: the centralized UKF exhibits approximately cubic growth with network size, whereas the distributed estimator scales approximately linearly. Sequential single-core measurements and projected parallel execution with one core per subsystem are shown for the distributed estimator; the parallel projection is obtained by dividing the aggregate sequential cost by the number of subsystem filters.}
\label{fig:ieee}
\end{figure}

To test whether probabilistic compositional inference remains accurate, uncertainty-aware, and computationally tractable at infrastructure scale, we evaluate the framework on standard IEEE power-network benchmarks ranging from 9 to 300 buses~\cite{Thurner2018Pandapower}. Each bus is modeled as a second-order Kuramoto oscillator~\cite{Filatrella2008Analysis} with inertia and damping, coupled to neighboring buses through the electrical admittance matrix. The resulting network dynamics define a joint dynamic state-and-parameter inference problem in which the phase and frequency states at all buses, together with the unknown natural frequencies, are inferred from noisy phase and frequency measurements. We compare two Kalman-based estimators directly on this task: a centralized UKF on the full augmented state, and the proposed distributed estimator on a partitioned network (Fig.~\ref{fig:ieee}a).  Weighted least squares (WLS)~\cite{Cohen2017Optimal} is included only as a reference for static state estimation; it does not infer the dynamic natural frequencies and therefore does not support posterior calibration analysis. Observations, initialization, and filter hyperparameters are matched across the two Kalman-based methods.

The distributed estimator matches the centralized UKF across all benchmark sizes on both state and parameter NRMSE, with maximum absolute deviations of $2.67\times10^{-3}$ on state NRMSE and $3.62\times10^{-2}$ on parameter NRMSE from 9 to 300 buses (Fig.~\ref{fig:ieee}c and Supporting Information); representative frequency trajectories at three buses of the 300-bus network confirm that the distributed and centralized estimates are visually indistinguishable (Fig.~\ref{fig:ieee}b). Empirical coverage at the nominal 95\% level remains close to nominal for both estimators across all network sizes, with values of $0.97$ for the distributed estimator and $0.98$ for the centralized UKF in the 300-bus case. WLS remains competitive for static state estimation, consistent with its established role for operating quantities such as phase angles, but it does not address the full dynamic joint state-and-parameter inference problem. The distributed formulation therefore scales to infrastructure-sized networks without sacrificing the accuracy or calibration of the centralized reference.

Computational behavior separates the two Kalman-based estimators sharply (Fig.~\ref{fig:ieee}d). The centralized UKF exhibits approximately cubic scaling with network size, reflecting covariance factorization and sigma-point propagation on the full augmented state. The distributed estimator scales approximately linearly: each subsystem filter operates on a bounded local augmented state of at most \(3S_{\max}=15\) variables, so total work grows with the number of subsystems rather than with the cube of the global state dimension. Even under sequential single-core execution, the distributed estimator is already approximately $19\times$ faster than the centralized UKF in the 300-bus case; under parallel execution with one core per subsystem, the projected speedup rises to approximately three orders of magnitude, because each Jacobi iteration is then governed by the slowest subsystem filter rather than by the sum of all subsystem runtimes. The gain arises from exploiting subsystem structure rather than altering the underlying estimator. Probabilistic compositional inference therefore delivers calibrated joint dynamic state-and-parameter inference while reducing computational scaling from cubic to near-linear.

\begin{figure}[tbp]
\includegraphics[width=\textwidth]{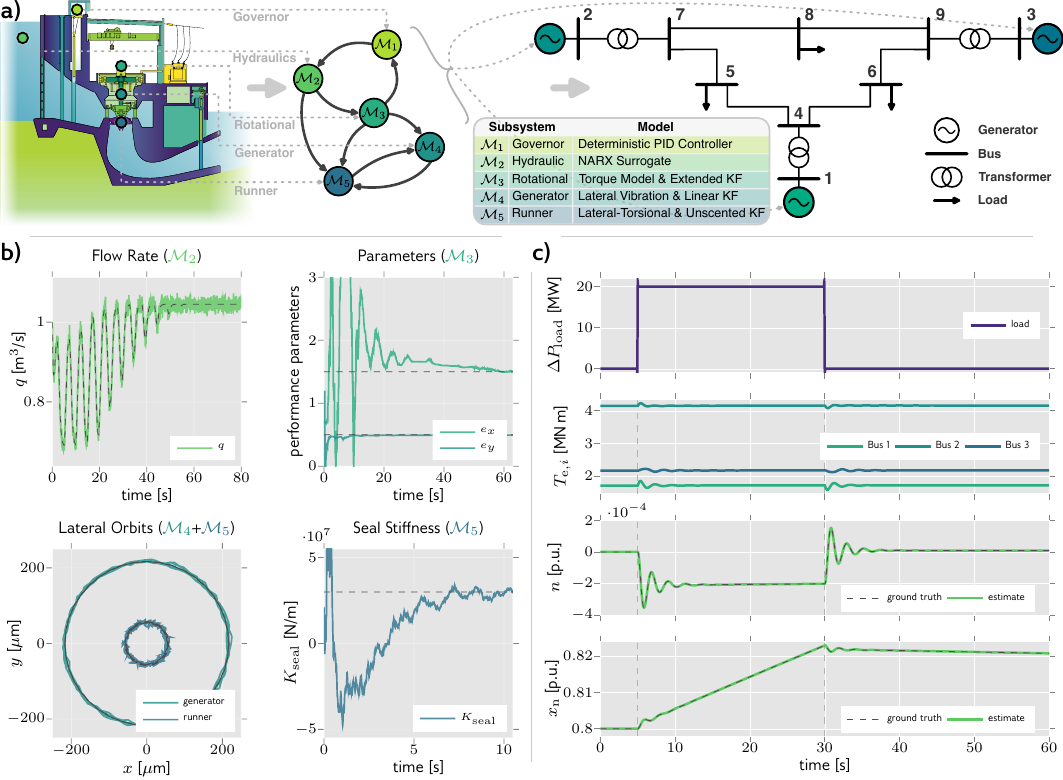}
\caption{\textbf{Heterogeneous and hierarchical composition across physics and scale.}
(\textit{a})~System architecture. Left: the hydroelectric turbine-generator unit is decomposed into five interacting subsystems exchanging directed interface messages, with deliberately heterogeneous local models: a deterministic proportional-integral-derivative (PID) governor ($\mathcal{M}_1$), a nonlinear autoregressive with exogenous inputs (NARX) hydraulic surrogate ($\mathcal{M}_2$), an extended Kalman filter (EKF) rotational subsystem ($\mathcal{M}_3$), a Kalman filter generator lateral-vibration subsystem ($\mathcal{M}_4$), and a UKF runner lateral-torsional subsystem ($\mathcal{M}_5$). Right: hierarchical embedding in which three turbine instances replace the generator buses of the IEEE 9-bus network; inter-cluster messages are exchanged through the same Jacobi schedule that operates within each turbine.
(\textit{b})~Standalone turbine inference: flow rate $q$ tracked by the hydraulic surrogate ($\mathcal{M}_2$); cross-physics torque coefficients $\tilde e_x$, $\tilde e_y$ ($\mathcal{M}_3$) and seal stiffness $\tilde K_{\text{seal}}$ ($\mathcal{M}_5$) converging from biased initial estimates toward their true values (dashed); and generator and runner lateral orbits ($\mathcal{M}_4{+}\mathcal{M}_5$) reconstructed by the two structural filters.
(\textit{c})~Cross-scale disturbance propagation in the hierarchical configuration: a $+20$~MW load step at bus~9, applied at $t=5$~s and removed at $t=30$~s, drives the grid-induced electrical torque $T_{e,i}$ at the three generator buses and the resulting turbine-level speed deviation $n$ and needle position $x_n$ at the most-affected generator, where the embedded estimates (solid) overlap the ground truth (dashed).}
\label{fig:turbine}
\end{figure}

\subsection*{Heterogeneous and hierarchical composition across physics and scale}

To test whether probabilistic compositional inference can combine fundamentally different local models within a single architecture, we apply the framework to a hydroelectric turbine-generator unit~\cite{Xu2017Sensitivity} decomposed into five interacting subsystems spanning control, hydraulics, rotational dynamics, generator vibration, and runner dynamics (Fig.~\ref{fig:turbine}a).  The subsystem models are deliberately heterogeneous: they combine a deterministic proportional-integral-derivative (PID) controller for the governor, a learned nonlinear autoregressive with exogenous inputs (NARX)~\cite{Schar2025Surrogate, Tcheumchoua2022Torque} surrogate for hydraulics, and probabilistic estimators of different classes for the rotational and structural subsystems. Despite this heterogeneity, the coupled inference proceeds through the same Jacobi message-passing schedule used in the previous case studies, without architectural changes or special-case logic for learned, deterministic, or probabilistic components. Starting from biased initial parameter estimates, the joint estimator recovers latent dynamics and unknown parameters across all five subsystems: the aggregate estimation RMSE on the lateral displacements (\(x_G\), \(y_G\), \(x_R\), \(y_R\)) is \(9.14\times10^{-6}\)~m, the seal stiffness and the cross-physics torque coefficients \(\tilde e_x\) and \(\tilde e_y\) converge from biased initializations toward their true values with trajectory NRMSEs of \(5.25\times10^{-1}\), \(1.19\), and \(8.23\times10^{-2}\) dominated by the initial transients, and the hydraulic surrogate maintains a one-step-ahead RMSE of \(2.42\times10^{-3}\) during online inference (Fig.~\ref{fig:turbine}b). Learned, deterministic, and probabilistic components thus operate within one coupled inferential scaffold, each contributing in its native form.

We next test whether the same architecture composes across scales by embedding the multi-physics turbine within an electromechanical grid network. The three generator buses of the IEEE 9-bus~\cite{Thurner2018Pandapower} system are replaced by full five-subsystem turbine networks, while the six non-generator buses retain reduced first-order dynamics, yielding a hierarchical system with three turbine subnetworks nested within an outer grid graph (Fig.~\ref{fig:turbine}a). A load step at bus 9 raises demand by 20 MW at \(t = 5\) s and restores it at \(t = 30\) s, driving a disturbance that propagates from the grid into the embedded turbine rotational subsystems and then through the governor, hydraulic, and structural components before feeding back into the network through the generator-bus power injections (Fig.~\ref{fig:turbine}c). The same Jacobi schedule operates at both levels: every interface variable, from a coupling force within a runner subsystem to an electrical power exchange between generator buses, is communicated through the same mechanism. The compositional architecture therefore operates without distinction across local multi-physics coupling and network-level interaction.

To test whether this hierarchical embedding degrades local inference, we compare each embedded turbine estimate with a standalone reference driven by a grid-induced torque trajectory at the corresponding generator bus, obtained from a separate grid-only simulation. Aggregated across all three generators, the embedded estimator matches the standalone reference on speed deviation \(n\), needle position \(x_n\), and flow rate \(q\) with an aggregate trajectory RMSE of \(1.02\times10^{-3}\), and the embedded estimates remain visually indistinguishable from the ground-truth trajectories at the most-affected generator (Fig.~\ref{fig:turbine}c). Each subsystem retains its own state, covariance, and parameter inference throughout; composition is realized through the exchanged interface messages rather than through collapse into a single global estimator.

\subsection*{Subsystem structure as the organizing principle}

Across the three case studies, probabilistic compositional inference exploits the same architectural principle in progressively more demanding regimes. Under sparse boundary sensing, subsystem decomposition suffices to recover hidden dynamics at the level of point estimates, and probabilistic messaging is what restores calibration and accommodates learned interaction laws. At infrastructure scale, the same subsystem structure that preserves inferential accuracy also reduces computational scaling from cubic to near-linear. In the hierarchical multi-physics setting, that structure further allows learned, deterministic, and probabilistic components to compose across scales without collapsing local posteriors into a global estimate. None of these capabilities is delivered by the alternatives. Monolithic data-assimilation methods deliver calibrated joint inference but do not compose heterogeneous components and scale unfavorably with system size; modular co-simulation frameworks preserve heterogeneity and scale but do not carry uncertainty across subsystem interfaces; graph-based learned dynamics absorb coupling structure but replace rather than compose with mechanistic components and do not by themselves produce calibrated posteriors for hidden states and unknown parameters. What the three case studies establish jointly is that subsystem structure, rather than any particular estimator or surrogate, is the organizing principle on which scalable, heterogeneous, uncertainty-aware inverse inference in coupled engineered systems can be built. The contribution is architectural: probabilistic compositional inference does not introduce a new filter, surrogate, or optimization method, but reorganizes how coupled inverse problems are formulated so that existing components can be assembled into one coherent inference whose structure mirrors the system being inferred.

\section*{Discussion}
 
Probabilistic compositional inference is not a refinement of any one existing inference technique, but a different way of organizing inverse modeling in coupled systems. Across a canonical mechanical testbed, infrastructure-scale power networks, and a multi-physics turbine embedded within a power-grid network, the same architecture supports sparse-sensing inference with calibrated uncertainty, scales to networks of several hundred coupled components with near-linear computational growth, and composes mechanistic, deterministic, and learned components hierarchically without forcing them into a common representation. The framework introduces no new filter, surrogate, or optimization method. Instead, it provides a way to assemble existing components into a single coupled inference whose graph structure mirrors that of the system being inferred. Because the architecture constrains only how subsystems exchange interface information, rather than which local models or estimators they must use, it can absorb advances in any component method, whether a more accurate filter, a more expressive surrogate, or a more efficient distributed schedule, without changing the overall design. Probabilistic compositional inference is therefore best understood not as a particular inference algorithm, but as an organizing principle for inverse modeling in coupled systems.
 
Two broader research directions in inverse problems under uncertainty are especially relevant to this work. The first is the exploitation of mathematical and physical structure as a route to scalable and well-posed inference, an idea that underlies reduced-order modeling~\cite{Benner2015Survey}, multifidelity methods~\cite{Peherstorfer2018Survey}, and structure-preserving inference more broadly~\cite{Yang2020BPINNs}. Probabilistic compositional inference engages this direction at the architectural level: rather than searching for exploitable structure inside a monolithic state-space formulation, it treats the subsystem decomposition of an engineered system as the inferential structure itself, so that the interaction graph mirrors the physical organization of the problem. The second direction is the integration of heterogeneous data, models, and physics within a single inferential workflow, which has become increasingly important as inverse problems must combine measurements and simulators that differ in type, fidelity, and provenance~\cite{Carter2023Advanced}. Here the framework contributes by allowing mechanistic, learned, and deterministic components to coexist within one coupled inference without forcing them into a common mathematical representation, while preserving each component in its native form. In this sense, probabilistic compositional inference contributes to a class of inverse methods that treat problem structure not as an after-the-fact computational convenience, but as a primary design principle.
 
This positioning should not obscure several genuine limitations of the framework as presented here. Its design assumes that subsystem boundaries are physically identifiable and that the interface variables exchanged between subsystems are well defined. Coupled systems whose boundaries are implicit, diffuse, or not readily associated with physically meaningful interface variables are therefore harder to represent in this form, and extending the framework to such settings remains an open problem. In addition, the subsystem partition is taken as given throughout this work, whether derived from known physical structure or imposed by a fixed graph-partitioning rule. The present work does not address how such a partition should be learned from data or adapted as the system evolves. Finally, although the framework propagates uncertainty across subsystem interfaces, it does not create calibration on its own: the quality of the resulting coupled inference still depends on the calibration of the local estimators. A natural next step is therefore to move beyond fixed subsystem decompositions toward settings in which the partition itself is inferred jointly with states, parameters, and couplings, allowing the architecture to discover the structure on which it operates rather than requiring it as a precondition.

\section*{Materials and Methods}
\paragraph*{Graph-theoretic formulation}
 
We represent a coupled engineered system as a directed graph \(G=(V,E)\) in which each node \(i\in V\) carries a local model \(\mathcal{M}_i\) and a local estimator over subsystem states \(\mathbf{x}_i\), exogenous inputs \(\mathbf{u}_i\), unknown parameters \(\boldsymbol{\theta}_i\), and observable outputs \(\mathbf{y}_i\),
\begin{equation*}
\mathbf{y}_i = \mathcal{M}_i(\mathbf{x}_i,\mathbf{u}_i;\boldsymbol{\theta}_i,\mathbf{m}_i^{\mathrm{in}}),
\end{equation*}
where \(\mathbf{m}_i^{\mathrm{in}}\) is the aggregate incoming interface information. Each directed edge \((j\rightarrow i)\in E\) carries an interface law
\begin{equation*}
\mathbf{m}_{j\rightarrow i}=h_{ji}(\mathbf{s}_{j\rightarrow i},\mathbf{s}_{i\leftarrow j};\boldsymbol{\gamma}_{ji}),
\end{equation*}
with sender- and receiver-side interface variables \(\mathbf{s}\) and interaction parameters \(\boldsymbol{\gamma}_{ji}\). Interface variables are physically meaningful quantities exchanged across subsystem boundaries; in the three case studies considered below, they take the form of coupling forces across a spring-damper interface, electrical power exchanges between generator buses, and shaft torques transmitted between a turbine governor and its rotational subsystem. Local models may be mechanistic, data-driven, or hybrid, and different nodes need not share a model class. Inference proceeds by maintaining local posteriors over \((\mathbf{x}_i,\boldsymbol{\theta}_i)\) updated through message passing on \(G\), rather than assembling a global augmented state and covariance~\cite{Zechner2016Molecular, Singh2020Scalable}.

\paragraph*{Local subsystem estimators}
 
Each node is assigned an estimator class specified by its local dynamics, sensing configuration, and observability structure. For the canonical testbed, we compare three configurations under matched observations: a centralized UKF~\cite{Julier2004Unscented} on the full augmented state, a distributed Jacobi schedule in which each subsystem runs its own UKF and exchanges posterior means only, and a distributed Jacobi schedule in which each subsystem UKF additionally exchanges interface variances. For the IEEE benchmark networks, the bus set is partitioned into disjoint generator-anchored subsystems of size at most \(S_{\max}=5\) by a greedy expansion procedure defined on the admittance-weighted coupling graph (Supporting Information), so that each subsystem runs a UKF on a bounded augmented state of dimension at most \(3S_{\max}=15\). The centralized reference runs a UKF on the full network-level augmented state. WLS is included only as a scope-restricted static state-estimation benchmark. For the multi-physics turbine case study, the five subsystems are represented by estimators drawn from fundamentally different classes: a deterministic PID controller~\cite{ASTROM1995PID} for the governor \(\mathcal{M}_1\), a NARX neural-network surrogate~\cite{Yu2019Nonlinear} for the hydraulic subsystem \(\mathcal{M}_2\), an extended Kalman filter (EKF)~\cite{Wan2001Dual} for the rotational subsystem \(\mathcal{M}_3\), a Kalman filter~\cite{Kalman1960New} for the generator lateral-vibration subsystem \(\mathcal{M}_4\), and a UKF for the runner lateral-torsional subsystem \(\mathcal{M}_5\). Complete state-space definitions, observation models, parameter sets, filter hyperparameters, and the partitioning algorithm are provided in the Supporting Information.
 
\paragraph*{Probabilistic interface message construction}
 
The framework propagates uncertainty across subsystem boundaries by mapping local posterior uncertainty through the interface law. When the interface variables extracted from two neighboring subsystems carry posterior uncertainty, the induced uncertainty in the interface quantity is propagated into the receiving subsystem as an additional process-noise contribution to the local state component directly affected by that interface. For the canonical testbed, the two subsystems interact through the interface force \(F_b = k_3(x_2-x_3)+c_3(\dot{x}_2-\dot{x}_3) = \mathbf{a}^{\top}\delta\mathbf{z}\), with \(\mathbf{a}=[k_3,c_3]^{\top}\), exchanged as the message \(F_{\mathrm{msg}}\) in Fig.~\ref{fig:testbed}. Neglecting cross-covariances between subsystem interface states under the Jacobi exchange, the force variance reduces to
\begin{equation*}
\mathrm{var}(F_b)=\mathbf{a}^{\top}(\mathbf{P}_{s_1}+\mathbf{P}_{s_2})\mathbf{a},
\end{equation*}
where \(\mathbf{P}_{s_1}\) and \(\mathbf{P}_{s_2}\) are the interface-state covariance submatrices extracted from the two subsystem UKFs. This variance is added to the receiving subsystem's process-noise covariance on the interface-velocity component; to avoid repeated reinjection of previously accounted-for uncertainty, only the incremental change between successive filter steps is added at each step. The same construction applies in the IEEE benchmark networks and in the multi-physics turbine case study, with interface variables, affected state components, and process-noise updates defined by the local physics of each subsystem. The full derivation, incremental update rule, and numerical details are given in the Supporting Information.
 
\paragraph*{Distributed inference schedule}
 
Subsystem updates are coordinated by a Jacobi message-passing schedule~\cite{Ruozzi2013Messagepassing}, in which all subsystems are updated in parallel from interface messages computed at the previous iteration. At each global time step the schedule performs \(K\) inner iterations; \(K=1\) suffices for the canonical testbed and the IEEE benchmark networks, while \(K=2\) is used in the hierarchical turbine-grid case study to resolve the fast hydraulic-rotational coupling within each sweep. Because all subsystem updates within an iteration depend only on messages from the previous iteration, they are conditionally independent and can be executed in parallel. Sequential Gauss-Seidel and multi-step Adams-Bashforth schedules~\cite{Yedidia2011MessagePassing} are compatible with the same framework; an empirical comparison is reported in the Supporting Information.
 
\paragraph*{Learned interaction laws}
 
When an analytical interface law is unavailable or only partially specified, the framework identifies that law from interface observations using sparse identification of nonlinear dynamics (SINDy)~\cite{Brunton2016Discovering}. For the canonical testbed, the relative interface displacement and velocity are reconstructed from interface acceleration data and regressed against a candidate library of linear, cubic, quadratic-dissipative, mixed, and constant terms using sequentially thresholded least squares, which retains only the dominant contributions. The preprocessing pipeline, explicit library, training protocol, and recovered coefficients are reported in the Supporting Information. Once identified, the learned interaction law replaces the analytical edge law in the graph and is used by the same probabilistic interface construction and Jacobi schedule described above, without modification to the remainder of the distributed estimator.
 
\paragraph*{Benchmarking protocol}
 
All numerical experiments are implemented in Python. The canonical testbed consists of two coupled subsystems with synthetic boundary acceleration measurements at masses 1 and 4, hidden internal states at masses 2 and 3, and one unknown stiffness parameter \(k_4\) inferred jointly with the states. The IEEE benchmark networks range from 9 to 300 buses, modeled as second-order Kuramoto oscillators with noisy phase and frequency measurements at every bus and unknown natural frequencies inferred jointly with the dynamic states; partitioning uses PYPOWER case data~\cite{Thurner2018Pandapower}. The multi-physics turbine case study uses a hydroelectric turbine-generator unit decomposed into five subsystems; in the hierarchical configuration, three turbine instances replace the generator buses of the IEEE 9-bus network and a \(+20/-20\) MW load step at bus 9 is used as the disturbance.
 
Baselines are matched per case study: a centralized UKF and a deterministic Jacobi scheme for the canonical testbed; a centralized UKF and a scope-restricted WLS reference for the IEEE benchmarks; and a standalone turbine estimator driven by a grid-induced torque trajectory from a separate grid-only simulation for the hierarchical case. Within each case study, observations, initialization, simulation horizon, and estimator hyperparameters are held fixed across the compared methods so that performance differences reflect the inference formulation rather than tuning choices.
 
State accuracy is reported as RMSE on unmeasured or target variables, aggregated over states and time steps when multiple variables are reported jointly. When comparisons are made across quantities with different physical units or magnitudes, we report NRMSE, computed as the RMSE divided by the range of the corresponding ground-truth signal; for constant parameters, whose ground-truth range is zero, the RMSE is divided by the true parameter value instead. Parameter accuracy is reported analogously over the reported parameter trajectories or terminal estimates; where trajectory NRMSEs are reported, they reflect the initial bias and transient as well as the converged estimate. Calibration is assessed through empirical credible-interval coverage and time-averaged predictive negative log-likelihood. Runtime is reported as wall-clock time measured under sequential single-core execution; for the distributed framework, parallel one-core-per-subsystem execution is additionally projected by dividing the aggregate sequential cost by the number of subsystem filters. All reported metrics are computed from a single noise realization with a fixed random seed per case study (Supporting Information).

\section*{Data Availability}
All study data, simulation code, and analysis scripts are available at \url{https://github.com/cisgroup/compositional-inference}.

\section*{Acknowledgments}
This work has been supported by a grant from the Fund for Energy Research with Corporate Partners administered by the Andlinger Center for Energy and the Environment at Princeton University.

\putbib[references]
\end{bibunit}

\clearpage
\setcounter{figure}{0}\renewcommand{\thefigure}{S\arabic{figure}}
\setcounter{table}{0}\renewcommand{\thetable}{S\arabic{table}}
\setcounter{equation}{0}\renewcommand{\theequation}{S\arabic{equation}}

\begin{center}
  {\LARGE\bfseries Supplementary Information}\\[1.5ex]
  {\large Subsystem Structure as an Inferential Resource for Coupled Engineered Systems}
\end{center}
\bigskip

\begin{bibunit}[unsrtnat]

\section*{Introduction}
This supporting information accompanies the main manuscript and provides the theoretical and mathematical
foundations required to reproduce the results reported there, together with additional studies that further
probe the proposed probabilistic compositional inference architecture.
The notation follows the main manuscript throughout: each subsystem $i$ carries a local model $\mathcal{M}_i$
over states $\mathbf{x}_i$, exogenous inputs $\mathbf{u}_i$, and parameters $\boldsymbol{\theta}_i$, and each
directed edge $(j \rightarrow i)$ carries an interface law $h_{ji}$ that maps the interface variables
$\mathbf{s}_{j \rightarrow i}$ and $\mathbf{s}_{i \leftarrow j}$ into the message $\mathbf{m}_{j \rightarrow i}$
exchanged between the subsystems.

The document is organized in two parts.
The first part presents the theoretical background: the message-passing schedules (Jacobi, Gauss--Seidel, and
Adams--Bashforth) reinterpreted as communication protocols among subsystems, the local estimator classes
(linear, extended, and unscented Kalman filters), the NARX surrogate-modeling framework, and the diffusion
models used for fast uncertainty propagation on subsystem graphs.
The second part documents the three case studies of the main manuscript in their order of appearance: the
canonical mass--spring testbed under sparse sensing, the IEEE power-network scalability benchmarks, and the
multi-physics turbine--generator system together with its hierarchical embedding in the IEEE 9-bus network.
For each case study, the complete state-space models, observation models, parameter sets, filter
hyperparameters, and message-passing protocols are specified, and the quantitative results summarized in the
main text are substantiated and extended.

Beyond the experiments of the main manuscript, the supplement reports complementary studies, including an
empirical comparison of alternative message-passing schedules, scalability experiments on extended
spring--mass chains, and a six-DOF chain with diffusion-based uncertainty propagation.
All reported metrics are computed from a single noise realization with a fixed random seed per case study, and
references are provided throughout to guide the reader to more comprehensive treatments of the underlying
methods.

\section*{Theoretical background}
This section presents the algorithmic building blocks used throughout this work: the message-passing schemes
that coordinate the subsystem updates, the Kalman-filter family used for local state and parameter estimation,
the NARX surrogate class, and the diffusion models used for fast uncertainty propagation.

\subsection*{Message passing algorithms}

In this study, we employ classical iterative and time-integration techniques (Jacobi, Gauss--Seidel, and Adams--Bashforth) not in their conventional roles, but rather as structured message-passing schemes~\cite{Ascher1995ImplicitExplicit}. 
These methods provide well-established convergence and stability properties that naturally map onto our message-passing framework, where subsystems exchange information asynchronously or synchronously and update their states according to the chosen update strategy. 
By reinterpreting these classical methods as communication and update protocols, we leverage their mathematical maturity while maintaining compatibility with the decomposed system architecture.
These formulations enable consistent propagation of both deterministic states and uncertainty across coupled
systems.
Although we only introduce three message-passing schemes, many other variants and extensions exist in the
literature~\cite{Yedidia2011MessagePassing,Gilmer2017Neural}. 
To formalize the three update schemes introduced above, we first define a common node-level model and message notation.

We assume that each node \(i\) carries a local model \(\mathcal{M}_{i}\) which maps subsystem states
\(\mathbf{x}_{i}\) and exogenous inputs \(\mathbf{u}_{i}\) to output quantities \(\mathbf{y}_{i}\), given model
parameters \(\boldsymbol{\theta}_i\), consistent with the graph-theoretic formulation in the main manuscript.
This formulation accommodates both static maps and dynamic state transitions (\textit{e.g.}, where $\mathbf{y}_i$
represents the state at the next time step). The ideal system evolution is given by:
\begin{equation}
\mathbf{y}_{i} = \mathcal{M}_i \left( \mathbf{x}_{i}, \mathbf{u}_{i} ; \boldsymbol{\theta}_i, \mathbf{m}^{\mathrm{in}}_{i} \right),
\end{equation}
where $\mathbf{m}^{\mathrm{in}}_{i}$ collects the incoming interaction messages from the neighborhood
$\mathcal{N}_i$ of node $i$, \textit{i.e.}, the set of subsystems whose edges are directed into $i$.
We explicitly account for the fact that local models are often approximations (surrogates) of the true physics or
an underlying model.
Therefore, the actual output is governed by the surrogate model $\tilde{\mathcal{M}}_i$ augmented with uncertainty
terms:
\begin{equation}
\mathbf{y}_i
= \tilde{\mathcal{M}}_i\!\left(\mathbf{x}_i,\mathbf{u}_i;\boldsymbol{\theta}_i,\mathbf{m}_i^{\mathrm{in}}\right)
+ \mathbf{w}_i + \boldsymbol{\epsilon}_i,
\end{equation}
where $\mathbf{w}_{i}$ represents aleatoric process noise (\textit{e.g.}, discretization error, random forcing) and
$\boldsymbol{\epsilon}_{i}$ explicitly captures epistemic model-form uncertainty (the discrepancy between the
surrogate $\tilde{\mathcal{M}}_i$ and the true physics or underlying model).

\subsubsection*{Jacobi message passing}

In the Jacobi message-passing scheme~\cite{Ding2025New}, all subsystems are updated in parallel using messages
computed from the previous iteration.
At iteration $k$, each subsystem $i$ receives incoming messages $\mathbf{m}_{j \to i}^{(k-1)}$ from its neighboring
subsystems $j \in \mathcal{N}_i$, which are constructed based on the interface states from the previous iteration.
Following the interaction model defined in the main manuscript, the message from subsystem $j$ to subsystem $i$ is
given by
\begin{equation}
\mathbf{m}_{j \to i}^{(k-1)} = h_{ji}\big(\mathbf{s}_{j \to i}^{(k-1)}, \mathbf{s}_{i \leftarrow j}^{(k-1)}; \boldsymbol{\gamma}_{ji}\big),
\label{eq:mp-form}
\end{equation}
where $\mathbf{s}_{j \to i}$ collects the sender-side interface variables of subsystem $j$ (\textit{e.g.}, boundary
displacements and velocities), extracted from its local state by a projection operator, $\mathbf{s}_{i \leftarrow j}$
collects the corresponding receiver-side interface variables of subsystem $i$, $h_{ji}$ is the interface law
attached to the directed edge $(j \rightarrow i)$, $\boldsymbol{\gamma}_{ji}$ denotes its interaction parameters,
and the total incoming message to node $i$ is obtained through aggregation of all neighboring messages.
Each subsystem then updates its state independently by solving its local model
\begin{equation}
\mathbf{y}_i^{(k)} = \mathcal{M}_i\big(\mathbf{x}_i, \mathbf{u}_i; \boldsymbol{\theta}_i, \mathbf{m}_i^{\mathrm{in},(k-1)}\big),
\end{equation}
where $\mathbf{m}_i^{\mathrm{in},(k-1)} = \mathcal{A}\big(\{\mathbf{m}_{j \to i}^{(k-1)}\}_{j \in
\mathcal{N}_i}\big)$ denotes the aggregated incoming messages.
This formulation corresponds to a fully parallel update scheme, where all subsystems use only information from the
previous iteration, ensuring no intra-iteration dependency between subsystems.
As a result, the Jacobi scheme is naturally suited for distributed and parallel implementations, as illustrated in
Algorithm~\ref{AL:jacobi-mp-algorithm}, where interface forces are computed using lagged states and each subsystem
is solved independently at every iteration.
\begin{algorithm}[H]
\caption{Jacobi message passing}
\label{AL:jacobi-mp-algorithm}
\begin{algorithmic}[1]
\For{$k = 1,2,\dots,\texttt{max\_iter}$}

    \Statex \hspace{1em} \textit{// exchange previous interface variables}
    \For{each subsystem $i$}
        \State send $\mathbf{s}_{i \to j}^{(k-1)}$ to all $j \in \mathcal{N}_i$
    \EndFor

    \Statex \hspace{1em} \textit{// construct incoming messages using previous iteration values}
    \For{each subsystem $i$}
        \For{each $j \in \mathcal{N}_i$}
            \State receive $\mathbf{s}_{j \to i}^{(k-1)}$
            \State compute
            \[
            \mathbf{m}_{j \to i}^{(k-1)}
            =
            h_{ji}\big(
            \mathbf{s}_{j \to i}^{(k-1)},
            \mathbf{s}_{i \leftarrow j}^{(k-1)};
            \boldsymbol{\gamma}_{ji}
            \big)
            \]
        \EndFor
        \State aggregate incoming messages
        \[
        \mathbf{m}_{i}^{\mathrm{in},(k-1)}
        =
        \mathcal{A}\big(\{\mathbf{m}_{j \to i}^{(k-1)}\}_{j \in \mathcal{N}_i}\big)
        \]
    \EndFor

    \Statex \hspace{1em} \textit{// update all subsystems independently in parallel}
    \For{each subsystem $i$}
        \State solve
        \[
        \mathbf{y}_i^{(k)}
        =
        \mathcal{M}_i\big(
        \mathbf{x}_i, \mathbf{u}_i;
        \boldsymbol{\theta}_i,
        \mathbf{m}_i^{\mathrm{in},(k-1)}
        \big)
        \]
    \EndFor

\EndFor
\end{algorithmic}
\end{algorithm}

In Algorithm~\ref{AL:jacobi-mp-algorithm}, $\mathcal{M}_i(\cdot)$ denotes the local dynamical model of subsystem
$i$, and $h_{ji}(\cdot)$ denotes the interface law that constructs the message from subsystem $j$ to subsystem $i$.
The quantity $\mathbf{s}_{i \to j}$ represents the interface variables transmitted from subsystem $i$ to subsystem
$j$, $\mathbf{y}_i$ denotes the state or output vector of subsystem $i$, and $\mathbf{m}_{j \to i}$ is the message
sent from subsystem $j$ to subsystem $i$.
The operator $\mathcal{A}(\cdot)$ aggregates all incoming messages from neighboring subsystems.
The parameter \texttt{max\_iter} denotes the maximum number of inner message-passing iterations performed at each
global time step to resolve the coupling among subsystems and obtain an approximately consistent subsystem
response.
\subsubsection*{Gauss--Seidel message passing}

In the Gauss--Seidel message-passing scheme~\cite{Shang2009distributed}, subsystems are updated sequentially, and
each subsystem uses the most recently available messages from previously updated neighboring subsystems within the
same iteration.
In contrast to the Jacobi scheme, this introduces an intra-iteration dependency, since the update of one subsystem
may immediately affect the update of the next.
Following the interaction model defined in Eq.~\ref{eq:mp-form}, the message from subsystem $j$ to subsystem $i$ at
iteration $k$ is written as
\begin{equation}
\mathbf{m}_{j \to i}^{(k)}
=
h_{ji}\big(
\mathbf{s}_{j \to i}^{(k_j)},
\mathbf{s}_{i \leftarrow j}^{(k-1)};
\boldsymbol{\gamma}_{ji}
\big),
\end{equation}
where $k_j = k$ if subsystem $j$ has already been updated in the current iteration, and $k_j = k-1$ otherwise.
The aggregated incoming message for subsystem $i$ is then computed as
\begin{equation}
\mathbf{m}_i^{\mathrm{in},(k)}
=
\mathcal{A}\big(\{\mathbf{m}_{j \to i}^{(k)}\}_{j \in \mathcal{N}_i}\big),
\end{equation}
and the subsystem is updated immediately according to
\begin{equation}
\mathbf{y}_i^{(k)}
=
\mathcal{M}_i\big(
\mathbf{x}_i, \mathbf{u}_i;
\boldsymbol{\theta}_i,
\mathbf{m}_i^{\mathrm{in},(k)}
\big).
\end{equation}
Because newly updated subsystem states are immediately reused during the same iteration, the Gauss--Seidel scheme
typically yields faster convergence than the Jacobi scheme for strongly coupled systems, although it is less
naturally parallelizable.
\begin{algorithm}[H]
\caption{Gauss--Seidel message passing}
\label{AL:gauss-seidel-mp-algorithm}
\begin{algorithmic}[1]
\For{$k = 1,2,\dots,\texttt{max\_iter}$}

    \Statex \hspace{1em} \textit{// update subsystems sequentially}
    \For{each subsystem $i$ in the prescribed update order}

        \Statex \hspace{2em} \textit{// construct incoming messages using the most recent neighbor information}
        \For{each $j \in \mathcal{N}_i$}
            \If{subsystem $j$ has already been updated in iteration $k$}
                \State set $k_j = k$
            \Else
                \State set $k_j = k-1$
            \EndIf
            \State receive $\mathbf{s}_{j \to i}^{(k_j)}$
            \State compute
            \[
            \mathbf{m}_{j \to i}^{(k)}
            =
            h_{ji}\big(
            \mathbf{s}_{j \to i}^{(k_j)},
            \mathbf{s}_{i \leftarrow j}^{(k-1)};
            \boldsymbol{\gamma}_{ji}
            \big)
            \]
        \EndFor

        \State aggregate incoming messages
        \[
        \mathbf{m}_{i}^{\mathrm{in},(k)}
        =
        \mathcal{A}\big(\{\mathbf{m}_{j \to i}^{(k)}\}_{j \in \mathcal{N}_i}\big)
        \]

        \State solve
        \[
        \mathbf{y}_i^{(k)}
        =
        \mathcal{M}_i\big(
        \mathbf{x}_i, \mathbf{u}_i;
        \boldsymbol{\theta}_i,
        \mathbf{m}_i^{\mathrm{in},(k)}
        \big)
        \]

        \State send updated $\mathbf{s}_{i \to j}^{(k)}$ to all $j \in \mathcal{N}_i$

    \EndFor

\EndFor
\end{algorithmic}
\end{algorithm}

In Algorithm~\ref{AL:gauss-seidel-mp-algorithm}, $k_j = k$ if neighboring subsystem $j$ has already been updated in
the current iteration and $k_j = k-1$ otherwise, which makes the Gauss--Seidel update sequential.

\subsubsection*{Adams--Bashforth 2 (AB2) message passing}

In the Adams--Bashforth 2 (AB2) message-passing scheme~\cite{Hairer1993RungeKutta,Borodulin2025Accuracy}, subsystem
interactions are communicated through local messages at each global time step, while the subsystem response is
advanced using the standard second-order explicit multi-step Adams--Bashforth integration rule.
At time step $n$, each subsystem receives the interface variables of its neighboring subsystems,
constructs the corresponding incoming messages through the interface laws, and aggregates them to form the total
coupling input.

For notational clarity, we use $k$ to denote inner message-passing iterations (Jacobi and Gauss--Seidel), whereas
$n$ denotes global physical time steps in the AB2 time-integration scheme.
Following the interaction model defined in Eq.~\ref{eq:mp-form}, the message from subsystem $j$ to subsystem $i$ at
time step $n$ is written as
\begin{equation}
\mathbf{m}_{j \to i}^{(n)}
=
h_{ji}\big(
\mathbf{s}_{j \to i}^{(n)},
\mathbf{s}_{i \leftarrow j}^{(n)};
\boldsymbol{\gamma}_{ji}
\big),
\end{equation}
and the aggregated incoming message is given by
\begin{equation}
\mathbf{m}_i^{\mathrm{in},(n)}
=
\mathcal{A}\big(\{\mathbf{m}_{j \to i}^{(n)}\}_{j \in \mathcal{N}_i}\big).
\end{equation}
The subsystem evolution is then advanced using the AB2 update rule,
\begin{equation}
\mathbf{y}_i^{(n+1)}
=
\mathbf{y}_i^{(n)}
+
\Delta t
\left[
\frac{3}{2}\,\mathbf{f}_i^{(n)}
-
\frac{1}{2}\,\mathbf{f}_i^{(n-1)}
\right],
\end{equation}
where $\mathbf{f}_i^{(n)}$ denotes the local time derivative or evolution operator of subsystem $i$, evaluated
using $\mathbf{x}_i$, $\mathbf{u}_i$, $\boldsymbol{\theta}_i$, and the aggregated incoming message
$\mathbf{m}_i^{\mathrm{in},(n)}$.
Compared with single-step message-passing formulations, the AB2 scheme incorporates information from both the
current and previous time steps, which improves temporal accuracy while preserving an explicit update structure.
\begin{algorithm}[H]
\caption{AB2 message passing}
\label{AL:ab2-mp-algorithm}
\begin{algorithmic}[1]
\For{$n = 1,2,\dots,\texttt{time\_steps}$}

    \Statex \hspace{1em} \textit{// exchange current interface variables}
    \For{each subsystem $i$}
        \State send $\mathbf{s}_{i \to j}^{(n)}$ to all $j \in \mathcal{N}_i$
    \EndFor

    \Statex \hspace{1em} \textit{// construct incoming messages at time step $n$}
    \For{each subsystem $i$}
        \For{each $j \in \mathcal{N}_i$}
            \State receive $\mathbf{s}_{j \to i}^{(n)}$
            \State compute
            \[
            \mathbf{m}_{j \to i}^{(n)}
            =
            h_{ji}\big(
            \mathbf{s}_{j \to i}^{(n)},
            \mathbf{s}_{i \leftarrow j}^{(n)};
            \boldsymbol{\gamma}_{ji}
            \big)
            \]
        \EndFor
        \State aggregate incoming messages
        \[
        \mathbf{m}_{i}^{\mathrm{in},(n)}
        =
        \mathcal{A}\big(\{\mathbf{m}_{j \to i}^{(n)}\}_{j \in \mathcal{N}_i}\big)
        \]
    \EndFor

    \Statex \hspace{1em} \textit{// evaluate subsystem evolution operator}
    \For{each subsystem $i$}
        \State compute
        \[
        \mathbf{f}_i^{(n)}
        =
        \mathcal{F}_i\big(
        \mathbf{x}_i, \mathbf{u}_i;
        \boldsymbol{\theta}_i,
        \mathbf{m}_i^{\mathrm{in},(n)}
        \big)
        \]
    \EndFor

    \Statex \hspace{1em} \textit{// advance subsystem states using AB2 integration}
    \For{each subsystem $i$}
        \State update
        \[
        \mathbf{y}_i^{(n+1)}
        =
        \mathbf{y}_i^{(n)}
        +
        \Delta t
        \left[
        \frac{3}{2}\,\mathbf{f}_i^{(n)}
        -
        \frac{1}{2}\,\mathbf{f}_i^{(n-1)}
        \right]
        \]
    \EndFor

\EndFor
\end{algorithmic}
\end{algorithm}
In Algorithm~\ref{AL:ab2-mp-algorithm}, $\mathbf{y}_i^{(n)}$ denotes the state vector of subsystem $i$ at time step
$n$, $\mathcal{F}_i(\cdot)$ denotes its local evolution operator, and $\Delta t$ is the time step size.
Unlike the Jacobi and Gauss--Seidel formulations, where $\mathcal{M}_i(\cdot)$ represents an iterative subsystem
update map, the AB2 scheme advances $\mathbf{y}_i$ explicitly in time using its state derivative
$\mathbf{f}_i^{(n)} = \mathcal{F}_i\big(\mathbf{x}_i,\mathbf{u}_i;\boldsymbol{\theta}_i,\mathbf{m}_i^{\mathrm{in},(n)}\big)$.

\subsection*{Kalman filters}
In this study, different Kalman filtering techniques~\cite{Kalman1960New, Ghorbani2021Nonlinear, Khodarahmi2023Review} are employed at the subsystem
level to estimate local states, identify unknown parameters, and quantify the associated uncertainties using
incoming messages and available measurements.
Each subsystem is equipped with its own state-space model, which may be physics-based, data-driven, or a hybrid
combination of both.
The transition and measurement models may differ across subsystems, reflecting the heterogeneous nature of the
decomposed system.
Each subsystem $i$ is described by a discrete state-space model of the form
\begin{equation}
\mathbf{x}_{i,n+1} = \boldsymbol{\phi}_i(\mathbf{x}_{i,n}, \mathbf{u}_{i,n}) + \mathbf{v}_{i,n}, 
\quad
\mathbf{y}_{i,n} = \boldsymbol{\psi}_i(\mathbf{x}_{i,n}) + \mathbf{w}_{i,n},
\label{eq:state-space-model}
\end{equation}
where $\mathbf{x}_{i,n}$ is the state vector, $\mathbf{y}_{i,n}$ is the measurement vector, and $\mathbf{v}_{i,n}
\sim \mathcal{N}(0,\mathbf{Q}_i)$ and $\mathbf{w}_{i,n} \sim \mathcal{N}(0,\mathbf{R}_i)$ denote the process and
measurement noise, respectively.
The functions $\boldsymbol{\phi}_i(\cdot)$ and $\boldsymbol{\psi}_i(\cdot)$ denote the subsystem-specific
transition and measurement models used in the Kalman filtering framework.
In the Kalman filter, the state and measurement variables are assumed to follow Gaussian distributions, and
the transition and measurement models are represented in matrix form.
Regardless of the specific Kalman filtering variant, the estimation procedure is initialized using prior knowledge
of the initial state mean $\mathbf{x}_{i,0}$ and the initial state covariance matrix $\mathbf{P}_{i,0}$.
\subsubsection*{Linear Kalman filter}
For linear subsystems, the transition and measurement models are written as
\begin{equation}
\mathbf{x}_{i,n+1} = \mathbf{M}_i \mathbf{x}_{i,n} + \mathbf{v}_{i,n}, 
\quad
\mathbf{y}_{i,n} = \mathbf{H}_i \mathbf{x}_{i,n} + \mathbf{w}_{i,n}.
\end{equation}
Here, $\mathbf{M}_i$ and $\mathbf{H}_i$ denote the transition and measurement matrices of subsystem $i$,
respectively, and correspond to the linear forms of the general nonlinear functions $\boldsymbol{\phi}_i(\cdot)$
and $\boldsymbol{\psi}_i(\cdot)$ introduced above.
The Kalman filter, Algorithm~\ref{AL:linear-kalman-filter}, recursively performs a prediction step followed
by a measurement update to estimate the subsystem state and its covariance.
\begin{algorithm}[H]
\caption{Linear Kalman filter for subsystem $i$}
\label{AL:linear-kalman-filter}
\begin{algorithmic}[1]
\For{$n = 1,2,\dots$}
    \Statex \hspace{1em} \textit{// prediction}
    \State $\hat{\mathbf{x}}_{i,n+1} = \mathbf{M}_i \mathbf{x}_{i,n}$
    \State $\mathbf{P}_{i,n+1|n} = \mathbf{M}_i \mathbf{P}_{i,n} \mathbf{M}_i^\top + \mathbf{Q}_i$
    \Statex \hspace{1em} \textit{// update}
    \State $\mathbf{K}_{i,n+1} = \mathbf{P}_{i,n+1|n} \mathbf{H}_i^\top (\mathbf{H}_i \mathbf{P}_{i,n+1|n} \mathbf{H}_i^\top + \mathbf{R}_i)^{-1}$
    \State $\mathbf{x}_{i,n+1} = \hat{\mathbf{x}}_{i,n+1} + \mathbf{K}_{i,n+1}\left(\mathbf{y}_{i,n+1} - \mathbf{H}_i \hat{\mathbf{x}}_{i,n+1}\right)$
    \State $\mathbf{P}_{i,n+1} = \mathbf{P}_{i,n+1|n} - \mathbf{K}_{i,n+1} \mathbf{H}_i \mathbf{P}_{i,n+1|n}$
\EndFor
\end{algorithmic}
\end{algorithm}
In Algorithm~\ref{AL:linear-kalman-filter}, the notation $\hat{\mathbf{x}}_{i,n+1}$ denotes the predicted (a
priori) state estimate before incorporating the measurement.
At each iteration, the Kalman filter recursively updates both the mean and the covariance of the state vector,
providing an estimate of the subsystem state along with its associated uncertainty.
\subsubsection*{Extended Kalman filter (EKF)}
For nonlinear subsystems, the transition and measurement models $\boldsymbol{\phi}_i(\cdot)$ and
$\boldsymbol{\psi}_i(\cdot)$ in Eq.~\ref{eq:state-space-model} cannot be expressed in matrix form, which prevents
the direct application of the Kalman filter.
The EKF addresses this limitation by approximating the nonlinear models using first-order
Taylor expansion, resulting in locally linear transition and measurement matrices at each time
step~\cite{Jazwinski2007Stochastic, Wan2001Dual}.
Specifically, the Jacobian matrices
\[
\mathbf{M}_{i,n} = \frac{\partial \boldsymbol{\phi}_i}{\partial \mathbf{x}}\bigg|_{\mathbf{x}_{i,n}},
\qquad
\mathbf{H}_{i,n+1} = \frac{\partial \boldsymbol{\psi}_i}{\partial \mathbf{x}}\bigg|_{\hat{\mathbf{x}}_{i,n+1}},
\]
are used in place of $\mathbf{M}_i$ and $\mathbf{H}_i$ within the prediction and update steps.
The algorithmic implementation of the EKF is shown in Algorithm~\ref{AL:extended-kalman-filter}.
While the EKF enables state estimation for nonlinear systems, its accuracy depends on the validity of the local
linear approximation and may degrade for strongly nonlinear dynamics or when the system exhibits significant
higher-order effects.
\begin{algorithm}[H]
\caption{EKF for subsystem $i$}
\label{AL:extended-kalman-filter}
\begin{algorithmic}[1]
\For{$n = 1,2,\dots$}

    \Statex \hspace{1em} \textit{// prediction}
    \State $\hat{\mathbf{x}}_{i,n+1} = \boldsymbol{\phi}_i(\mathbf{x}_{i,n}, \mathbf{u}_{i,n})$
    \State evaluate $\mathbf{M}_{i,n} = \dfrac{\partial \boldsymbol{\phi}_i}{\partial \mathbf{x}}\big|_{\mathbf{x}_{i,n}}$
    \State $\mathbf{P}_{i,n+1|n} = \mathbf{M}_{i,n} \mathbf{P}_{i,n} \mathbf{M}_{i,n}^\top + \mathbf{Q}_i$

    \Statex \hspace{1em} \textit{// update}
    \State $\hat{\mathbf{y}}_{i,n+1} = \boldsymbol{\psi}_i(\hat{\mathbf{x}}_{i,n+1})$
    \State evaluate $\mathbf{H}_{i,n+1} = \dfrac{\partial \boldsymbol{\psi}_i}{\partial \mathbf{x}}\big|_{\hat{\mathbf{x}}_{i,n+1}}$
    \State $\mathbf{K}_{i,n+1} = \mathbf{P}_{i,n+1|n} \mathbf{H}_{i,n+1}^\top (\mathbf{H}_{i,n+1} \mathbf{P}_{i,n+1|n} \mathbf{H}_{i,n+1}^\top + \mathbf{R}_i)^{-1}$
    \State $\mathbf{x}_{i,n+1} = \hat{\mathbf{x}}_{i,n+1} + \mathbf{K}_{i,n+1}\left(\mathbf{y}_{i,n+1} - \hat{\mathbf{y}}_{i,n+1}\right)$
    \State $\mathbf{P}_{i,n+1} = \mathbf{P}_{i,n+1|n} - \mathbf{K}_{i,n+1} \mathbf{H}_{i,n+1} \mathbf{P}_{i,n+1|n}$

\EndFor
\end{algorithmic}
\end{algorithm}

\subsubsection*{Unscented Kalman filter (UKF)}
The UKF avoids local linearization by propagating a set of deterministically chosen sigma
points through the nonlinear transition and measurement functions $\boldsymbol{\phi}_i(\cdot)$ and
$\boldsymbol{\psi}_i(\cdot)$~\cite{Wan2000unscented, Julier2004Unscented}.
For each subsystem, $2L+1$ sigma points are generated from the current state mean $\hat{\mathbf{x}}$ and
covariance $\mathbf{P}$, where $L$ is the dimension of the state vector. With scaling parameters $\alpha$,
$\beta$, and $\kappa$, define the composite scaling
\begin{equation}
  \lambda \;=\; \alpha^{2}\,(L + \kappa) \;-\; L,
  \label{eq:ukf_lambda}
\end{equation}
and choose the symmetric sigma points
\begin{align}
  \boldsymbol{\chi}_{0}   &= \hat{\mathbf{x}}, \notag\\
  \boldsymbol{\chi}_{i}   &= \hat{\mathbf{x}} + \bigl[\sqrt{(L+\lambda)\,\mathbf{P}}\bigr]_{i},
    \quad i=1,\dots,L, \notag\\
  \boldsymbol{\chi}_{i+L} &= \hat{\mathbf{x}} - \bigl[\sqrt{(L+\lambda)\,\mathbf{P}}\bigr]_{i},
    \quad i=1,\dots,L,
  \label{eq:ukf_sigma}
\end{align}
where $\bigl[\sqrt{(L+\lambda)\,\mathbf{P}}\bigr]_{i}$ is the $i$-th row (or column) of a matrix square root
of $(L+\lambda)\,\mathbf{P}$ (typically obtained by Cholesky factorization). The corresponding mean and
covariance weights are
\begin{equation}
  w_{m}^{0} = \frac{\lambda}{L+\lambda},\qquad
  w_{c}^{0} = \frac{\lambda}{L+\lambda} + (1-\alpha^{2}+\beta),\qquad
  w_{m}^{j} = w_{c}^{j} = \frac{1}{2(L+\lambda)},\;\; j=1,\dots,2L.
  \label{eq:ukf_weights}
\end{equation}
The standard choice $\alpha=1$, $\beta=2$, $\kappa=0$ gives $\lambda=0$ and a symmetric, unit-spread sigma-point
set. In our code $\gamma$ is an implementation flag, not the sigma-point radius; $\gamma=0$ selects the standard
$(\alpha,\beta,\kappa)=(1,2,0)$ setting above.
These sigma points are propagated through the nonlinear transition and measurement models to compute the predicted
state mean, measurement mean, and the associated covariance matrices required for the Kalman update.

\begin{algorithm}[H]
\caption{UKF for subsystem $i$}
\label{AL:unscented-Kalman-filter}
\begin{algorithmic}[1]
\For{$n = 1,2,\dots$}
    \Statex \hspace{1em} \textit{// generate sigma points}
    \State Generate $\{\chi_{i,n}^j,\, w_m^j,\, w_c^j\}_{j=0}^{2L}$ from $\mathbf{x}_{i,n}$ and $\mathbf{P}_{i,n}$
    \Statex \hspace{1em} \textit{// state prediction}
    \State $\hat{\chi}_{i,n+1}^j = \boldsymbol{\phi}_i(\chi_{i,n}^j,\mathbf{u}_{i,n}), \quad j=0,\dots,2L$
    \State $\hat{\mathbf{x}}_{i,n+1} = \sum_{j=0}^{2L} w_m^j \hat{\chi}_{i,n+1}^j$
    \State $\mathbf{P}_{i,n+1|n}^{xx} = \sum_{j=0}^{2L} w_c^j (\hat{\chi}_{i,n+1}^j-\hat{\mathbf{x}}_{i,n+1})(\hat{\chi}_{i,n+1}^j-\hat{\mathbf{x}}_{i,n+1})^\top + \mathbf{Q}_i$
    \Statex \hspace{1em} \textit{// measurement prediction}
    \State $\hat{\mathbf{y}}_{i,n+1}^j = \boldsymbol{\psi}_i(\hat{\chi}_{i,n+1}^j), \quad j=0,\dots,2L$
    \State $\hat{\mathbf{y}}_{i,n+1} = \sum_{j=0}^{2L} w_m^j \hat{\mathbf{y}}_{i,n+1}^j$
    \State $\mathbf{P}_{i,n+1|n}^{yy} = \sum_{j=0}^{2L} w_c^j (\hat{\mathbf{y}}_{i,n+1}^j-\hat{\mathbf{y}}_{i,n+1})(\hat{\mathbf{y}}_{i,n+1}^j-\hat{\mathbf{y}}_{i,n+1})^\top + \mathbf{R}_i$
    \State $\mathbf{P}_{i,n+1|n}^{xy} = \sum_{j=0}^{2L} w_c^j (\hat{\chi}_{i,n+1}^j-\hat{\mathbf{x}}_{i,n+1})(\hat{\mathbf{y}}_{i,n+1}^j-\hat{\mathbf{y}}_{i,n+1})^\top$
    \Statex \hspace{1em} \textit{// update}
    \State $\mathbf{K}_{i,n+1} = \mathbf{P}_{i,n+1|n}^{xy} \left(\mathbf{P}_{i,n+1|n}^{yy}\right)^{-1}$
    \State $\mathbf{x}_{i,n+1} = \hat{\mathbf{x}}_{i,n+1} + \mathbf{K}_{i,n+1}\left(\mathbf{y}_{i,n+1} - \hat{\mathbf{y}}_{i,n+1}\right)$
    \State $\mathbf{P}_{i,n+1} = \mathbf{P}_{i,n+1|n}^{xx} - \mathbf{K}_{i,n+1}\mathbf{P}_{i,n+1|n}^{yy}\mathbf{K}_{i,n+1}^\top$
\EndFor
\end{algorithmic}
\end{algorithm}
In Algorithm~\ref{AL:unscented-Kalman-filter}, $\chi_{i,n}^j$ denotes the sigma points associated with subsystem
$i$, while $w_m^j$ and $w_c^j$ are the corresponding weights used to compute the predicted means and covariance
matrices. More details about the UKF can be found in~\cite{Ghorbani2018iterated}.
\subsection*{Nonlinear autoregressive with exogenous inputs (NARX)}
The NARX model is a data-driven approach for modeling nonlinear
dynamical systems, where the current output is expressed as a function of past outputs and external
inputs~\cite{Leontaritis1985Inputoutput, Yu2019Nonlinear}.
In discrete time, the model can be expressed in vector form as
\begin{equation}
y_{n+1} = \mathcal{F}\big(\mathbf{y}_{n}^{(n_y)},\, \mathbf{u}_{n}^{(n_u)},\, \boldsymbol{\eta}\big),
\end{equation}
where $\mathbf{y}_{n}^{(n_y)} = [y_n, y_{n-1}, \dots, y_{n-n_y+1}]$ collects the past outputs, $\mathbf{u}_{n}^{(n_u)} = [u_n, u_{n-1},
\dots, u_{n-n_u+1}]$ the current and past external inputs, and $\boldsymbol{\eta}$ denotes the model parameters learned from data.
Additional exogenous or static features can be incorporated as auxiliary inputs to the mapping $\mathcal{F}(\cdot)$
to account for system-specific characteristics.
The NARX formulation is particularly suitable for systems where the governing physical relationships are partially
unknown or difficult to model explicitly, but sufficient input--output data are available.
By combining historical system responses, external inputs, and auxiliary variables, the model captures temporal
dependencies and nonlinear interactions in a compact and computationally efficient form.
This makes it well-suited for integration within the proposed framework, where subsystem dynamics or coupling terms
can be learned directly from data while remaining compatible with the overall message passing and state estimation
structure.

\subsection*{Diffusion models}

Diffusion models are used in this study to propagate local perturbations or uncertainty across coupled subsystems
in a computationally efficient manner.
This technique complements the message-passing inference of the main manuscript; it is exercised in the
additional six-DOF study reported below and is not used in the main-text case studies.
Within the proposed framework, diffusion operates on the graph induced by subsystem interactions, where nodes
represent subsystems and weighted edges encode the strength of coupling between them.
Given a local perturbation or defect signal at a subsystem, diffusion provides a mechanism to distribute its
influence to neighboring subsystems without requiring a full re-solution of the global system.

Let the subsystem interaction graph be represented by a weighted adjacency matrix $\mathbf{W}$ and the
corresponding graph Laplacian $\mathbf{L}$.
A scalar or vector perturbation $\delta \mathbf{f}$ defined at a given subsystem can then be propagated through the
network using a diffusion operator, resulting in a distributed influence across all subsystems.
In this study, two diffusion strategies are considered: a local 1-hop propagation
model~\cite{Biswas2024Generalized} and a global heat-kernel-based diffusion model~\cite{Lafferty2005Diffusion,Patel2025HKDMGIN}.
After constructing the subsystem interaction graph, diffusion models are used to approximate how a local
perturbation or physics modification in one subsystem propagates across the network without re-solving the fully
coupled system.
In this setting, subsystems are represented as graph nodes, interfaces are represented as weighted edges, and the
perturbation is introduced as a defect signal at a source node.
The edge weights are constructed from interaction quantities exchanged through message passing, allowing the
diffusion process to remain consistent with the underlying subsystem coupling structure.
\subsubsection*{1-hop technique}
The 1-hop diffusion model approximates the propagation of a local perturbation by distributing it only to immediate
neighboring subsystems.
Given a perturbation magnitude $q = |\delta \mathbf{f}|$ at a source subsystem, the influence is partitioned
between the source and its neighbors based on normalized edge weights.
Specifically, the propagated quantities are computed as
\begin{equation}
s_i = \frac{q}{1+\alpha},
\qquad
s_j = \eta_{ij} \frac{\alpha q}{1+\alpha},
\label{eq:appendix_1hop_scores}
\end{equation}
where $\alpha$ controls the total amount of propagation. %
The parameter $s_i$ denotes the score retained at the source subsystem where the perturbation is introduced, while $s_j$
denotes the score assigned to each immediate neighboring subsystem $j \in \mathcal{N}_i$.
The normalized coupling weight $\eta_{ij}$ is defined as $\eta_{ij} = w_{ij} / \sum_{k \in \mathcal{N}_i}
w_{ik}$, ensuring that the propagated portion of the perturbation is distributed among neighboring subsystems
according to their relative interaction strength.
Here, $w_{ij}$ denotes the coupling weight between subsystems $i$ and $j$, which quantifies the strength of their
interaction and is typically constructed from interface quantities such as exchanged forces, energy, or other
interaction measures.
This approach provides a computationally inexpensive approximation of influence propagation, as it relies only on
local connectivity and avoids global matrix operations.
However, it is limited to first-order interactions and does not capture long-range effects across the network.
\subsubsection*{Heat kernel}
To capture multi-hop and global propagation effects, a diffusion process based on the graph heat kernel is
employed.
Given the graph Laplacian $\mathbf{L}$, the diffusion operator $\mathbf{H}(\beta)$ is defined as
\begin{equation}
\mathbf{H}(\beta) = \exp(-\beta \mathbf{L}),
\label{eq:appendix_heat_kernel}
\end{equation}
where $\beta$ is a diffusion parameter controlling the extent of propagation.
The distributed influence of a perturbation $\delta \mathbf{f}$ is then obtained as
\begin{equation}
\mathbf{s} = \mathbf{H}(\beta)\, \delta \mathbf{f},
\label{eq:appendix_hk_scores}
\end{equation}
which naturally accounts for all paths in the interaction graph.

Compared to the 1-hop approach, the heat-kernel diffusion captures higher-order interactions and provides a
smoother and more globally consistent propagation of perturbations.
The parameter $\beta$ controls the balance between localized and global effects, enabling a continuous transition
between short-range and long-range propagation.

\section*{Case studies}

The proposed framework is now demonstrated on the three case studies of the main manuscript, presented in
their order of appearance, together with the complementary studies introduced above.
Each case study opens with the claim of the main text that it supports, followed by the complete model
specifications, estimation protocols, and quantitative results.

\subsection*{Chain of damped mass--spring systems}

This section provides the full formulation and protocols for the canonical testbed of the main manuscript,
which isolates how uncertainty should be communicated across subsystem interfaces during joint
state-and-parameter estimation under sparse boundary sensing.
The presentation progressively transitions from forward simulation to inverse estimation, followed by the
deterministic, probabilistic, and learned message-passing strategies and their scalability properties.

\subsubsection*{Four-degree-of-freedom (DOF) damped mass--spring system}
We consider a one-dimensional chain of four lumped masses connected through linear springs and viscous dampers.
Each mass interacts only with its immediate neighbors, resulting in a nearest-neighbor coupled dynamical system.
The equations of motion are governed by the linear second-order system
\begin{equation}
M \ddot{x}(t) + C \dot{x}(t) + K x(t) = f(t),
\end{equation}
where $M$, $C$, and $K$ denote the global mass, damping, and stiffness matrices, respectively, and $f(t)$ is the
external forcing vector.
For the four-degree-of-freedom chain considered here, under the lumped-mass assumption, the mass matrix is
diagonal, while the stiffness and damping matrices have a tridiagonal structure induced by the local mechanical
couplings between adjacent masses.
The global system matrices are given by
\begin{equation}
M=\begin{bmatrix} m_1 & 0 & 0 & 0 \\ 0 & m_2 & 0 & 0 \\ 0 & 0 & m_3 & 0 \\ 0 & 0 & 0 & m_4 \end{bmatrix}, 
K=\begin{bmatrix} k_1+k_2 & -k_2 & 0 & 0 \\ -k_2 & k_2+k_3 & -k_3 & 0 \\ 0 & -k_3 & k_3+k_4 & -k_4 \\ 0 & 0 & -k_4 & k_4 \end{bmatrix},  
C=\begin{bmatrix} c_1+c_2 & -c_2 & 0 & 0 \\ -c_2 & c_2+c_3 & -c_3 & 0 \\ 0 & -c_3 & c_3+c_4 & -c_4 \\ 0 & 0 & -c_4 & c_4 \end{bmatrix}. 
\end{equation}

For the simulations considered in this study, the masses are taken as $m_1=m_2=m_3=m_4=500~\mathrm{kg}$.
The spring constants are chosen as $k_1=k_2=k_3=k_4=5\times10^4~\mathrm{N/m}$, and the damping coefficients are set
to $c_1=c_2=c_3=c_4=300~\mathrm{N\,s/m}$.
The displacement vector is defined as $x(t) = [x_1(t), x_2(t), x_3(t), x_4(t)]^\top$, and the corresponding
velocity vector is $\dot{x}(t) = [\dot{x}_1(t), \dot{x}_2(t), \dot{x}_3(t), \dot{x}_4(t)]^\top$.
Introducing displacement and velocity as state variables leads to an equivalent first-order state-space
representation of dimension eight.
The monolithic model is used as the reference representation of the full coupled system throughout this study.

\begin{figure*}[t]
\centering
\includegraphics[width=\linewidth]{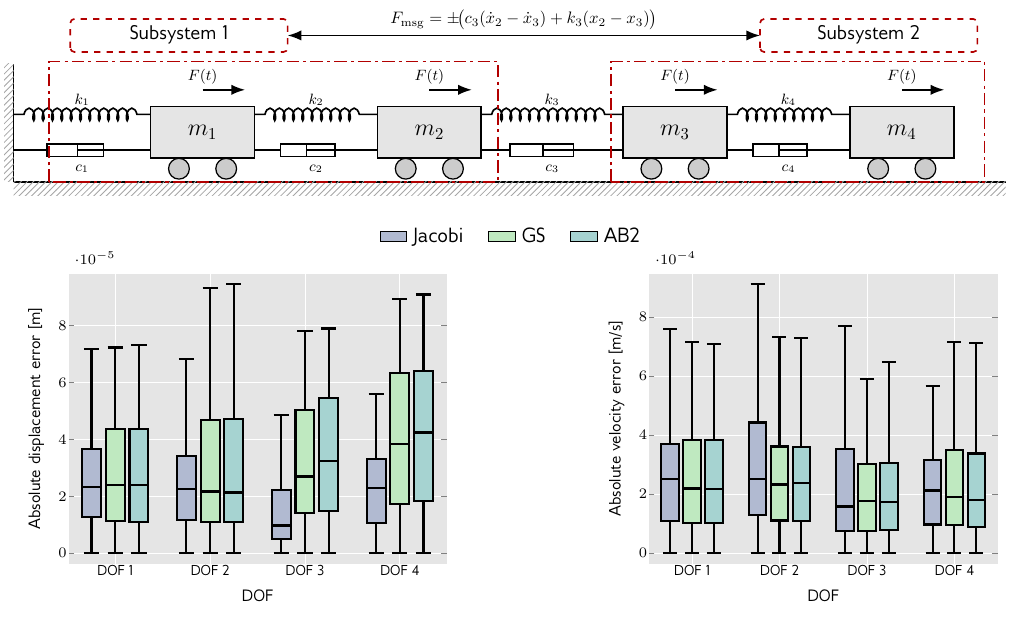}
\caption{Top: Four-DOF mass--spring--damper system decomposed into two coupled subsystems exchanging the
interface force $F_{\mathrm{msg}}$.
Bottom: Boxplots of absolute displacement (left) and velocity (right) errors relative to the monolithic solution
for the Jacobi, Gauss--Seidel, and AB2 schemes. Jacobi attains the lowest displacement errors, most visibly at
the interface-adjacent DOFs~3 and~4, while the velocity errors remain comparable across the three schemes.
}
\label{fig:4dof_error_boxplots}
\end{figure*}

\paragraph{Forward problem using the distributed formulation}
The forward problem investigates whether a decomposed representation of the system can reproduce the monolithic
solution when the coupling between subsystems is enforced through a known message-passing function.
The system is partitioned into two subsystems, coupled through the interface force
\begin{equation}
F_{\mathrm{msg}} = k_3 (x_2 - x_3) + c_3 (\dot{x}_2 - \dot{x}_3).
\end{equation}
Three coupling strategies are considered: Jacobi (parallel updates), Gauss--Seidel (sequential updates), and an
Adams--Bashforth 2 (AB2) extrapolation-based scheme.
Figure~\ref{fig:4dof_error_boxplots} presents the error distributions relative to the monolithic solution.
Contrary to common expectations, the Jacobi scheme yields the lowest displacement errors, most visibly at the
interface-adjacent DOFs~3 and~4, while the velocity errors remain comparable across the three schemes.
This behavior is attributed to the fact that Jacobi enforces a consistent time level across subsystems, as both
rely on interface states from the same time step.
In contrast, the Gauss--Seidel and AB2 schemes introduce an update asymmetry, where one subsystem uses partially
updated information, leading to a bias in the coupling.
Importantly, the Jacobi scheme matches or outperforms the sequential schemes while remaining fully parallelizable.

\paragraph{Inverse problem using the distributed formulation, deterministic message passing}
We next investigate whether the inverse problem for the same four-DOF mass--spring--damper system can be solved
accurately in a distributed manner under sparse acceleration measurements.
More specifically, whether the latent states and the unknown stiffness parameter can be recovered using a
distributed formulation in which the two subsystems exchange only the mean values of their interface-state
estimates through a deterministic message-passing rule with known coupling parameters.
In the present setup, the measurements are sparse and are taken only at the two boundary degrees of freedom.
Subsystem~1 uses the acceleration measurement at DOF~1, denoted by $a_1$, while Subsystem~2 uses the acceleration
measurement at DOF~4, denoted by $a_4$.
For fairness, the centralized estimator is also provided with the same sparse measurement set, namely the pair
$(a_1,a_4)$.
The synthetic measurements are generated by adding Gaussian noise to the true accelerations computed from the
monolithic model.
Measurements are available at every time step, with an effective sampling frequency of $1000\,\mathrm{Hz}$.

In this inverse setting, both the centralized and distributed estimators use the same discrete-time propagation
rule using a first-order Euler update, such that $q_{k+1}=q_k+\Delta t\,v_k$ and $v_{k+1}=v_k+\Delta t\,a_k$.
The distributed estimator employs deterministic Jacobi-type message passing to reconstruct the coupling force
transmitted through the interface quantities $(x_2,v_2)$ and $(x_3,v_3)$ and known stiffness and damping values.
The message is defined as
\begin{equation}
F_{\mathrm{msg}} = k_3(x_2-x_3) + c_3(v_2-v_3).
\end{equation}
This force is then used to modify the local subsystem inputs during the UKF prediction step.
No covariance information is exchanged between the subsystems, and the coupling is therefore entirely deterministic
and mean-based.
The UKF is adopted because the problem is formulated as a joint state--parameter
estimation task.
In addition to the latent displacement and velocity states, the stiffness parameter $k_4$ is treated as an unknown
quantity and estimated online as part of the augmented state vector.

In the distributed formulation, Subsystem~1 estimates the state vector $[x_1,x_2,v_1,v_2]^\top$, whereas
Subsystem~2 estimates the augmented state vector $[x_3,x_4,v_3,v_4,k_4]^\top$.
In the centralized formulation, the full augmented state is given by $[x_1,x_2,x_3,x_4,v_1,v_2,v_3,v_4,k_4]^\top$.
The UKF is particularly suitable here because both the state transition map and the measurement map depend
nonlinearly on the augmented state through the stiffness-dependent acceleration terms.
To ensure a fair comparison, both the centralized and distributed estimators use the same time step, the same
sigma-point scaling setting (denoted $\gamma=0$ in our implementation, corresponding to the standard symmetric
sigma-point spread with $\alpha=1$, $\beta=2$, $\kappa=0$), and the same process-noise scale of $10^{-8}$.
The initial guess for the unknown stiffness is also chosen identically in both formulations as
$k_4^{(0)}=30000\,\mathrm{N/m}$.
Likewise, the same reference uncertainty scale is used for the parameter initialization, with
$k_4^{\mathrm{ref}}=50000\,\mathrm{N/m}$.
The initial state conditions are also matched between the two approaches, with $x_1(0)=0.01$, $v_1(0)=0.01$, and
all remaining states initialized at zero.
As in the other case studies, all reported testbed metrics are computed from a single noise realization generated with a fixed random seed.

\begin{figure*}[t]
\centering
\includegraphics[width=\linewidth]{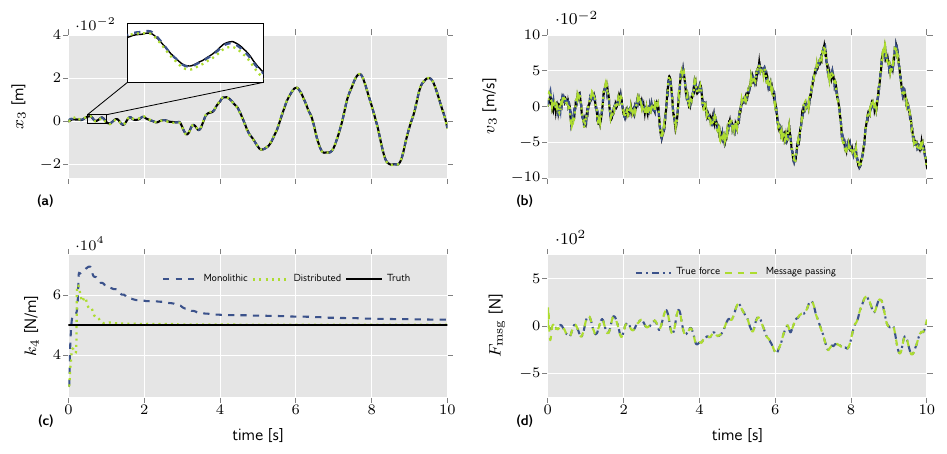}
\caption{
Inverse estimation results for the four-DOF mass--spring--damper system under sparse boundary sensing using
acceleration measurements at DOFs~1 and~4 only. (a) Reconstructed displacement response $x_3$ at the unmeasured
internal DOF. (b) Reconstructed velocity response $v_3$ at the same unmeasured DOF. (c) Online estimation of the
unknown stiffness parameter $k_4$ for the monolithic and distributed estimators, compared with the true
value. (d) Comparison between the true interface force and the deterministic message-passing force reconstructed
from the subsystem mean estimates. The results show that sparse boundary measurements, together with deterministic
mean-based message passing, are sufficient to recover both latent internal states and the unknown parameter with
good accuracy.
}
\label{fig:4dof-inverse-summary}
\end{figure*}

Figure~\ref{fig:4dof-inverse-summary} summarizes the inverse estimation results obtained under sparse sensing.
Panel~(a) shows the reconstructed displacement response $x_3$, while panel~(b) shows the reconstructed velocity
response $v_3$, both at DOF~3, which is not directly instrumented.
The close agreement between the true response, the monolithic UKF estimate, and the distributed estimate
indicates that both approaches can successfully recover unmeasured internal states using only boundary acceleration
measurements.
Panel~(c) presents the online estimate of the unknown stiffness parameter $k_4$.
Both the monolithic and distributed estimators converge from the underestimated initial guess toward the true
stiffness value; in this configuration, the distributed estimate corrects its initial transient more quickly,
whereas the monolithic estimate overshoots more strongly and settles toward the true value more slowly.
This behavior reflects how the parameter is updated in the two formulations: in the distributed estimator,
$k_4$ is corrected inside the low-dimensional subsystem-2 filter directly from the local boundary measurement,
whereas the monolithic estimator adjusts $k_4$ jointly with all nine components of the augmented state, which
slows the transient correction.
Both estimates settle close to the true value, showing that the parameter
remains identifiable under sparse sensing when the interface model is known.
Panel~(d) compares the true interface force with the message-passing force reconstructed from the subsystem mean
estimates.
The close agreement between these two curves confirms that the exchanged deterministic message accurately captures
the interface interaction between the two subsystems.

Overall, the results demonstrate that the inverse problem can still be solved accurately in a distributed manner
under sparse measurements, provided that the coupling parameters are known and that the interface message is
properly constructed from the estimated subsystem states.

\paragraph{Uncertainty propagation through message passing}

In the proposed distributed formulation, each subsystem is equipped with an independent UKF that can
propagate both the mean and covariance of its local state estimate at every time step.
Accordingly, the UKF associated with subsystem \(i\) provides a Gaussian approximation of the form
\((\mathbf{\mu}_i,\mathbf{P}_i)\), where \(\mathbf{\mu}_i\) denotes the state mean and \(\mathbf{P}_i\) denotes the
corresponding state covariance matrix.
This is the key ingredient that enables probabilistic message passing, because the exchanged interface quantities
are not treated as deterministic values only, but as random quantities whose uncertainty can be propagated across
subsystem boundaries.
For the four-degree-of-freedom chain considered here, the two subsystems interact through the interface force
acting between masses \(2\) and \(3\). 
\begin{equation}
F_b = k_3(x_2-x_3)+c_3(\dot{x}_2-\dot{x}_3),
\label{eq:Fb_def_appendix}
\end{equation}
where \(x_2,\dot{x}_2\) are the interface displacement and velocity states of subsystem \(1\), and
\(x_3,\dot{x}_3\) are the corresponding interface states of subsystem \(2\).
To express this relation more compactly, define the relative interface state
\begin{equation}
\delta \mathbf{z}
=
\begin{bmatrix}
\delta x \\
\delta \dot{x}
\end{bmatrix}
=
\begin{bmatrix}
x_2-x_3 \\
\dot{x}_2-\dot{x}_3
\end{bmatrix},
\qquad
\mathbf{a}
=
\begin{bmatrix}
k_3 \\
c_3
\end{bmatrix}.
\end{equation}
Then, \eqref{eq:Fb_def_appendix} can be written in the linear form
\begin{equation}
F_b = \mathbf{a}^\top \delta \mathbf{z}.
\label{eq:Fb_linear_appendix}
\end{equation}
Therefore, uncertainty propagation across subsystems begins by computing the variance of this interface force from
the state covariances delivered by the two subsystem UKFs. %

Let \(\mathbf{P}_{\delta z}=\mathrm{cov}(\delta \mathbf{z})\). For any deterministic vector \(\mathbf{a}\) and
random vector \(\mathbf{x}\), the variance of the linear form \(\mathbf{a}^\top \mathbf{x}\) is given by
\begin{equation}
\mathrm{var}(\mathbf{a}^\top \mathbf{x})=\mathbf{a}^\top \mathrm{cov}(\mathbf{x})\,\mathbf{a}.
\label{eq:linear_var_identity_appendix}
\end{equation}
Applying~\eqref{eq:linear_var_identity_appendix} to~\eqref{eq:Fb_linear_appendix} yields
\begin{equation}
\mathrm{var}(F_b)=\mathbf{a}^\top \mathbf{P}_{\delta z}\,\mathbf{a}.
\label{eq:varFb_basic_appendix}
\end{equation}

Next, write
\begin{equation}
\mathbf{z}_2=
\begin{bmatrix}
x_2 \\
\dot{x}_2
\end{bmatrix},
\qquad
\mathbf{z}_3=
\begin{bmatrix}
x_3 \\
\dot{x}_3
\end{bmatrix},
\qquad
\delta \mathbf{z}=\mathbf{z}_2-\mathbf{z}_3.
\end{equation}
Then the covariance of the relative interface state is
\begin{equation}
\mathrm{cov}(\delta \mathbf{z})
=
\mathrm{cov}(\mathbf{z}_2-\mathbf{z}_3)
=
\mathrm{cov}(\mathbf{z}_2)
+
\mathrm{cov}(\mathbf{z}_3)
-
\mathrm{cov}(\mathbf{z}_2,\mathbf{z}_3)
-
\mathrm{cov}(\mathbf{z}_3,\mathbf{z}_2).
\label{eq:cov_delta_expand_appendix}
\end{equation}
Under the standard Jacobi message-passing assumption, the subsystem estimates are treated as conditionally
independent during the interface exchange step, so the cross-covariance terms are neglected. 
\begin{equation}
\mathrm{cov}(\mathbf{z}_2,\mathbf{z}_3)\approx 0,
\qquad
\mathrm{cov}(\mathbf{z}_3,\mathbf{z}_2)\approx 0.
\end{equation}
As a result,
\begin{equation}
\mathbf{P}_{\delta z}
\approx
\mathbf{P}_{s_1}+\mathbf{P}_{s_2},
\end{equation}
where \(\mathbf{P}_{s_1}\) and \(\mathbf{P}_{s_2}\) denote the covariance submatrices associated with the interface
states \([x_2,\dot{x}_2]^\top\) and \([x_3,\dot{x}_3]^\top\), respectively. 
Substituting this result into~\eqref{eq:varFb_basic_appendix} gives the variance of the message-passing force:
\begin{equation}
\mathrm{var}(F_b)
=
\begin{bmatrix}
k_3 & c_3
\end{bmatrix}
\left(\mathbf{P}_{s_1}+\mathbf{P}_{s_2}\right)
\begin{bmatrix}
k_3 \\
c_3
\end{bmatrix}.
\label{eq:varFb_final_appendix}
\end{equation}

This expression shows that uncertainty in the interface force is jointly determined by the uncertainty in the
interface states of both subsystems.
Therefore, the interface force acts as the channel through which estimation uncertainty is transferred from one
subsystem to the other.
To determine which subsystem states are directly affected by this uncertainty, consider the continuous-time
equation of motion for the interface mass \(M_3\) in subsystem \(2\):
\begin{equation}
M_3\ddot{x}_3 = F_b + \cdots,
\end{equation}
so that
\begin{equation}
\ddot{x}_3 = \frac{1}{M_3}F_b + \cdots.
\label{eq:xddot3_cont_appendix}
\end{equation}
This shows that the message-passing force enters subsystem \(2\) through the acceleration equation of mass \(3\). %
Using a forward Euler discretization over a time step \(\Delta t\), the velocity update becomes
\begin{equation}
\dot{x}_{3,k+1}
=
\dot{x}_{3,k}
+
\Delta t\,\ddot{x}_{3,k}.
\end{equation}
Substituting~\eqref{eq:xddot3_cont_appendix} gives
\begin{equation}
\dot{x}_{3,k+1}
=
\dot{x}_{3,k}
+
\frac{\Delta t}{M_3}F_{b,k}
+
\cdots.
\label{eq:xdot3_update_appendix}
\end{equation}

Now decompose the force into its mean and zero-mean fluctuation components:
\begin{equation}
F_{b,k}=\bar{F}_{b,k}+\tilde{F}_{b,k},
\qquad
\mathbb{E}[\tilde{F}_{b,k}]=0,
\qquad
\mathrm{var}(\tilde{F}_{b,k})=\mathrm{var}(F_b).
\end{equation}
Substituting this decomposition into~\eqref{eq:xdot3_update_appendix} yields
\begin{equation}
\dot{x}_{3,k+1}
=
\dot{x}_{3,k}
+
\frac{\Delta t}{M_3}\bar{F}_{b,k}
+
\frac{\Delta t}{M_3}\tilde{F}_{b,k}
+
\cdots.
\end{equation}
Hence, the stochastic contribution induced by the uncertain message-passing force is
\begin{equation}
w_{\dot{x}_3,k}
=
\frac{\Delta t}{M_3}\tilde{F}_{b,k}.
\label{eq:wxdot3_appendix}
\end{equation}
Using the variance scaling identity
\begin{equation}
\mathrm{var}(aX)=a^2\mathrm{var}(X),
\end{equation}
it follows from~\eqref{eq:wxdot3_appendix} that
\begin{equation}
\mathrm{var}(w_{\dot{x}_3,k})
=
\left(\frac{\Delta t}{M_3}\right)^2 \mathrm{var}(F_b).
\label{eq:var_wxdot3_basic_appendix}
\end{equation}
Finally, inserting~\eqref{eq:varFb_final_appendix} into~\eqref{eq:var_wxdot3_basic_appendix} gives
\begin{equation}
\mathrm{var}(w_{\dot{x}_3,k})
=
\left(\frac{\Delta t}{M_3}\right)^2
\begin{bmatrix}
k_3 & c_3
\end{bmatrix}
\left(\mathbf{P}_{s_1}+\mathbf{P}_{s_2}\right)
\begin{bmatrix}
k_3 \\
c_3
\end{bmatrix}.
\label{eq:var_wxdot3_final_appendix}
\end{equation}
Therefore, the uncertainty associated with the message-passing force enters subsystem \(2\) as an additive process-noise contribution on the \(\dot{x}_3\) state. 
The displacement state \(x_3\) is affected only indirectly through time integration of the uncertain velocity, while parameter estimates are influenced indirectly through the UKF update and state-parameter coupling. %
Accordingly, the effective process-noise covariance for subsystem \(2\) is modified as
\begin{equation}
Q_{2,\mathrm{eff}}(\dot{x}_3,\dot{x}_3)
=
Q_2(\dot{x}_3,\dot{x}_3)
+
\left(\frac{\Delta t}{M_3}\right)^2
\begin{bmatrix}
k_3 & c_3
\end{bmatrix}
\left(\mathbf{P}_{s_1}+\mathbf{P}_{s_2}\right)
\begin{bmatrix}
k_3 \\
c_3
\end{bmatrix}.
\label{eq:Qeff_appendix}
\end{equation}

In implementation, the same mechanism is applied using the mean and covariance outputs of the two subsystem UKFs at
each filtering step.
The interface states \((x_2,\dot{x}_2)\) from subsystem \(1\) and \((x_3,\dot{x}_3)\) from subsystem \(2\) are
extracted from the local UKF estimates, and the mean force is evaluated as
\begin{equation}
\bar{F}_b = k_3(x_2-x_3)+c_3(\dot{x}_2-\dot{x}_3).
\end{equation}

The corresponding covariance submatrices are then used to compute \(\mathrm{var}(F_b)\) according to~\eqref{eq:varFb_final_appendix}. 
The force mean is passed to the neighboring subsystem as the deterministic message, while the force variance is injected into the receiving subsystem through the augmented process-noise term
in~\eqref{eq:Qeff_appendix}. 
Thus, the probabilistic message-passing strategy exchanges both first-order and second-order statistical information across the subsystem interface. 

To avoid artificial over-inflation of uncertainty due to repeated accumulation of the same interface-force variance
over time, an incremental uncertainty propagation strategy is employed.
Specifically, only the newly introduced variance at time step \(k\) is injected into the receiving subsystem:
\begin{equation}
\mathrm{var}(F_b)^{(k)}_{\mathrm{used}}
=
\max\!\left(
0,\,
\mathrm{var}(F_b)^{(k)}-\mathrm{var}(F_b)^{(k-1)}
\right).
\label{eq:incremental_var_appendix}
\end{equation}
This correction prevents repeated reinjection of previously accounted-for interface uncertainty and improves the
numerical stability of the distributed estimation procedure.

\begin{figure}[t]
\centering
\includegraphics[width=0.95\linewidth]{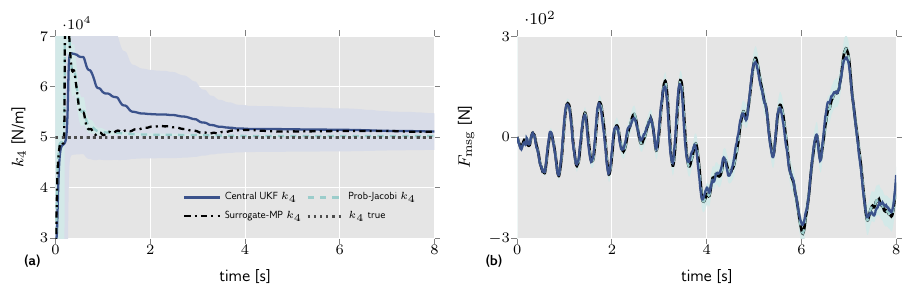}
\caption{Comparison of parameter estimation and interface-force reconstruction for the four-degree-of-freedom chain model. (a) Estimates of \(k_4\) obtained from the centralized UKF, the probabilistic Jacobi message-passing UKF, and the learned surrogate message-passing model, together with their uncertainty bounds where applicable. (b) Reconstructed interface message ($F_{\mathrm{msg}}$), including the probabilistic Jacobi mean and uncertainty envelope, the surrogate reconstruction, and the ground truth.}
\label{fig:prob_message_passing_results}
\end{figure}

We implemented this probabilistic message-passing update on the same four-degree-of-freedom inverse problem used above.
The resulting estimates are shown in Fig.~\ref{fig:prob_message_passing_results}: the dashed curves and shaded uncertainty envelopes labeled Prob-Jacobi correspond to the probabilistic Jacobi message-passing estimator. 
For the unknown stiffness parameter, the method converges toward the true value while retaining an uncertainty band, and for the interface force it reconstructs the transmitted message together with the propagated force uncertainty. 
The corresponding state estimates are not shown in this figure, but they remain accurate; the figure focuses on the parameter and interface-force quantities most directly affected by the uncertainty-propagation mechanism derived here.

Overall, the probabilistic message-passing formulation extends deterministic interface exchange by allowing each subsystem UKF to communicate not only a mean interaction force but also the uncertainty associated with that force.
This makes the coupling uncertainty-aware and provides a consistent mechanism for propagating estimation uncertainty across subsystem boundaries.

\paragraph{Quantitative comparison and calibration}

Table~\ref{tab:testbed_calibration} summarizes the accuracy and calibration metrics reported in the main
manuscript for the three matched estimators: the centralized UKF on the full augmented state, the distributed
Jacobi scheme with deterministic mean-valued messages, and the distributed Jacobi scheme with probabilistic
interface messages.
State RMSE is computed on the hidden internal states (DOFs~2 and~3) and aggregated over states and time steps;
the parameter NRMSE for the constant stiffness $k_4$ is the RMSE of the parameter trajectory divided by the true
value, so that the initial bias and the convergence transient contribute to the reported number.
Empirical coverage is the fraction of time steps at which the true hidden state lies inside the estimator's
central credible interval at the nominal level (95\% or 68\%), and the predictive negative log-likelihood (NLL)
is the time-averaged Gaussian negative log-likelihood of the true state under the filter's predictive
distribution, which penalizes both mean error and variance miscalibration.
All three estimators recover the latent states and the unknown parameter, but only the centralized UKF and the
probabilistic Jacobi scheme remain calibrated: deterministic messaging undercovers at both nominal levels, and
its NLL deteriorates by several orders of magnitude, because the receiving subsystem conditions on a single
interface value and omits the uncertainty transmitted through the coupling.
When the analytical interface law is replaced by the learned interaction law identified below, the probabilistic
Jacobi estimator attains a state RMSE of $2.38\times10^{-3}$, a parameter NRMSE of $2.39\times10^{-2}$, and a
95\% coverage of $1.00$, leaving the calibration mechanism intact.

\begin{table}[htbp]
\centering
\caption{%
  Accuracy and calibration of the three matched estimators on the four-DOF testbed.
  State RMSE is reported on the hidden internal states and parameter NRMSE on the $k_4$ trajectory; coverage is
  the empirical credible-interval coverage at the nominal 95\% and 68\% levels, and NLL is the time-averaged
  predictive negative log-likelihood.
}
\label{tab:testbed_calibration}
\footnotesize
\setlength{\tabcolsep}{4pt}
\begin{tabular}{@{}lccccc@{}}
\toprule
\textbf{Estimator} & \textbf{State RMSE} & \textbf{Param.\ NRMSE} & \textbf{Cov.\ 95\%} & \textbf{Cov.\ 68\%} & \textbf{NLL} \\
\midrule
Centralized UKF      & $4.07\times10^{-5}$ & $2.47\times10^{-2}$ & 1.00 & 1.00 & $-4.29\times10^{1}$ \\
Deterministic Jacobi & $2.02\times10^{-4}$ & $4.16\times10^{-3}$ & 0.83 & 0.66 & $4.98\times10^{3}$ \\
Probabilistic Jacobi & $1.05\times10^{-4}$ & $4.65\times10^{-3}$ & 1.00 & 1.00 & $-5.25\times10^{1}$ \\
\bottomrule
\end{tabular}
\end{table}

\paragraph{Learning the message-passing function}

Up to this point, the message-passing function has been assumed to be known, and the focus has been on how to propagate information across subsystems using either deterministic or probabilistic message passing within the distributed UKF framework.
In particular, the interface force was explicitly defined as a function of the relative displacement and velocity, and its mean and variance were propagated between subsystems.
This naturally raises the question of how the message-passing function can be identified when the coupling law is not known a priori.

To address this problem, we adopt a data-driven system identification strategy based on the sparse identification of nonlinear dynamics (SINDy) framework~\cite{Brunton2016Discovering}.
SINDy seeks to identify governing equations directly from data by constructing a library of candidate nonlinear functions and selecting a sparse subset that best explains the observed dynamics.
Given a library matrix \(\boldsymbol{\Theta}(\mathbf{z})\) composed of candidate functions of the state variables, SINDy assumes a representation of the form
\begin{equation}
y = \boldsymbol{\Theta}(\mathbf{z})\,\boldsymbol{\xi},
\end{equation}
where \(\boldsymbol{\xi}\) is a sparse vector of coefficients.
The sparse solution is obtained using a sequentially thresholded least-squares (STLSQ) procedure, which alternates between least-squares regression and coefficient thresholding to eliminate insignificant terms while retaining the dominant structure of the dynamics~\cite{Kaiser2018Sparse}. More details on the SINDy algorithm and its implementation can be found in the original paper~\cite{Brunton2016Discovering,Forootani2025robust}.
This approach yields parsimonious and interpretable models that are robust to noise and suitable for physical systems.

In the present setting, the objective is to learn the message-passing force as a function of the interface kinematic quantities.
We assume that acceleration measurements are available for the interface masses, and that the true interface force is either measured or reconstructed from high-fidelity simulations.
Displacement and velocity signals at the interface are obtained by integrating the measured accelerations.
To mitigate low-frequency drift and accumulated numerical errors, the integration is performed using a trapezoidal scheme followed by high-pass filtering, ensuring stable reconstruction of \(x_2, x_3, \dot{x}_2,\) and \(\dot{x}_3\).
The relative interface states are then formed as
\begin{equation}
\Delta x = x_2 - x_3,
\qquad
\Delta v = \dot{x}_2 - \dot{x}_3,
\end{equation}
which serve as the input variables for the identification problem. 

A candidate library is constructed to include linear and nonlinear terms of the relative states, including polynomial and dissipative contributions.
The library takes the form
\begin{equation}
\boldsymbol{\Theta} =
\begin{bmatrix}
\Delta x &
\Delta v &
(\Delta x)^3 &
|\Delta v|\,\Delta v &
\Delta x\,\Delta v &
1
\end{bmatrix},
\end{equation}
which allows the identification of both linear stiffness and damping effects as well as potential nonlinear corrections.
The SINDy regression is then applied to the interface force data using the STLSQ algorithm, with appropriate normalization and thresholding to promote sparsity.
The implementation follows a standard least-squares initialization followed by iterative pruning of small coefficients and refitting of the remaining terms, ensuring convergence to a sparse and stable model.

The resulting surrogate message-passing function is obtained as
\begin{equation}
\hat{F}_b = \hat{k}\,\Delta x + \hat{c}\,\Delta v + \alpha_3 (\Delta x)^3 + \alpha_4 |\Delta v|\,\Delta v,
\end{equation}
where the dominant contributions correspond to the linear stiffness and damping terms, while higher-order terms capture mild nonlinear effects. 
The identified coefficients are \(\hat{k}=5.613034\times 10^{4}\,\mathrm{N/m}\),
\(\hat{c}=3.308257\times 10^{2}\,\mathrm{Ns/m}\), \(\alpha_3=-2.863096\times 10^{8}\,\mathrm{N/m^3}\), and
\(\alpha_4=6.696752\times 10^{3}\,\mathrm{Ns^2/m^2}\), showing that the learned message-passing law is dominated by the expected linear stiffness and damping terms, with additional cubic stiffness and nonlinear damping corrections; the leading coefficients lie within $12.2\%$ and $10.3\%$ of the true interface stiffness $k_3$ and damping $c_3$, as reported in the main manuscript.
In practice, the identified coefficients are extracted directly from the sparse coefficient vector
\(\boldsymbol{\xi}\), with \(\hat{k}\) and \(\hat{c}\) corresponding to the leading linear terms.
The identification procedure is implemented in Python using a custom SINDy routine based on sequential thresholded least squares, as illustrated in the provided implementation~\cite{Kaptanoglu2022PySINDy}. %

The effect of using the learned message-passing function is summarized in
Fig.~\ref{fig:prob_message_passing_results}. 
In panel~(a), the learned surrogate message-passing estimator is shown together with the centralized and probabilistic Jacobi estimators for the unknown stiffness parameter \(k_4\). 
Although the surrogate is identified from interface data rather than specified analytically, its parameter estimate settles close to the true value, indicating that the learned coupling law preserves the information needed for state-parameter inference. 
Panel~(b) compares the reconstructed interface message with the true force; the surrogate message-passing curve tracks the main temporal structure of the interface force, showing that the learned law can replace the analytical spring-damper coupling in the distributed estimator.
As in the probabilistic case, the figure emphasizes the parameter and interface-message quantities most directly associated with the learned interaction law; the corresponding state estimates are not shown here.

This learned message-passing function can then be used in place of the analytical coupling law within the distributed UKF framework, enabling a fully data-driven representation of subsystem interactions while preserving the probabilistic uncertainty propagation mechanism described in the previous section.

\subsubsection*{Scalability}

\begin{figure}
\centering
\includegraphics[width=\linewidth]{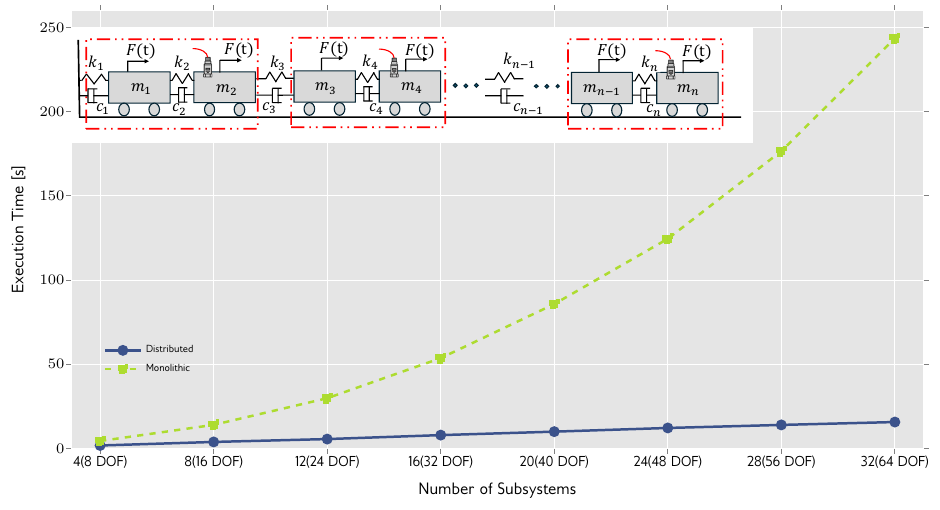}
\caption{
Execution-time scaling of centralized vs distributed UKF as the chain size increases from 8 to 64 DOFs (4–32 subsystems).
The monolithic UKF shows rapidly increasing cost with DOF due to growth of the augmented state and global covariance operations, whereas the distributed UKF scales approximately linearly because estimation is performed on fixed-size subsystem filters coupled only through low-dimensional interface-force messages (schematic inset).
}
\label{fig:comp_time_scal}
\end{figure}

To assess scalability, we use a 64-DOF damped mass--spring chain decomposed into 32 interacting 2-DOF subsystems coupled through a single spring--damper interface and exchanging an equivalent interaction-force message.
Building on this template, we scale the model by keeping the leftmost 2-DOF subsystem fixed and appending additional 2-DOF subsystems on the right per step.
Each added subsystem introduces one additional unknown stiffness parameter (local to that subsystem), and we assume one acceleration measurement per subsystem, preserving sparse, local sensing as the network grows.
The inverse problem is performed at the subsystem level, posed as state estimation with known physical parameters and a single acceleration measurement at DOF~2 for the first subsystem and as joint state--parameter inference under stochastic excitation using a UKF for the remaining subsystems.
The same subsystem partitioning, discretization, and Jacobi message-passing technique is used for all system sizes, and the same UKF settings are applied to each subsystem, regardless of the total number of subsystems in the chain.
Figure~\ref{fig:comp_time_scal} compares the computational cost of monolithic vs distributed formulations as the total number of DOFs increases.
The monolithic solution becomes rapidly computationally prohibitive as the system size increases, since enlarging the augmented state expands both the number of sigma points and the dimension of the covariance matrices; the associated covariance propagation, Cholesky factorization, and matrix inversion operations scale as $\mathcal{O}(n^3)$ with the total number of states $n$.
In contrast, the distributed formulation exhibits approximately linear scaling with respect to the number of subsystems, as each filter operates on a fixed-size local state and interconnections are handled through low-dimensional interface messages, thereby avoiding the construction, factorization, or inversion of a global $n \times n$ covariance matrix.

\subsubsection*{Six-DOF damped mass--spring system}

\begin{figure}[htbp]
    \centering
    \includegraphics[width=0.8\linewidth]{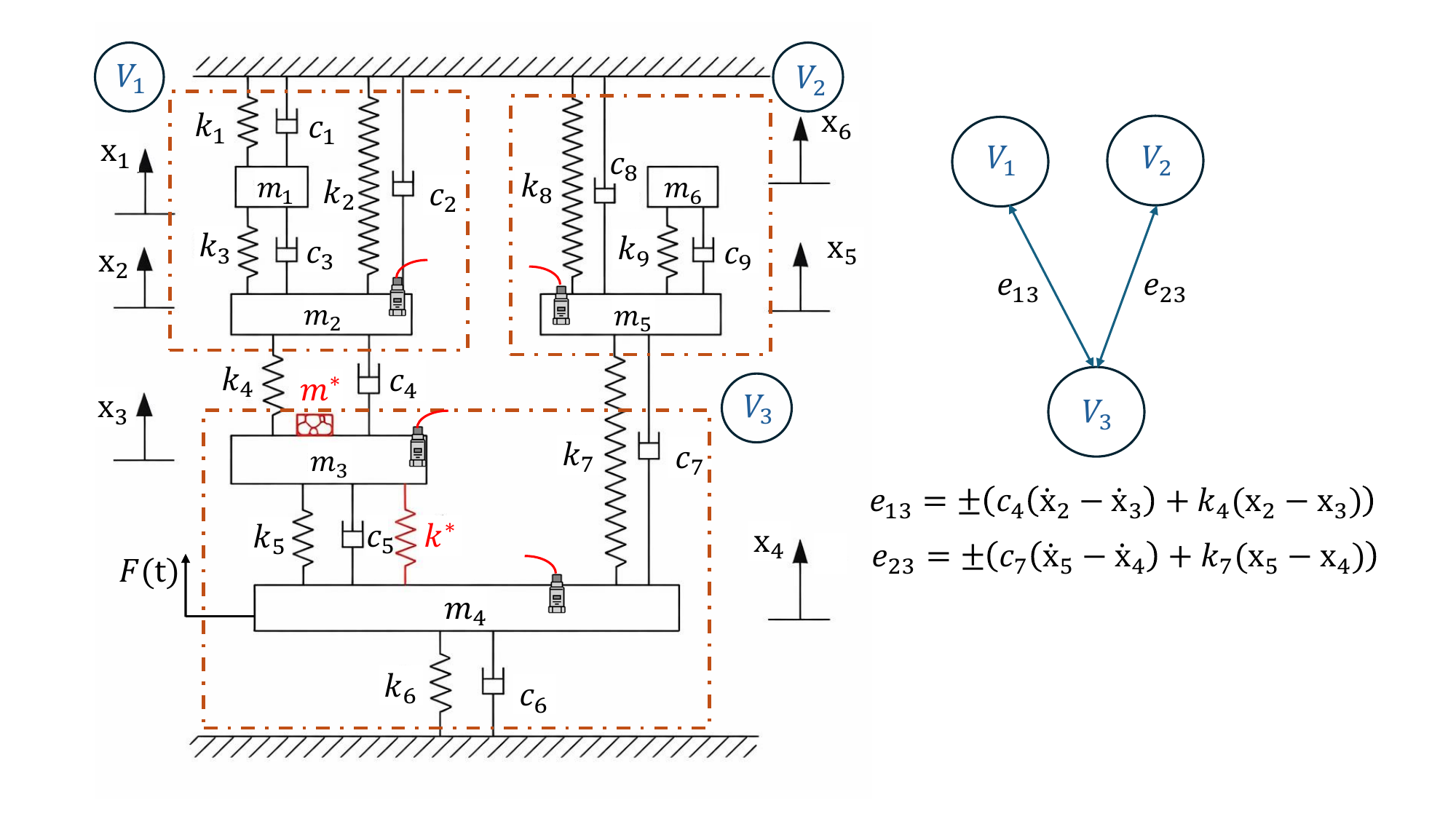}
    \caption{
    Six-DOF mass--spring--damper system decomposed into the subsystems $V_1$ (masses 1--2), $V_3$ (masses 3--4),
    and $V_2$ (masses 5--6), coupled through the interface forces $e_{13}$ and $e_{23}$ (right: interaction
    graph and interface laws).
    Accelerometer symbols mark the measured masses 2--5; the added mass $m^{*}$ and the internal stiffness
    $k^{*}$ of subsystem $V_3$, highlighted in red, are among the unknown parameters estimated online.
    }
    \label{fig:6dof-system}
\end{figure}

As an additional study beyond the experiments reported in the main manuscript, we consider a one-dimensional chain of six lumped masses connected through linear springs and viscous dampers~\cite{Zhu2019Maintaining}, as shown in Fig.~\ref{fig:6dof-system}.
This larger chain provides a second inverse-estimation example and serves as the testbed for the diffusion-based
uncertainty propagation introduced in the theoretical background.
The system follows the standard second-order dynamics
\begin{equation}
M \ddot{x}(t) + C \dot{x}(t) + K x(t) = f(t),
\end{equation}
where the displacement vector is defined as $x(t) = [x_1,x_2,x_3,x_4,x_5,x_6]^\top$.
All masses are taken as $m_1=\cdots=m_6=500~\mathrm{kg}$, and an additional unknown lumped mass $m^*=100~\mathrm{kg}$ is attached to the third mass.
The system is excited by an external force applied at DOF~4, and noisy acceleration measurements are collected at selected locations. 

The inverse problem is to recover the latent displacement and velocity states of the six-DOF chain while simultaneously identifying the unknown physical parameters from sparse acceleration measurements.
The available measurements are the accelerations \(a_2\), \(a_3\), \(a_4\), and \(a_5\), corresponding to DOFs~2--5; the displacement and velocity states are not measured directly.
The system is decomposed into three interacting subsystems, as shown in Fig.~\ref{fig:6dof-system}: Subsystem~V1
contains $(x_1,x_2)$, Subsystem~V3 contains $(x_3,x_4)$, and Subsystem~V2 contains $(x_5,x_6)$.

The unknown parameters are the local stiffness and damping pair \((k_3,c_3)\) in subsystem V1, the added-mass and internal V3 parameters \((m^*,k_6,c_6,k^*)\), and the right-subsystem (V2) stiffness and damping pair \((k_9,c_9)\).
Thus, the task is a joint state--parameter estimation problem under sparse sensing, in which the measurements are distributed across the chain rather than available as a full-system observation.

Each subsystem evolves according to its local dynamics, while the coupling between them is enforced through interface forces.

Subsystem~V1 is governed by
\begin{align}
m_1 \ddot{x}_1 &= -k_1 x_1 - c_1 \dot{x}_1 - k_3(x_1-x_2) - c_3(\dot{x}_1-\dot{x}_2), \\
m_2 \ddot{x}_2 &= -k_2 x_2 - c_2 \dot{x}_2 + k_3(x_1-x_2) + c_3(\dot{x}_1-\dot{x}_2) - e_{13},
\end{align}
where $(k_3,c_3)$ are unknown and estimated online.

Subsystem~V3 incorporates the additional mass and nonlinear coupling through
\begin{align}
(m_3+m^*) \ddot{x}_3 &= -k^*(x_3-x_4) - c_5(\dot{x}_3-\dot{x}_4) + e_{13}, \\
m_4 \ddot{x}_4 &= k^*(x_3-x_4) + c_5(\dot{x}_3-\dot{x}_4) 
- k_6 x_4 - c_6 \dot{x}_4 + f(t) - e_{23},
\end{align}
where $(m^*,k_6,c_6,k^*)$ are treated as unknown parameters.

Subsystem~V2 is described by
\begin{align}
m_5 \ddot{x}_5 &= -k_8 x_5 - c_8 \dot{x}_5 - k_9(x_5-x_6) - c_9(\dot{x}_5-\dot{x}_6) + e_{23}, \\
m_6 \ddot{x}_6 &= k_9(x_5-x_6) + c_9(\dot{x}_5-\dot{x}_6),
\end{align}
with $(k_9,c_9)$ being unknown.

The interaction between subsystems is captured through deterministic message passing using the interface forces
\begin{align}
e_{13} &= k_4(x_2-x_3) + c_4(v_2-v_3), \\
e_{23} &= k_7(x_4-x_5) + c_7(v_4-v_5).
\end{align}
The signs of $e_{13}$ and $e_{23}$ entering the equations of motion follow the action--reaction convention shown
schematically in Fig.~\ref{fig:6dof-system}: $e_{13}$ enters subsystem~V1 with a negative sign and subsystem~V3
with a positive sign, while $e_{23}$ enters subsystem~V3 with a negative sign and subsystem~V2 with a positive
sign.
A Jacobi-type update is employed, in which all subsystems are propagated in parallel using interface quantities from the same time step, ensuring consistency across the distributed system.

Operationally, the sparse measurements are assigned to the local filters according to the subsystem decomposition:
Subsystem~V1 uses $a_2$, Subsystem~V3 uses $(a_3,a_4)$, and Subsystem~V2 uses $a_5$.
The measurements are generated by adding Gaussian noise with variance $10^{-3}$ (or $10^{-2}$ for V3) to the true accelerations.

Each subsystem performs joint state--parameter estimation using an independent UKF.
The augmented states include both kinematic variables and unknown parameters, which are modeled as random walks.
A common time step of $\Delta t = 2\times10^{-3}\,\mathrm{s}$ is used across all subsystems.
Using the same hyperparameters for all subsystems, the initial parameter guesses are intentionally biased at $70\%$ of their true values to assess convergence behavior.

\begin{figure}[htbp]
    \centering
    \includegraphics[width=\linewidth]{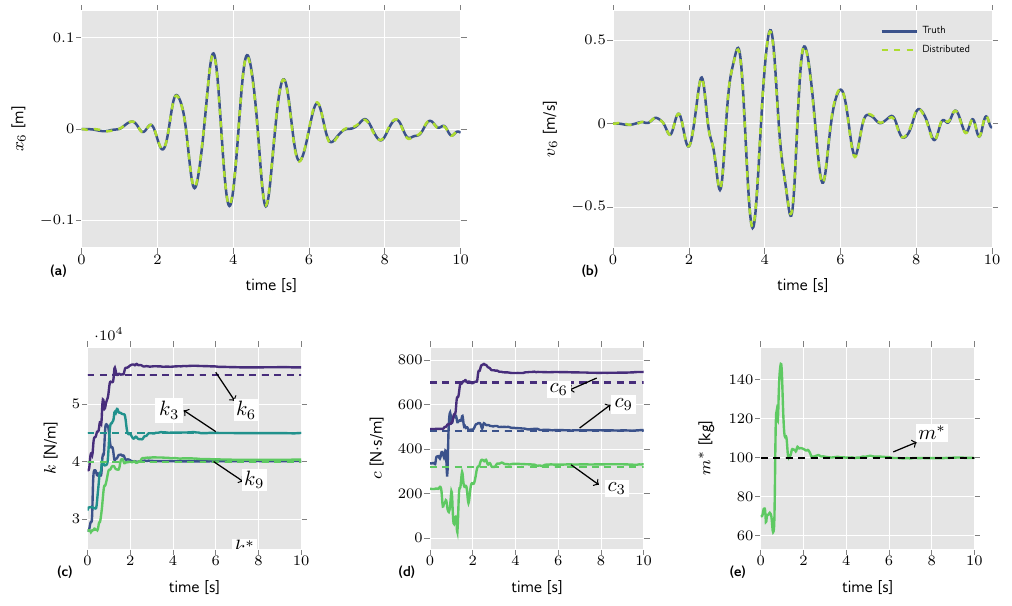}
    \caption{
    Distributed estimation results for the six-DOF chain under sparse sensing.
    (a) Reconstructed displacement $x_6$ and (b) reconstructed velocity $v_6$ at the unmeasured DOF~6 (solid:
    truth; dashed: distributed estimate).
    (c) Identified stiffness parameters ($k_3$, $k_6$, $k_9$, $k^{*}$),
    (d) identified damping parameters ($c_3$, $c_6$, $c_9$), and
    (e) identified equivalent mass $m^{*}$, each converging from biased initializations; dashed lines mark the
    true values.
    }
    \label{fig:6dof-parameter-estimation-results}
\end{figure}

The resulting estimation performance is shown in Fig.~\ref{fig:6dof-parameter-estimation-results}.
Panels~(a) and~(b) show that the distributed estimator accurately reconstructs the unmeasured
internal displacement and velocity responses, $x_6$ and $v_6$.
Panels~(c)--(e) show the corresponding parameter recovery: the stiffness parameters, damping parameters, and
equivalent mass converge toward their true values despite biased initialization and limited sensing.
These results indicate that the inverse problem remains well-posed under distributed estimation with sparse measurements.

In this study, only the mean values of the subsystem states are exchanged through the message-passing mechanism.
However, the framework naturally allows for the propagation of uncertainty across subsystems, enabling future extensions in which full covariance information is communicated to improve robustness and consistency. We next investigate how local modifications in subsystem $V_3$ propagate through the system and affect remote responses.

\paragraph{Uncertainty propagation through diffusion models}

In this section, rather than recomputing the full dynamics for each modification in one subsystem, we leverage the subsystem structure, take advantage of the diffusion process, and reuse the message-passing variables to construct a diffusion-based sensitivity model.

The three-subsystem decomposition introduced previously naturally defines a graph:
\[
V_1 \longleftrightarrow V_3 \longleftrightarrow V_2,
\]
where the connections correspond to the physical interface forces already used in the distributed estimator.
Specifically, the coupling between $V_1$ and $V_3$ is given by
\begin{equation}
e_{13}(t)=k_4\big(x_2-x_3\big)+c_4\big(v_2-v_3\big),
\end{equation}
and the coupling between $V_3$ and $V_2$ is
\begin{equation}
e_{23}(t)=k_7\big(x_4-x_5\big)+c_7\big(v_4-v_5\big).
\end{equation}

These interface forces are not introduced artificially; they are already computed during the distributed UKF updates and therefore provide a natural mechanism to quantify how information flows across subsystems.

To translate physical modifications into graph signals, we introduce the concept of a defect force.
A local change in subsystem $V_3$ is interpreted as a missing internal contribution, which is expressed as an equivalent forcing term.

For example, removing the internal stiffness $k^{*}$ leads to the defect
\begin{equation}
\Delta f_{k^{*}}(t)=-\,k^{*}\big(x_3-x_4\big),
\end{equation}
while removing the added mass $m^{*}$ induces an inertia-based defect
\begin{equation}
\Delta f_{m^{*}}(t)=-\,m^{*}\,a_3^{\mathrm{base}}(t),
\end{equation}
where the baseline acceleration $a_3^{\mathrm{base}}(t)$ is reconstructed from the subsystem dynamics.
In both cases, the driving signal for the sensitivity analysis is taken as the scalar magnitude
\begin{equation}
q_\mathrm{def}(t)=|\Delta f(t)|,
\end{equation}
which represents the instantaneous strength of the local perturbation.

The key idea is that this defect does not remain localized: it propagates through the subsystem graph according to the strength of the interface couplings.
To quantify this, we construct edge weights using the root-mean-square (RMS) magnitude of the interface forces,
\begin{equation}
w_{13} \propto \mathrm{RMS}(e_{13}), 
\qquad 
w_{23} \propto \mathrm{RMS}(e_{23}),
\end{equation}
which are then normalized to obtain diffusion affinities $\eta_{13}$ and $\eta_{23}$.
These weights encode how strongly subsystem $V_3$ communicates with its neighbors.

Two complementary propagation models are considered.

In the first model, a local 1-hop approximation is used, where the defect is partitioned between the source subsystem and its immediate neighbors.
Using a leakage parameter $\alpha=0.6$, the defect is split as
\begin{equation}
s_{V3}(t)=\frac{q_\mathrm{def}(t)}{1+\alpha}, \qquad
s_{V1}(t)=\eta_{13}\frac{\alpha\,q_\mathrm{def}(t)}{1+\alpha}, \qquad
s_{V2}(t)=\eta_{23}\frac{\alpha\,q_\mathrm{def}(t)}{1+\alpha}.
\end{equation}
This model preserves locality and provides a direct interpretation of how perturbations are distributed across adjacent subsystems.

In the second model, a global diffusion mechanism is introduced through the graph Laplacian.
The weighted adjacency matrix constructed from the interface forces defines a Laplacian operator $\mathbf{L}$, and
the propagation is governed by the heat kernel
\begin{equation}
\mathbf{H}(\beta)=\exp(-\beta \mathbf{L}),
\end{equation}
with diffusion scale $\beta=0.9$.
The scalar source $q_\mathrm{def}(t)$ is embedded as a node-source vector
$\mathbf{q}_\mathrm{def}(t) = q_\mathrm{def}(t)\,\mathbf{e}_{V_3}\in\mathbb{R}^{N_s}$, where $\mathbf{e}_{V_3}$
is the unit vector with value $1$ at the source subsystem $V_3$ and $0$ at all other subsystems, and $N_s$ is
the number of subsystems.
The propagated node-level signal is then obtained as
\begin{equation}
\mathbf{s}(t)=\mathbf{H}(\beta)\,\mathbf{q}_\mathrm{def}(t),
\end{equation}
which accounts for all paths in the graph and produces a smoother, globally consistent redistribution of the defect.

The resulting node-level sensitivities are not yet expressed in physical units.
To map them back to displacements, they are scaled using the uncertainty of the distributed estimator.
Specifically, the standard deviations $\sigma_{x_1}=\sqrt{P_1(1,1)}$ and $\sigma_{x_6}=\sqrt{P_2(2,2)}$, extracted
from the terminal state covariances $P_1$ and $P_2$ of the $V_1$ and $V_2$ subsystem filters, are used to construct
envelopes around the baseline responses,
\begin{align}
x_1(t) &\in x_1^{\mathrm{base}}(t) \pm s_{V1}(t)\,\sigma_{x_1}, \\
x_6(t) &\in x_6^{\mathrm{base}}(t) \pm s_{V2}(t)\,\sigma_{x_6}.
\end{align}

\begin{figure}[t]
\centering
\includegraphics[width=\linewidth]{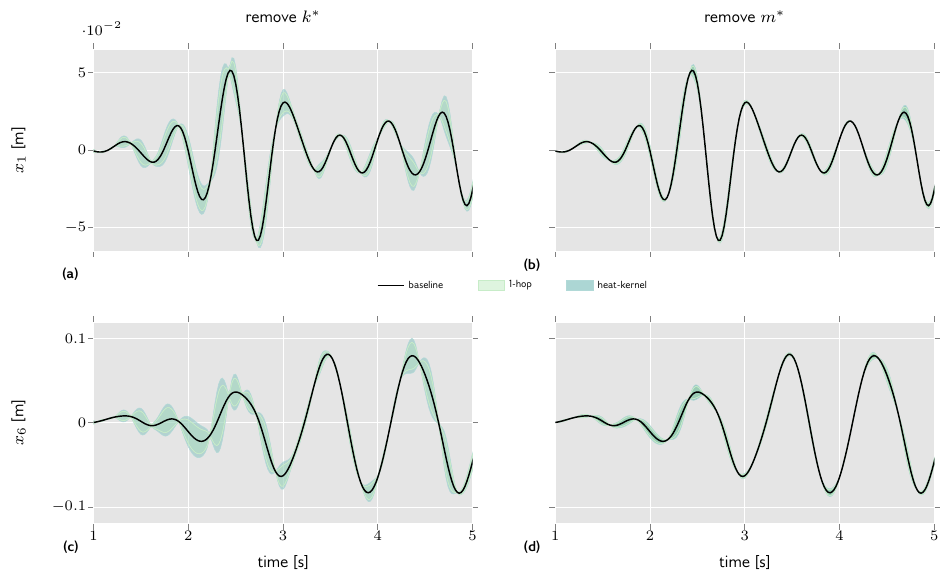}
\caption{%
Diffusion-based sensitivity envelopes for the six-DOF system under local physics removal in subsystem $V_3$.
Panels~(a) and~(c) show removal of the internal stiffness $k^{*}$; panels~(b) and~(d) show removal of the added mass
$m^{*}$.
Panels~(a) and~(b) report the response at $x_1$, whereas panels~(c) and~(d) report the response at $x_6$.
In each panel, the black curve denotes the baseline response obtained from the distributed UKF, the light-green band denotes the 1-hop diffusion envelope defined by~\eqref{eq:appendix_1hop_scores}, and the teal band denotes the heat-kernel envelope defined by Eqs.~(\ref{eq:appendix_heat_kernel})\textendash(\ref{eq:appendix_hk_scores}).
The envelopes are obtained by converting the local parameter removal into a defect force, propagating that defect on the subsystem graph, and scaling the resulting node scores by the terminal subsystem uncertainties.
}
\label{fig:appendix_diffusion_envelope}
\end{figure}

Figure~\ref{fig:appendix_diffusion_envelope} summarizes how the network structure of the six-DOF system can be used to perform fast post-hoc sensitivity analysis after a local physics modification in subsystem $V_3$, without re-solving the full dynamical system.
Panels~(a) and~(c) show the effect of removing the internal stiffness $k^{*}$, while panels~(b) and~(d) show the
effect of removing the added mass $m^{*}$.
Panels~(a) and~(b) report the induced sensitivity envelopes at $x_1$, and panels~(c) and~(d) report the corresponding
envelopes at $x_6$.
In each panel, the black curve denotes the baseline response obtained from the distributed UKF, while the colored bands represent the sensitivity envelopes obtained by converting the local parameter change into an equivalent defect force, propagating that defect through the subsystem graph, and mapping the resulting node-level sensitivity back to the displacement response using the terminal subsystem uncertainty.

The main message of the figure is not to compare diffusion operators, but to show that once the system has been decomposed into interacting subsystems, the same message-passing variables used for estimation can also be reused to propagate local modifications through the network at negligible computational cost.
In this way, the network serves not only as a scalable representation of the full system, but also as a mechanism for rapid sensitivity screening.
For the present example, the distributed UKF requires $17.4813\,\mathrm{s}$, whereas the sensitivity propagation requires only $1.3264\,\mathrm{ms}$ using the 1-hop diffusion model and $3.2720\,\mathrm{ms}$ using the heat-kernel model, corresponding to speedups of approximately $13{,}180\times$ and $5{,}343\times$, respectively.
This demonstrates that local subsystem edits can be assessed almost instantly once the subsystem graph and interface-message structure have been identified.

\subsection*{Power-grid scalability benchmarks}
\label{App:power_grid}

This section supports the second case study of the main manuscript, which shows that subsystem structure
enables near-linear computational scaling without loss of accuracy or calibration.
The benchmarks use the standard IEEE power-network test cases~\cite{Thurner2018Pandapower} ranging from 9 to
300 buses.
These benchmarks represent canonical models of regional and national transmission grids, and are widely used in the
power-systems community for various applications, including evaluating the computational performance of analysis
and estimation algorithms.
All network data (bus parameters, branch impedances, and generator locations) are retrieved from the PYPOWER
library~\cite{Zimmerman2011MATPOWER}, which provides Python implementations of the standard IEEE test cases. 
Our goal is to implement the distributed approach on these benchmarks, compare the computational performance with the monolithic methods, and demonstrate that the message-passing estimator can be applied to a large-scale network with realistic topology and coupling structure, and that the computational cost scales favorably with system size.

\begin{figure}[htbp]
\centering
\includegraphics[width=0.82\linewidth]{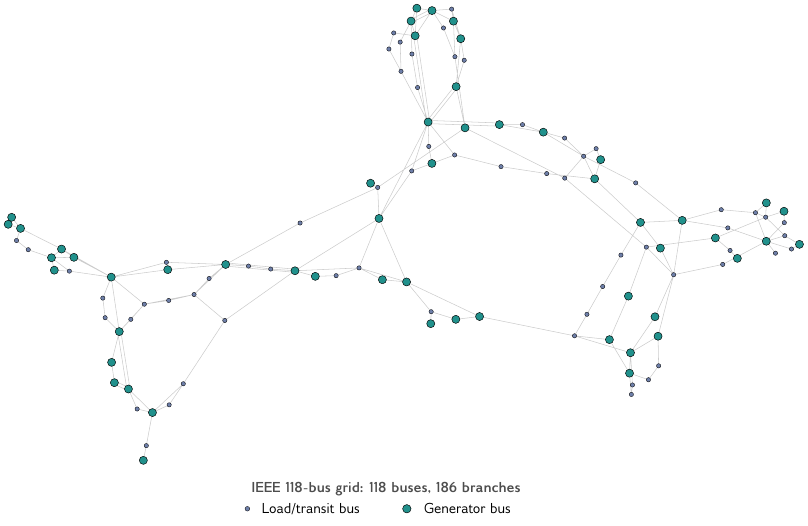}
\caption{%
  Representative IEEE power-network topology used in the scalability benchmarks.
  The IEEE 118-bus grid is shown with transmission branches as gray edges, load/transit buses as small dark-blue
  nodes, and generator buses as larger teal nodes.
  This graph structure determines the sparse electrical coupling pattern used to construct the bus admittance matrix.
}
\label{fig:ieee_grid_topology}
\end{figure}

For each benchmark case, the electrical network topology, illustrated for the IEEE 118-bus system in
Fig.~\ref{fig:ieee_grid_topology}, is encoded in the bus admittance matrix $\mathbf{Y}_\mathrm{bus} \in
\mathbb{C}^{N\times N}$, constructed using the standard nodal admittance formulation from the branch impedance and
shunt data~\cite{Kundur2012Power}.
The off-diagonal entry $Y_{ij}$ ($i \neq j$) is the negative of the series admittance of the branch connecting
buses $i$ and $j$; the diagonal entries include the sum of all admittances incident to bus $i$.
For the dynamical coupling model, the coupling coefficient between nodes $i$ and $j$ is taken as the magnitude of
the corresponding off-diagonal admittance entry:
\begin{equation}
  K_{ij} = |Y_{ij}|, \quad i \neq j,
  \qquad
  K_{ii} = 0,
  \label{eq:K_from_Ybus}
\end{equation}
so that more strongly coupled branches in the electrical network produce larger synchronizing forces in the
dynamical model.

In a real power grid, the evolution of phase angles $\delta_i$ at each bus in time is governed by the
electromechanical swing dynamics of the synchronous generators: when a disturbance shifts the power balance at any
bus, the rotor angles begin to oscillate and the network either recovers a common synchronous frequency or loses
synchrony entirely.
Capturing this behavior requires a dynamical model that couples the phase angles of all buses through the network
topology, reproduces the nonlinear sinusoidal power-angle relationship, and remains tractable enough to simulate at
scale~\cite{Dorfler2013Synchronization}.
The Kuramoto model satisfies all three requirements~\cite{Kuramoto1984Chemical, Kuramoto2005Selfentrainment}.
It assigns a phase $\theta_i(t)$ and a natural frequency $\Omega_i$ to each bus, and couples them through exactly
the sinusoidal term $K_{ij}\sin(\theta_j - \theta_i)$ that appears in the AC power-flow equations. The
Kuramoto/swing formulation therefore provides a reduced dynamical representation that preserves the
sinusoidal power-angle coupling structure of AC power-flow models while abstracting away the algebraic
network constraints.
Depending on whether the inertia of the synchronous machines is significant, two variants of the Kuramoto model are
used: a first-order (overdamped) form appropriate for load buses or quasi-static regimes, and a second-order
(inertial) swing form appropriate for generator buses where rotor mass effects cannot be neglected.
Both are defined in the following subsections.
\subsubsection*{First-order Kuramoto model}

The classical Kuramoto model~\cite{Acebron2005Kuramoto} describes a network of $N$ coupled phase oscillators in
which each node $i$ has a scalar phase $\theta_i(t)$ that evolves as
\begin{equation}
  \dot{\theta}_i
  =
  \Omega_i
  +
  \sum_{j=1}^{N} K_{ij} \sin(\theta_j - \theta_i),
  \quad i = 1, \ldots, N,
  \label{eq:kuramoto1}
\end{equation}
where $\Omega_i$ is the natural frequency of node $i$ and $K_{ij}$ is the coupling strength between nodes $i$ and
$j$.
In the power-systems context, $\theta_i$ represents the rotor phase angle, $\Omega_i$ encodes the mismatch between
mechanical input power and electrical load, and the coupling term models the synchronizing power exchanged over
transmission lines.
The first-order model~\eqref{eq:kuramoto1} assumes that inertia and damping effects are negligible, so that the
phase adjusts instantaneously to the net torque, and it is appropriate for studying synchronization phenomena in
the overdamped or quasi-static regime.
The system state is $\boldsymbol{\theta} = [\theta_1, \ldots,
\theta_N]^\top \in \mathbb{R}^N$, and the network
topology is fully encoded in the coupling matrix $\mathbf{K} \in \mathbb{R}^{N \times N}$.

\subsubsection*{Second-order Kuramoto (swing) model}

For large-scale power grids, the inertia and damping of the synchronous generators are non-negligible, and a more
physically accurate representation is the second-order Kuramoto model, also known as the swing
equation~\cite{Filatrella2008Analysis, Dorfler2014Synchronization}:
\begin{equation}
  m_i \ddot{\theta}_i + d_i \dot{\theta}_i
  =
  \Omega_i
  +
  \sum_{j=1}^{N} K_{ij} \sin(\theta_j - \theta_i),
  \quad i = 1, \ldots, N,
  \label{eq:kuramoto2}
\end{equation}
where $m_i > 0$ is the per-unit inertia coefficient, $d_i > 0$ is the damping coefficient, $\omega_i =
\dot{\theta}_i$ is the angular frequency deviation from the synchronous reference, and all other quantities are as
in~\eqref{eq:kuramoto1}.
Equation~(\ref{eq:kuramoto2}) reduces to the first-order model~(\ref{eq:kuramoto1}) when inertia is negligible
($m_i \to 0$) and the system is overdamped.
Defining $\omega_i = \dot{\theta}_i$, the second-order model is written as two first-order equations with state vector
at each node $\mathbf{x}_i = [\theta_i, \omega_i]^\top$:
\begin{align}
  \dot{\theta}_i &= \omega_i, \label{eq:sw_theta}\\
  \dot{\omega}_i
    &= \frac{1}{m_i}
       \left(
         -d_i\,\omega_i
         + \Omega_i
         + \sum_{j=1}^{N} K_{ij}\sin(\theta_j - \theta_i)
       \right).
  \label{eq:sw_omega}
\end{align}
\paragraph{Simulation setup}
In all cases, we assume that all the nodes follow the second-order Kuramoto model and the inertia is set to the uniform value $m_i = 1$ (absorbed into the
damping and coupling coefficients), so that the equations simplify to
\begin{align}
  \dot{\theta}_i &= \omega_i,\\
  \dot{\omega}_i &= -d_i\,\omega_i + \Omega_i
    + \sum_{j=1}^{N} K_{ij}\sin(\theta_j - \theta_i),
  \label{eq:swing_simple}
\end{align}
which is the form integrated numerically using the Heun method~\cite{Hairer1993RungeKutta} with time step $\Delta t = 0.01$\,s over a horizon
of $T = 3.0$\,s (300 steps total).
The damping coefficients are drawn independently from a uniform distribution, $d_i \sim \mathcal{U}(0.1, 0.30)$,
rounded to two decimal places.
The natural frequencies are drawn independently from a discrete uniform distribution over the set $\{-1.0, -0.9,
\ldots, 0.9, 1.0\}$ rad/s.
All phase angles are wrapped to the interval $(-\pi, \pi]$ to prevent unbounded growth.

\subsubsection*{Joint state and parameter estimation in power-grid systems}

The goal of this section is to estimate the states and natural frequencies of the power grid for all the network layouts and compare the computational performance between centralized and distributed approaches.
The identification task is formulated as a joint state--parameter estimation problem in which the network state
$(\boldsymbol{\theta}, \boldsymbol{\omega})$ and the unknown natural frequencies $\boldsymbol{\Omega} = [\Omega_1,
\ldots,
\Omega_N]^\top$ are simultaneously estimated from noisy
measurements of phase and frequency at every bus.
The damping coefficients $d_i$ are treated as known.
We use the UKF for joint state--parameter identification.

\paragraph{Augmented state vector.}
The natural frequencies are appended to the dynamical state to form the augmented state vector for the full
network:
\begin{equation}
  \mathbf{x}
  =
  \bigl[
    \boldsymbol{\theta}^\top,\;
    \boldsymbol{\omega}^\top,\;
    \boldsymbol{\Omega}^\top
  \bigr]^\top
  \in \mathbb{R}^{3N},
  \label{eq:aug_state_central}
\end{equation}
where the parameter dynamics follow a random-walk model, $\dot{\boldsymbol{\Omega}} = \mathbf{0}$, driven entirely
by process noise.
This is the standard augmented-state approach for the UKF, which
allows the filter to track slow parameter variations without explicit knowledge of the parameter dynamics.

\paragraph{Process model.}
Writing the augmented state as $\mathbf{x} = [\boldsymbol{\theta},\,\boldsymbol{\omega},\,\boldsymbol{\Omega}]^\top$
and the right-hand side of~\eqref{eq:swing_simple} as
$\mathbf{F}(\mathbf{x})$,
each UKF sigma point is advanced by one step using the explicit Heun (improved Euler) predictor--corrector
\begin{align}
  \tilde{\mathbf{x}}_{k+1} &= \mathbf{x}_k + \Delta t\,\mathbf{F}(\mathbf{x}_k),
    \label{eq:heun_pred}\\
  \mathbf{x}_{k+1}         &= \mathbf{x}_k
    + \tfrac{\Delta t}{2}\bigl[\mathbf{F}(\mathbf{x}_k) + \mathbf{F}(\tilde{\mathbf{x}}_{k+1})\bigr].
    \label{eq:heun_corr}
\end{align}

The natural-frequency block $\boldsymbol{\Omega}$ has zero deterministic drift in $\mathbf{F}$, so its only
evolution comes from the augmented-state random walk.

\paragraph{Observation model.}
Phase angles and angular frequencies are measured at every bus,
\begin{equation}
  \mathbf{y}(t_k)
  =
  \begin{bmatrix}
    \boldsymbol{\theta}(t_k)\\
    \boldsymbol{\omega}(t_k)
  \end{bmatrix}
  +
  \begin{bmatrix}
    \mathbf{v}_\theta\\
    \mathbf{v}_\omega
  \end{bmatrix},
  \quad
  \mathbf{v}_\theta,\,\mathbf{v}_\omega
  \sim \mathcal{N}(\mathbf{0},\, \sigma^2 \mathbf{I}_N),
  \label{eq:obs_grid}
\end{equation}
with measurement noise standard deviation $\sigma = 0.02$\,rad (or rad/s) for both channels.
The observation vector has dimension $2N$, covering all buses.

\paragraph{Filter settings.}
The UKF scaling parameter is $\gamma = 1$.
The initial error covariance, process noise covariance, and measurement noise covariance matrices are diagonal:
\begin{align}
  \mathbf{P}_0
  &=
  \mathrm{diag}\!\bigl(
    0.25\,\mathbf{1}_N,\;
    0.25\,\mathbf{1}_N,\;
    1.0\,\mathbf{1}_N
  \bigr),
  \label{eq:P0_central}
  \\
  \mathbf{Q}
  &=
  \mathrm{diag}\!\bigl(
    10^{-4}\,\mathbf{1}_N,\;
    10^{-4}\,\mathbf{1}_N,\;
    10^{-4}\,\mathbf{1}_N
  \bigr),
  \label{eq:Q_central}
  \\
  \mathbf{R}
  &=
  \mathrm{diag}\!\bigl(
    \sigma^2\,\mathbf{1}_N,\;
    \sigma^2\,\mathbf{1}_N
  \bigr)
  = 4\times10^{-4}\,\mathbf{I}_{2N}.
  \label{eq:R_central}
\end{align}
Initial state estimates are perturbed from the true initial conditions: phase angles are offset by
$\mathcal{N}(0,\,0.04)$\,rad, frequencies by $\mathcal{N}(0,\,0.04)$\,rad/s, and natural frequencies are
initialized at zero (no prior information).

\subsubsection*{Scalability: distributed estimation and network partitioning}

The centralized UKF applied to the augmented state~\eqref{eq:aug_state_central} requires propagating $2 \times 3N +
1 = 6N + 1$ sigma points through the full $N$-node process model at every time step.
The dominant cost is the Cholesky decomposition of the $3N \times 3N$ error covariance matrix, which scales as
$\mathcal{O}(N^3)$, followed by the cross-covariance and gain computations~\cite{Golub2013Matrixa}.
For $N = 300$, the augmented state has 900 components, making the centralized UKF computationally prohibitive for
near-real-time applications.

\paragraph{Generator-seeded graph partitioning.}
To exploit the modular structure of the power grid, the $N$ buses are partitioned into disjoint subsystems, each
anchored at a generator bus.
Generator buses are identified from the PYPOWER database using the generator dispatch table.
The partitioning algorithm proceeds as follows:
\begin{enumerate}
  \item \textbf{Initialization.} Each generator bus $g_c$ seeds an
    independent cluster $\mathcal{C}_c = \{g_c\}$.
    All other buses are initially unassigned.
  \item \textbf{Greedy expansion.} The clusters are grown
    iteratively.
    At each iteration, for every cluster $\mathcal{C}_c$ with
    $|\mathcal{C}_c| < S_{\max}$, the set of unassigned
    candidate nodes is
    \begin{equation}
      \mathcal{V}_c
      = \bigl\{
        v \notin \bigcup_{c'} \mathcal{C}_{c'}
        \;\mid\;
        \textstyle\sum_{u \in \mathcal{C}_c} K_{vu} > 0
      \bigr\}.
    \end{equation}
    The best candidate is the node that maximizes the internal
    coupling weight to the cluster,
    \begin{equation}
      v^*
      = \arg\max_{v \in \mathcal{V}_c}
        \Bigl(
          w_\mathrm{int}(c, v),\;
          \rho(c, v)
        \Bigr),
      \label{eq:best_cand}
    \end{equation}
    where the primary objective is
    $w_\mathrm{int}(c,v) = \sum_{u \in \mathcal{C}_c} K_{vu}$
    (total coupling weight from $v$ to the existing cluster),
    and the secondary tie-breaking objective is the
    internal-to-cut ratio
    \begin{equation}
      \rho(c, v)
      =
      \frac{w_\mathrm{int}(c, v)}{
        \sum_{u \notin \mathcal{C}_c} K_{vu} + \varepsilon},
      \quad \varepsilon = 10^{-12}.
      \label{eq:int_cut_ratio}
    \end{equation}
    The process repeats until no cluster can grow further.
  \item \textbf{Residual assignment.} Any buses still unassigned
    after the expansion phase are assigned to the feasible cluster
    (with $|\mathcal{C}_c| < S_{\max}$) that maximizes the same
    two-objective criterion~\eqref{eq:best_cand}.
    If all clusters are full, a new singleton cluster is created.
\end{enumerate}
The maximum cluster size is set to $S_{\max} = 5$ throughout all benchmark experiments.
This bound ensures that each local UKF operates on an augmented state of dimension at most $3 S_{\max} = 15$,
keeping the per-subsystem Cholesky cost constant and independent of $N$.

In addition to the UKF, the benchmark includes two deterministic least-squares baselines: weighted least squares (WLS) and weighted nonlinear least squares (WNLS)~\cite{Cohen2017Optimal}.
As in the main manuscript, these baselines are scope-restricted references for static state estimation: they do
not infer the dynamic natural frequencies and therefore do not enter the posterior-calibration comparison.
The difference is that WLS performs only a single linearized correction around the propagated state, whereas WNLS
re-linearizes and iterates that correction several times, here using a mild damped Gauss--Newton update, until the
nonlinear residual is reduced.
WNLS is therefore typically more accurate than one-shot WLS when the measurement relation is strongly nonlinear,
but it is also more expensive computationally.
In our implementation, these WLS and WNLS estimators use exactly the same state vector, swing-equation predictor,
measurement model, noise weights, and centralized-versus-distributed decomposition as the UKF, so they serve as
like-for-like algorithmic baselines rather than as a different modeling setup.
For the distributed versions, each subsystem solves its own local least-squares problem and exchanges boundary
phase estimates through the same Jacobi-style message-passing mechanism used by the distributed UKF.

\paragraph{Distributed UKF with Jacobi message passing.}
Each subsystem $\mathcal{C}_s$ with $m = |\mathcal{C}_s|$ buses maintains its own augmented state vector
\begin{equation}
  \mathbf{x}_s
  =
  \bigl[
    \boldsymbol{\theta}_s^\top,\;
    \boldsymbol{\omega}_s^\top,\;
    \boldsymbol{\Omega}_s^\top
  \bigr]^\top
  \in \mathbb{R}^{3m},
  \label{eq:sub_state}
\end{equation}
with local covariance $\mathbf{P}_s \in \mathbb{R}^{3m \times 3m}$.
At each time step $t_k$, the subsystems communicate their current phase-angle estimates to a shared global register
$\hat{\boldsymbol{\theta}}_\mathrm{global}(t_{k-1})$.
Each subsystem UKF then evaluates the coupling forces from external (boundary) nodes using the frozen global
register, decoupling the local update from all other subsystems.
This is the single-sweep Jacobi scheme with $N_\mathrm{Jacobi} = 1$ iteration per time step.
Specifically, the local process function for subsystem $s$ reads
\begin{equation}
  \dot{\omega}_a
  = -d_a\,\omega_a
    + \Omega_a
    + \sum_{b \in \mathcal{C}_s,\, b \neq a}
        K_{ab}\sin(\theta_b - \theta_a)
    + \sum_{j \notin \mathcal{C}_s}
        K_{aj}\sin\!\bigl(\hat{\theta}_j^\mathrm{global} - \theta_a\bigr),
  \label{eq:sub_dynamics}
\end{equation}
for each local node $a \in \mathcal{C}_s$, where $\theta_b$ (for $b \in \mathcal{C}_s$) is a propagated sigma-point
component and $\hat{\theta}_j^\mathrm{global}$ (for $j \notin \mathcal{C}_s$) is a fixed boundary message taken
from the previous step's global register.
After each local UKF update, the estimated phases $\hat{\boldsymbol{\theta}}_s$ are written back into the global
register before the next step begins.

\paragraph{Process and measurement noise for subsystems.}
Each local UKF uses diagonal covariance matrices with entries
\begin{align}
  \mathbf{P}_{s,0}
  &= \mathrm{diag}\!\bigl(
    0.25\,\mathbf{1}_m,\;
    0.25\,\mathbf{1}_m,\;
    1.0\,\mathbf{1}_m
  \bigr),
  \label{eq:P0_sub}
  \\
  \mathbf{Q}_s
  &= \mathrm{diag}\!\bigl(
    10^{-4}\,\mathbf{1}_m,\;
    10^{-4}\,\mathbf{1}_m,\;
    10^{-9}\,\mathbf{1}_m
  \bigr),
  \label{eq:Q_sub}
  \\
  \mathbf{R}_s
  &= \sigma^2\,\mathbf{I}_{2m}.
  \label{eq:R_sub}
\end{align}
Note that the process noise on the natural-frequency parameters is reduced to $10^{-9}$ in the distributed
formulation (compared to $10^{-4}$ in the centralized formulation) to reflect the much tighter local
constraints imposed by the smaller subsystem size. These values are used as the distributed UKF settings for
all benchmark cases.
All cases, including IEEE 300, use the same generator-seeded partitioning with the default cluster cap
$S_{\max}=5$ and a single Jacobi sweep per step ($N_\mathrm{Jacobi}=1$); the full set of numerical parameters is
listed in Table~\ref{tab:grid_params}.
For the IEEE 300 case this partitioning produces $N_s = 100$ subsystems with an average size of 3.0 buses, so the
dimension of every local augmented state remains bounded by $3S_{\max}=15$ independently of the network size.

\paragraph{Computational complexity.}
For the centralized UKF, the augmented state has dimension $3N$, and the dominant cost per step, the Cholesky
factorization of the $(3N) \times (3N)$ covariance matrix followed by sigma-point propagation, scales as
$\mathcal{O}(N^3)$.
For the distributed formulation, each of the $N_s \approx N/S_{\max}$ subsystems contains at most $m \leq S_{\max}$ buses. Since each bus contributes three state variables, the local state dimension is $3m \leq 3S_{\max}$, so the per-subsystem Cholesky cost is $\mathcal{O}(S_{\max}^3)=\mathcal{O}(1)$.
The total cost over all subsystems scales as $\mathcal{O}(N_s \times S_{\max}^3) = \mathcal{O}(N)$, achieving
linear scaling with network size.
This theoretical prediction is confirmed by the benchmarks in Table~\ref{tab:benchmark_results}: at 300 buses,
the distributed estimator is already approximately $19\times$ faster than the centralized UKF under sequential
single-core execution (145.6\,s versus 2836.8\,s).
Because all subsystem filters within a Jacobi sweep are conditionally independent, the aggregate sequential cost
can moreover be divided across one core per subsystem; this projected parallel execution reduces the 300-bus
runtime to approximately 1.46\,s, a speedup of roughly $1950\times$, \textit{i.e.}, approximately three orders of
magnitude, consistent with the values reported in the main manuscript.

\begin{table}[htbp]
\centering
\caption{%
  Computational benchmark results for the centralized and
  distributed UKF, WLS, and WNLS applied to IEEE power-network
  test cases.
  All experiments use $\Delta t = 0.01$\,s, $T = 3.0$\,s, and random seed 42.
  All cases use the default distributed setting $S_{\max} = 5$ and
  $N_\mathrm{Jacobi} = 1$.
  The number of clusters $N_s$ is produced by the generator-seeded
  partitioning algorithm.
  In the Central, Distributed, and Speedup columns, triplets are reported
  in the order (UKF, WLS, WNLS); timings are measured on a single CPU
  core (sequential execution), and the speedup entries are the
  elementwise ratios of centralized to distributed wall-clock time.
  The last column reports the projected parallel runtime of the
  distributed UKF, obtained by dividing the aggregate sequential cost by
  the number of subsystem filters $N_s$ (one core per subsystem).
}
\label{tab:benchmark_results}
\footnotesize
\setlength{\tabcolsep}{3pt}
\begin{tabular}{@{}lccclllc@{}}
\toprule
\textbf{Case}
  & \textbf{Buses} $N$
  & \textbf{Clusters} $N_s$
  & \textbf{Avg.\ size} $\bar{m}$
  & \textbf{Central (s)}
  & \textbf{Distributed (s)}
  & \textbf{Speedup} $(\times)$
  & \textbf{Parallel (s)} \\
\midrule
IEEE 9   &  9  &  3  &  3.0 & (0.79, 0.03, 0.11)     & (0.43, 0.04, 0.14)     & (1.82, 0.74, 0.83) & 0.14 \\
IEEE 14  & 14  &  5  &  2.8 & (1.72, 0.04, 0.16)     & (0.82, 0.07, 0.21)     & (2.10, 0.58, 0.78) & 0.16 \\
IEEE 30  & 30  &  6  &  5.0 & (7.5, 0.1, 0.4)        & (2.4, 0.1, 0.4)        & (3.2, 1.0, 1.0) & 0.39 \\
IEEE 39  & 39  & 10  &  3.9 & (14.3, 0.5, 1.2)       & (3.0, 0.2, 0.6)        & (4.8, 3.0, 2.2) & 0.30 \\
IEEE 57  & 57  & 12  &  4.8 & (45.5, 1.5, 3.2)       & (5.8, 0.3, 0.8)        & (7.8, 5.7, 3.9) & 0.48 \\
IEEE 118 & 118 & 54  &  2.2 & (215.1, 3.4, 6.0)      & (12.0, 1.0, 2.3)       & (17.9, 3.5, 2.6) & 0.22 \\
IEEE 300 & 300 & 100 &  3.0 & (2836.8, 17.3, 33.5)   & (145.6, 11.4, 30.2)    & (19.5, 1.5, 1.1) & 1.46 \\
\bottomrule
\end{tabular}
\end{table}

\begin{figure}[htbp]
\centering
\includegraphics[width=\linewidth]{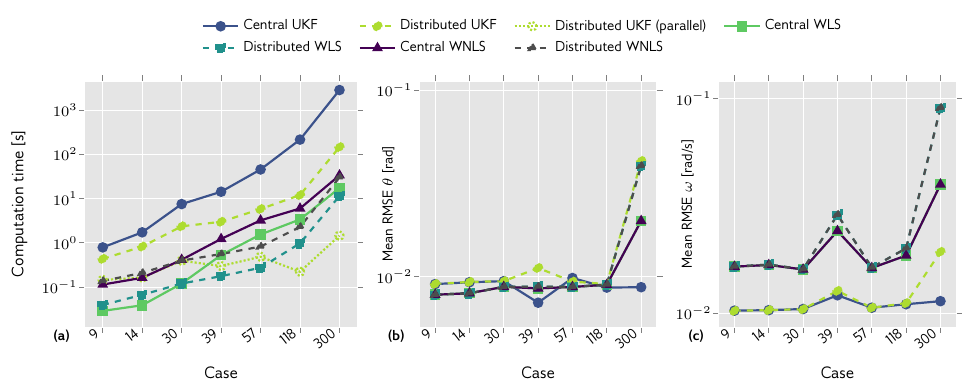}
\caption{%
  Scalability and accuracy benchmarks for the centralized and distributed estimators across the IEEE 9, 14, 30, 39, 57, 118, and 300-bus test cases. Each panel compares the centralized and distributed UKF, WLS, and WNLS.
  All quantities are plotted on logarithmic $y$-axes. (a)~Wall-clock computation time on a single CPU core, illustrating the transition from approximately cubic scaling for the centralized estimators to near-linear scaling for the distributed counterparts; the dotted curve additionally shows the projected parallel execution of the distributed UKF with one core per subsystem, obtained by dividing the aggregate sequential cost by the number of subsystem filters.
  (b)~Mean state-estimation RMSE for the phase angle $\theta$, showing that the centralized and distributed estimators remain within the same order of magnitude across the suite.
  (c)~Mean state-estimation RMSE for the rotor frequency $\omega$, exhibiting the same qualitative behavior as panel~(b).
}
\label{fig:ieee_benchmark_plots}
\end{figure}

Figure~\ref{fig:ieee_benchmark_plots} summarizes the resulting trends across the
benchmark suite.
Panel~(a) reports wall-clock computation time on a single
CPU core, panel~(b) reports the mean state-estimation RMSE for the phase
angle $\theta$, and panel~(c) reports the corresponding mean RMSE for the
rotor frequency $\omega$; all three panels use logarithmic $y$-axes so that
the centralized and distributed variants of the UKF, WLS, and WNLS
estimators can be compared on a common scale.
The timing comparison in Table~\ref{tab:benchmark_results} and
Fig.~\ref{fig:ieee_benchmark_plots}(a) shows that the distributed estimator
delivers a large computational advantage across the benchmark suite, with
the gap between the centralized and distributed curves widening as the
network size grows.
Accuracy degrades only mildly by comparison.
Panels~(b) and~(c) of Fig.~\ref{fig:ieee_benchmark_plots}
show that the mean state-estimation RMSE for $\theta$ and $\omega$ remains
of the same order for the centralized and distributed estimators across the
benchmark suite, with the centralized UKF usually only modestly more
accurate.
This modest loss in accuracy is small relative to the runtime savings.
Table~\ref{tab:benchmark_accuracy} extends the comparison to normalized state and parameter errors, as well as
parameter-coverage calibration, for all benchmark cases.
The distributed UKF typically has somewhat larger parameter normalized RMSE (NRMSE) and slightly lower 95\% coverage than the
centralized UKF, but these metrics remain in the same order of magnitude across all cases.
Across the suite, the largest deviation between the distributed and centralized estimators is
$2.67\times10^{-3}$ in state NRMSE (IEEE 300, averaged over $\theta$ and $\omega$) and $3.62\times10^{-2}$ in
parameter NRMSE (IEEE 39), matching the maximum deviations reported in the main manuscript; in the 300-bus case
the 95\% parameter coverage is $0.967$ for the distributed estimator and $0.982$ for the centralized UKF.
Taken together with the timing results, this shows that the distributed UKF trades only a modest loss in accuracy
for a very large reduction in computation time; the error scale stays comparable even when the centralized
estimator is slightly more accurate.

\begin{table}[htbp]
\centering
\caption{%
  Accuracy and calibration metrics for the centralized and distributed
  UKF on the IEEE benchmark suite.
  Entries are reported as pairs in the order
  (centralized, distributed).
}
\label{tab:benchmark_accuracy}
\footnotesize
\setlength{\tabcolsep}{4pt}
\begin{tabular}{@{}lcccc@{}}
\toprule
\textbf{Case}
  & \textbf{Mean NRMSE} $\theta$
  & \textbf{Mean NRMSE} $\omega$
  & \textbf{Mean NRMSE} $\Omega$
  & \textbf{95\% Cov.} $\Omega$ \\
\midrule
IEEE 9   & (0.0090, 0.0090) & (0.0026, 0.0026) & (0.036, 0.034) & (1.000, 1.000) \\
IEEE 14  & (0.0074, 0.0074) & (0.0018, 0.0018) & (0.040, 0.048) & (1.000, 0.990) \\
IEEE 30  & (0.0074, 0.0074) & (0.0018, 0.0018) & (0.034, 0.032) & (1.000, 1.000) \\
IEEE 39  & (0.0052, 0.0079) & (0.0005, 0.0005) & (0.180, 0.216) & (0.983, 0.921) \\
IEEE 57  & (0.0038, 0.0036) & (0.0015, 0.0015) & (0.050, 0.044) & (0.998, 0.992) \\
IEEE 118 & (0.0059, 0.0062) & (0.0008, 0.0008) & (0.076, 0.090) & (0.997, 0.958) \\
IEEE 300 & (0.0014, 0.0066) & (0.0000, 0.0001) & (0.171, 0.206) & (0.982, 0.967) \\
\bottomrule
\end{tabular}
\end{table}


\begin{table}[htbp]
\centering
\caption{%
  Numerical parameters for the power-grid scalability study.
}
\label{tab:grid_params}
\small
\begin{tabular}{lcc}
\toprule
\textbf{Parameter} & \textbf{Description} & \textbf{Value} \\
\midrule
\multicolumn{3}{l}{\textit{Simulation}} \\
$\Delta t$ & Integration time step & 0.01\,s \\
$T$ & Simulation horizon & 3.0\,s  \\
$N_\mathrm{steps}$ & Total time steps & 300  \\
\midrule
\multicolumn{3}{l}{\textit{True parameters}} \\
$d_i$ & Damping coefficient & $\mathcal{U}(0.10,\,0.30)$ rounded to 2 d.p.  \\
$\Omega_i$ & Natural frequency & $\mathcal{U}\{-1.0,\,-0.9,\ldots,0.9,\,1.0\}$\,rad/s \\
$K_{ij}$ & Coupling strength & $|Y_{ij}|$ from PYPOWER \\
\midrule
\multicolumn{3}{l}{\textit{Initial conditions}} \\
$\theta_0$ & True initial phase & $\mathcal{U}(-0.5,\,0.5)$\,rad \\
$\omega_0$ & True initial frequency & $\mathcal{U}(-0.2,\,0.2)$\,rad/s\\
\midrule
\multicolumn{3}{l}{\textit{Initial estimator perturbations}} \\
$\hat{\theta}_0$ & Estimated initial phase & $\theta_0 + \mathcal{N}(0,\,0.04)$ \\
$\hat{\omega}_0$ & Estimated initial frequency & $\omega_0 + \mathcal{N}(0,\,0.04)$ \\
$\hat{\Omega}_0$ & Estimated initial nat.\ freq. & $0$ (no prior)  \\
\midrule
\multicolumn{3}{l}{\textit{Measurement noise}} \\
$\sigma$ & Noise std (phase \& frequency) & 0.02\,rad (or rad/s) \\
\midrule
\multicolumn{3}{l}{\textit{UKF settings (all cases)}} \\
$\gamma$ & UKF scaling parameter & 1.0 \\
$\sigma_\theta^2$ & $\mathbf{P}_0$ entry (phase) & 0.25  \\
$\sigma_\omega^2$ & $\mathbf{P}_0$ entry (frequency) & 0.25  \\
$\sigma_\Omega^2$ & $\mathbf{P}_0$ entry (parameter) & 1.0 \\
$Q_\theta$ & Process noise (phase) & $10^{-4}$ \\
$Q_\omega$ & Process noise (frequency) & $10^{-4}$  \\
$Q_\Omega^{(\mathrm{c})}$ & Process noise (param., central) & $10^{-4}$  \\
$Q_\Omega^{(\mathrm{d})}$ & Process noise (param., dist.) & $10^{-9}$  \\
\midrule
\multicolumn{3}{l}{\textit{Partitioning}} \\
$S_{\max}$ & Maximum cluster size & 5 \\
$N_\mathrm{Jacobi}$ & Jacobi iterations per step & 1 \\
$\varepsilon$ & Internal/cut ratio regularizer & $10^{-12}$ \\
\midrule
\multicolumn{3}{l}{\textit{Random seed}} \\
-- & NumPy default RNG seed & 42  \\
\bottomrule
\end{tabular}
\end{table}

\subsection*{Multi-physics turbine--generator system}
\label{App:turbine}

\begin{figure}[htbp]
\centering
\includegraphics[width=0.58\linewidth]{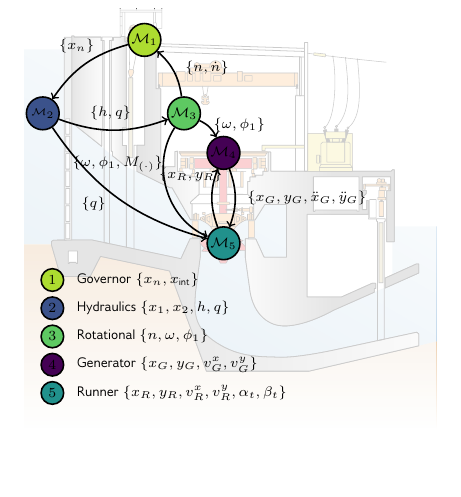}
\caption{%
  Subsystem decomposition of the hydro-turbine case study.
  The five subsystems $\mathcal{M}_1$--$\mathcal{M}_5$ are overlaid on the hydro-power-plant schematic, with directed
  interface messages exchanged between subsystems; each edge label lists the interface variables carried by
  that message.
  The legend lists the state variables assigned to each subsystem.
}
\label{fig:sos_turbine_schematic}
\end{figure}

This section provides the complete mathematical formulation of the heterogeneous turbine--generator unit of the
main manuscript's third case study, which demonstrates that fundamentally different local models (a
deterministic controller, a learned surrogate, and probabilistic estimators of different classes) compose
within a single distributed inference architecture.
The turbine--generator system is decomposed into five subsystems: a governor subsystem $\mathcal{M}_1$, a hydraulic subsystem $\mathcal{M}_2$, a rotational subsystem $\mathcal{M}_3$, a generator lateral-vibration subsystem $\mathcal{M}_4$, and a runner lateral--torsional subsystem $\mathcal{M}_5$.
Each subsystem is described by its state-space model, its observation model, the interface messages it receives from and sends to other subsystems, and, where applicable, the filter type used for inference.
Figure~\ref{fig:sos_turbine_schematic} gives a visual summary of this decomposition and the interface variables
exchanged between subsystems.
In the remainder of this section we describe the state, observation, and filtering structure of each subsystem in
turn; representative estimation results are discussed at the end of the section.

In particular, the governor subsystem $\mathcal{M}_1$ is represented by a deterministic control law, the hydraulic subsystem
$\mathcal{M}_2$ by a NARX surrogate, the rotational subsystem $\mathcal{M}_3$ by an EKF, the generator
lateral-vibration subsystem $\mathcal{M}_4$ by a Kalman filter, and the runner lateral--torsional subsystem
$\mathcal{M}_5$ by a UKF.
All parameter values and filter hyperparameters are collected in
Tables~\ref{tab:physical_params}--\ref{tab:filter_hyperparams} at the end of this section.

\subsubsection*{Global state vector and system decomposition}

The full turbine--generator model used in this study consists of a 19-state nonlinear dynamical system whose state vector is partitioned
according to the five subsystems as
\begin{equation}
  \mathbf{x}
  =
  \bigl[
    \underbrace{x_1,\, x_2,\, h,\, q}_{\mathcal{M}_2},\;
    \underbrace{x_n,\, x_{\mathrm{int}}}_{\mathcal{M}_1},\;
    \underbrace{n,\, \omega,\, \phi_1}_{\mathcal{M}_3},\;
	    \underbrace{x_G,\, y_G,\, \dot{x}_G,\, \dot{y}_G}_{\mathcal{M}_4},\;
	    \underbrace{x_R,\, y_R,\, \dot{x}_R,\, \dot{y}_R,\,
	                \alpha_t,\, \beta_t}_{\mathcal{M}_5}
  \bigr]^\top
  \in \mathbb{R}^{19}.
  \label{eq:global_state}
\end{equation}
The hydraulic and governor state variables are expressed in normalized per-unit form, whereas dimensional
parameters such as $H_\mathrm{rated}$, $Q_\mathrm{rated}$, $s_e$, and the time constants retain their physical
units. The structural subsystem is formulated in SI units.
Table~\ref{tab:state_decomp} summarizes the physical meaning and units of each state, identifies whether it is
directly measured or latent, and lists the augmented states appended during estimation.

\subsubsection*{Governor subsystem $\mathcal{M}_1$: deterministic PID controller}

The governor regulates the turbine speed by adjusting the needle actuator position $x_n$ via a PID control law~\cite{Yang2019Hydropower}.
No probabilistic filter is applied to this subsystem; the states are propagated deterministically using a
fixed-step Heun (second-order Runge--Kutta) integrator~\cite{Hairer1993RungeKutta}.

\paragraph{State vector.}
The governor state is
\begin{equation}
  \mathbf{x}_{\mathcal{M}_1} = [x_n,\; x_{\mathrm{int}}]^\top,
\end{equation}
where $x_n \in [0,1]$ is the normalized needle stroke (actuator position) and $x_{\mathrm{int}}$ is the integrator
accumulator of the speed error.

\paragraph{Process model.}
The speed-reference command is prescribed as a sinusoidal excitation,
\[
s_\mathrm{cmd}(t)=A_s\sin(2\pi f_s t),
\]
where \(A_s\) and \(f_s\) denote the imposed excitation amplitude and frequency, respectively.
In this study, \(A_s=0.1\) p.u. and \(f_s=0.2\) Hz are selected to generate a small-amplitude
periodic speed perturbation for evaluating the closed-loop response. The corresponding tracking
error is
\[
e(t)=s_\mathrm{cmd}(t)-n,
\]
where \(n\) is the estimated normalized speed deviation.
The integrator and the PID control signal are
\begin{align}
  \dot{x}_{\mathrm{int}} &= e,  \label{eq:gov_int}\\
  u_\mathrm{PID} &= k_p\, e + k_i\, x_{\mathrm{int}}
                    - k_d\, \dot{n}, \label{eq:gov_pid}
\end{align}
where $\dot{n}$ is the normalized speed-rate message from $\mathcal{M}_3$.
The PID output is saturated to the physical range of the actuator:
\begin{equation}
  u = \mathrm{sat}(u_\mathrm{PID})
  =
  \begin{cases}
    0, & u_\mathrm{PID} < 0,\\
    u_\mathrm{PID}, & 0 \le u_\mathrm{PID} \le 1,\\
    1, & u_\mathrm{PID} > 1.
  \end{cases}
  \label{eq:gov_sat}
\end{equation}
The needle actuator is modeled as a first-order servo:
\begin{equation}
  \dot{x}_n = \frac{u - x_n}{T_y},
  \label{eq:gov_servo}
\end{equation}
where $T_y$ is the servo time constant.

\paragraph{Interface.}
The governor receives $n$ and $\dot{n}$ from $\mathcal{M}_3$ and sends $x_n$ to $\mathcal{M}_2$:
\begin{equation}
  \text{(from } \mathcal{M}_1\text{):}\quad
  x_n \rightarrow \mathcal{M}_2.
\end{equation}

\subsubsection*{Hydraulic subsystem $\mathcal{M}_2$: NARX surrogate model}
The hydraulic subsystem governs the evolution of the elastic pressure wave in the penstock and the flow through
the Pelton nozzle. The underlying pressure--flow dynamics are motivated by the classical one-dimensional
elastic water-hammer formulation for pressurized conduits~\cite{Chaudhry2014Applieda,Gurecky2023Elastic}.
For the Pelton turbine system considered here, a normalized reduced-order state-space model is adopted following
the hydro-turbine governing-system formulation in~\cite{Xu2017Sensitivity}.
\begin{equation}
  \mathbf{x}_h =
  \begin{bmatrix}
    x_1 & x_2 & x_3 & q
  \end{bmatrix}^{\top},
  \label{eq:hyd_state}
\end{equation}
where $x_1$ and $x_2$ are auxiliary penstock states, $x_3=h$ is the normalized hydraulic-head deviation, and
$q$ is the normalized turbine flow.

The reduced-order penstock dynamics are written as
\begin{align}
  \dot{x}_1 &= x_2, \label{eq:hyd1}\\
  \dot{x}_2 &= x_3, \label{eq:hyd2}\\
  \dot{x}_3 &=
    -\frac{\pi^2}{T_e^2}x_2
    +\frac{1}{Z_nT_e^3}
    \left(
      h_0 - \frac{x_3}{y_r^2} - q^2
    \right),
  \label{eq:hyd3}
\end{align}
where $T_e$ is the elastic time constant of the penstock, $Z_n$ is the number of nozzles, $y_r$ is the rated
opening parameter, and $h_0$ is the normalized upstream head coefficient. This reduced-order representation is
formulated in normalized variables; the auxiliary states $x_1$ and $x_2$ do not correspond directly to dimensional
head or flow variables, but form an integrator chain whose third state is the normalized head, $x_3 \equiv h$, so
that~\eqref{eq:hyd1}--\eqref{eq:hyd3} realize a third-order approximation of the elastic water-column dynamics
of~\cite{Xu2017Sensitivity}.

The nozzle flow is represented by a first-order relaxation toward the quasi-steady nozzle flow,
\begin{equation}
  \dot{q}
  =
  \frac{q^\star(x_n,x_3)-q}{T_q},
  \label{eq:flow_ode}
\end{equation}
where $T_q$ is the flow-inertia time constant, $x_n$ is the normalized needle stroke supplied by the governor
subsystem, and $q^\star$ is the quasi-steady flow. The quasi-steady flow is modeled as
\begin{equation}
  q^\star(x_n,x_3)
  =
  C_q A_n(x_n)
  \sqrt{2gH_\mathrm{rated}(1+x_3)},
  \qquad 1+x_3>0,
  \label{eq:nozzle_flow}
\end{equation}
where $H_\mathrm{rated}$ is the rated head, $A_n(x_n)$ is the effective nozzle area, and $C_q$ is a normalization
coefficient that absorbs the division by the rated flow (so that $q^\star$ is expressed per-unit) and is selected
so that the nominal operating condition gives $q^\star=1$. The expressions for the effective
nozzle area $A_n(x_n)$ and its geometric coefficients follow the turbine nozzle formulation
in~\cite{Xu2017Sensitivity}; the first-order relaxation representation~\eqref{eq:flow_ode} is adopted in this
work as a tractable approximation of the full nozzle dynamics.

\paragraph{NARX surrogate.}
In the distributed estimation framework, the physics-based head ODE~\eqref{eq:hyd1}--\eqref{eq:hyd3} is
replaced by a data-driven NARX model that maps a window of past
needle-position and flow observations directly to the one-step-ahead hydraulic head.
This choice reflects the fact that $h$ is the only hydraulic quantity required by downstream subsystems, while
$x_1$ and $x_2$ are purely internal and not independently measurable.
The NARX surrogate is a feedforward neural network with the form
\begin{equation}
  \hat{h}(t+\Delta t)
  =
  f_\theta\!\bigl(
    \mathbf{u}_{\mathrm{dyn}}(t),\;
    \mathbf{u}_{\mathrm{stat}}
  \bigr),
  \label{eq:narx}
\end{equation}
where the dynamic input window
$\mathbf{u}_{\mathrm{dyn}}(t) = \bigl[x_n(t-(L-1)\Delta t_s),\ldots,x_n(t);\;
q(t-(L-1)\Delta t_s),\ldots,q(t)\bigr] \in\mathbb{R}^{2L}$
stacks $L = 200$ past samples of the needle position and the measured flow at sampling interval $\Delta t_s$, and
the static feature vector $\mathbf{u}_{\mathrm{stat}} = [T_e,\, T_q,\, y_r,\, h_0]$ encodes the known penstock
parameters.
The network $f_\theta$ is a three-layer multilayer perceptron with input dimension $2L + 4 = 404$, hidden dimension
128, ReLU activations, and a dropout rate of 0.05 applied after the first hidden layer.
All inputs and the target $h$ are normalized to zero mean and unit variance using statistics computed over the
training set.

\begin{figure}[t]
\centering
\includegraphics[width=\linewidth]{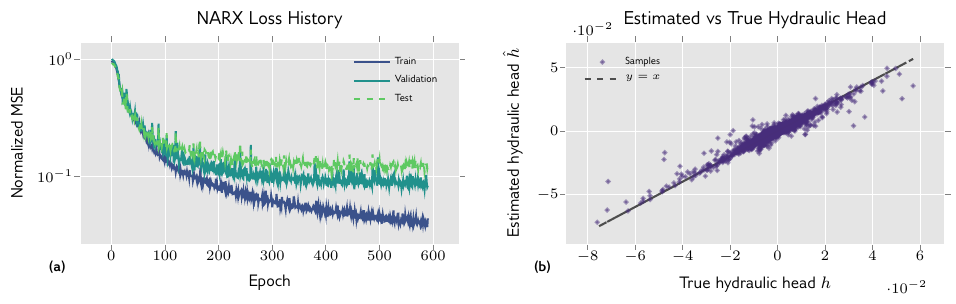}
\caption{%
Performance of the NARX surrogate for the hydraulic-head prediction
task.
(a) Training, validation, and test loss histories in terms of the
normalized mean-squared error.
(b) Predicted versus true hydraulic head $h$ on the test set, with
the dashed line indicating perfect agreement.
The final test metrics are 0.076930 normalized MSE, 0.277362
normalized RMSE, and 0.14 normalized MAE, corresponding to
$3.77\times 10^{-6}$ physical MSE, $1.94\times 10^{-3}$
physical RMSE, and $9.97\times 10^{-4}$ physical MAE.
}
\label{fig:narx_plot}
\end{figure}

\paragraph{Training.}
Training data are generated from multiple simulations of the monolithic model~\eqref{eq:hyd1}--\eqref{eq:flow_ode}
under varied operating scenarios.
Scenarios are split 70\,/\,15\,/\,15 (train\,/\,validation\,/\,test) at the scenario level to prevent temporal
leakage.
Sliding windows of length $L = 200$ are extracted with stride 20, yielding one training sample per stride step.
The network is trained by minimizing the mean squared error (MSE) between predicted and true $h(t+\Delta t_s)$
using the AdamW optimizer~\cite{Loshchilov2018Decoupled,Zhou2024Understanding} with learning rate $10^{-4}$, weight decay $10^{-6}$, and batch size 64.
Training runs for up to 650 epochs with early stopping triggered after 137 consecutive epochs without improvement
on the validation loss, as shown in Fig.~\ref{fig:narx_plot}(a).

\paragraph{Integration with the distributed estimator.}
At each time step $t_k$ of the estimation loop, the NARX predictor produces $\hat{h}(t_{k+1})$, which is stored as
the updated head estimate.
The flow $q$ is not predicted by the surrogate; instead, it is treated as a direct measurement $y_q(t_k)$ subject
to sensor noise (standard deviation $\sigma_q = 0.01$\,p.u.).
The interface variables passed to downstream subsystems are therefore
\begin{equation}
  \text{(from } \mathcal{M}_2\text{):}\quad
  y_q(t_k) \rightarrow \text{(to } \mathcal{M}_{3,5}\text{):}\quad q,
  \qquad
  \hat{h}(t_{k+1}) \rightarrow \text{(to } \mathcal{M}_{3}\text{):}\quad h,
  \label{eq:hyd_msg}
\end{equation}
The predictive accuracy of the trained NARX surrogate on the held-out test set is summarized in
Panel (b) of Fig.~\ref{fig:narx_plot}. In normalized units, the model achieves an MSE of 0.076, a root mean squared error (RMSE) of 0.277, and a
mean absolute error (MAE) of 0.14. When mapped back to physical units, these correspond to an MSE of $3.77\times 10^{-6}$, an
RMSE of $1.94\times 10^{-3}$, and an MAE of $9.97\times 10^{-4}$. These results indicate that the surrogate
reproduces the one-step-ahead hydraulic head with small absolute error while remaining suitable for integration
into the coupled distributed estimator.

\subsubsection*{Rotational subsystem $\mathcal{M}_3$: torque model and EKF}

The rotational subsystem models the angular dynamics of the shaft and serves as the primary coupling node between
the hydraulic, control, and structural subsystems.
An EKF is applied to jointly estimate the shaft speed, the shaft angle, and two uncertain
torque coefficients.

\paragraph{Augmented state vector.}
The process state is augmented with scaled versions of the hydraulic torque coefficients:
\begin{equation}
  \mathbf{x}_{\mathcal{M}_3}
  =
  \bigl[
    n,\; \omega,\; \phi_1,\;
    \tilde{e}_x,\; \tilde{e}_y
  \bigr]^\top
  \in\mathbb{R}^5,
  \label{eq:rot_state}
\end{equation}
where $n = (\omega - \omega_\mathrm{rated})/\omega_\mathrm{rated}$ is the per-unit speed deviation, $\omega$ is the
absolute angular velocity, $\phi_1$ is the generator rotor angle, and $\tilde{e}_x = e_x / e_{x,0}$, $\tilde{e}_y =
e_y / e_{y,0}$ are the scaled torque coefficients (ratio to nominal values $e_{x,0}$, $e_{y,0}$).
Scaling to unity at nominal values improves filter conditioning and allows a common prior for all estimated
parameters.

\paragraph{Process model.}
The hydraulic torque is computed using the linearized torque-deviation model~\cite{Xu2017Sensitivity}:
\begin{equation}
  M_t
  =
  M_{gB}
  \Bigl[
    1
    + e_x n
    + e_y(q-1)
    + e_h h
  \Bigr],
  \label{eq:Mt}
\end{equation}
where $M_{gB}=P_\mathrm{rated}/\omega_\mathrm{rated}$ is the rated mechanical torque at the generator,
$n$ is the normalized speed deviation, $q$ is the normalized turbine flow, and $h=x_3$ is the normalized
hydraulic-head deviation. Here, the torque coefficient $e_h$ is assumed to be known from the calibrated turbine model.
The electrical (generator) braking torque and the speed-deviation damping term are
\begin{equation}
  M_g = M_{gB}\,\frac{\omega}{\omega_\mathrm{rated}},
  \qquad
  M_\mathrm{damp} = D_\omega\,(\omega - \omega_\mathrm{rated}),
  \label{eq:Mg}
\end{equation}
where $D_\omega$ is the rotational damping coefficient that stabilizes synchronous operation.
The continuous-time process model for $\mathcal{M}_3$ is
\begin{align}
  \dot{\omega}
    &= \frac{M_t - M_g - M_\mathrm{damp}}{J_\mathrm{tot}},
    \label{eq:omega_dot}\\
  \dot{n}
    &= \frac{\dot{\omega}}{\omega_\mathrm{rated}},
    \label{eq:n_dot}\\
  \dot{\phi}_1
    &= \omega,
    \label{eq:phi1_dot}\\
  \dot{\tilde{e}}_x
    &= 0,
    \label{eq:ex_dot}\\
  \dot{\tilde{e}}_y
    &= 0,
    \label{eq:ey_dot}
\end{align}
where the total moment of inertia is $J_\mathrm{tot} = J_{1,\mathrm{eff}} + J_{2,\mathrm{eff}}$, with effective
inertias defined below.
Equation~\eqref{eq:n_dot} is obtained by differentiating the definition of the normalized speed deviation $n$ and
carries no independent dynamics; it keeps the redundant speed states consistent within the filter.
The torque parameters are treated as random-walk processes driven only by process noise; their nominal derivatives
are set to zero (\eqref{eq:ex_dot} and \eqref{eq:ey_dot}).

\paragraph{Observation model.}
The measured quantity is the shaft angular velocity from an encoder:
\begin{equation}
  y_\omega = \omega + v_\omega,
  \qquad
  v_\omega \sim \mathcal{N}(0, \sigma_\omega^2),\;
  \sigma_\omega = 0.02\;\mathrm{rad\,s}^{-1}.
  \label{eq:rot_obs}
\end{equation}
The observation function is $h_{\mathcal{M}_3}(\mathbf{x}) = \omega$ (the second element of
$\mathbf{x}_{\mathcal{M}_3}$).

\paragraph{Messages produced.}
After each EKF update, the following quantities are derived from the posterior mean
$\hat{\mathbf{x}}_{\mathcal{M}_3}$ and sent to other subsystems:
\begin{equation}
  \text{(from } \mathcal{M}_3\text{):}\quad
  n,\;\omega,\;\phi_1,\;\dot{n}
  \;\longrightarrow\;
  \mathcal{M}_1,\,\mathcal{M}_4,\,\mathcal{M}_5.
\end{equation}
The torque signals $M_t$ and $M_g$, computed from the posterior state using~\eqref{eq:Mt} and~\eqref{eq:Mg}, are
additionally passed to $\mathcal{M}_5$ for the torsional coupling term.

\subsubsection*{Generator lateral-vibration subsystem $\mathcal{M}_4$: Kalman filter}

The generator lateral-vibration subsystem describes the planar motion of the generator rotor mass $m_1$ in the $(x,
y)$ plane perpendicular to the shaft axis~\cite{Genta2005Dynamics}.
Because the driving forces depend linearly on the displacement and velocity states (the only nonlinearity, rubbing,
is small and occurs only when the air-gap clearance is exceeded), a Kalman filter is employed.

\paragraph{State vector.}
\begin{equation}
  \mathbf{x}_{\mathcal{M}_4}
  =
  [x_G,\; y_G,\; \dot{x}_G,\; \dot{y}_G]^\top
  \in\mathbb{R}^4.
\end{equation}

\paragraph{Process model.}
Treating the velocity components $\dot{x}_G$ and $\dot{y}_G$ as independent states, the equations of motion are
\begin{align}
  \frac{\mathrm{d}}{\mathrm{d}t}\,x_G &= \dot{x}_G, \label{eq:xGdot}\\
  \frac{\mathrm{d}}{\mathrm{d}t}\,y_G &= \dot{y}_G, \label{eq:yGdot}\\
  m_1\ddot{x}_G &= F_{x,G}, \label{eq:vxGdot}\\
  m_1\ddot{y}_G &= F_{y,G}, \label{eq:vyGdot}
\end{align}
where the total lateral forces in the $x$ and $y$ directions are
\begin{align}
  F_{x,G}
  &=
  \underbrace{-(k_{\mathrm{gen}} + K_{11})\,x_G - K_{12}\,x_R}_{
      \text{bearing + shaft stiffness}}
  \underbrace{-k_{\mathrm{cross}}\,y_G}_{
      \text{cross-coupling stiffness}}
  \underbrace{-c_{\mathrm{gen}}\,\dot{x}_G
              -c_{\mathrm{cross}}\,\dot{y}_G}_{
      \text{damping}}
  \nonumber\\
  &\quad
  +\underbrace{m_1 e_1 \omega^2 \cos\phi_1}_{
      \text{rotating unbalance}}
  +F_{\mathrm{rub},x}^G,
  \label{eq:FxG}
  \\[6pt]
  F_{y,G}
  &=
  -(k_{\mathrm{gen}} + K_{11})\,y_G - K_{12}\,y_R
  +k_{\mathrm{cross}}\,x_G
  -c_{\mathrm{gen}}\,\dot{y}_G
  +c_{\mathrm{cross}}\,\dot{x}_G
  \nonumber\\
  &\quad
  +m_1 e_1\omega^2\sin\phi_1
  +F_{\mathrm{rub},y}^G.
  \label{eq:FyG}
\end{align}
The shaft stiffness coefficients $K_{11}$ and $K_{12}$, which represent the elastic compliance of the hollow shaft between the generator bearing and the runner bearing, are derived in the following section.
The rubbing force activates only when the generator orbit radius $r_G = \sqrt{x_G^2 + y_G^2}$ exceeds the generator
air-gap clearance $\delta_0$:
\begin{equation}
  \begin{bmatrix}
    F_{\mathrm{rub},x}^G\\
    F_{\mathrm{rub},y}^G
  \end{bmatrix}
  =
  \begin{cases}
    -\dfrac{(r_G - \delta_0)\,k_r}{r_G}
    \begin{bmatrix}
      x_G - f_\mathrm{fric}\,y_G\\
      f_\mathrm{fric}\,x_G + y_G
    \end{bmatrix},
    & r_G > \delta_0,\\[10pt]
    \mathbf{0}, & r_G \le \delta_0,
  \end{cases}
  \label{eq:rub_G}
\end{equation}
where $k_r$ is the rubbing stiffness and $f_\mathrm{fric}$ is the dry-friction coefficient~\cite{Goldman1995Rotortostator}.

\paragraph{Observation model.}
Displacement sensors measure $x_G$ and $y_G$ directly:
\begin{equation}
  \mathbf{y}_G
  =
  \begin{bmatrix} x_G \\ y_G \end{bmatrix}
  +
  \begin{bmatrix} v_{x_G} \\ v_{y_G} \end{bmatrix},
  \qquad
  v_{x_G}, v_{y_G} \sim \mathcal{N}(0,\sigma_G^2),\;
  \sigma_G = 5\;\mu\mathrm{m}.
  \label{eq:gen_obs}
\end{equation}
The velocities $\dot{x}_G$ and $\dot{y}_G$ are latent states estimated by the Kalman filter.

\paragraph{Messages produced.}
The posterior mean displacements and the accelerations $\ddot{x}_G$, $\ddot{y}_G$ (computed from the force model
evaluated at the posterior state) are sent to $\mathcal{M}_5$ for the torsional coupling calculation:
\begin{equation}
  \text{(from } \mathcal{M}_4\text{):}\quad
  x_G,\; y_G,\; \ddot{x}_G,\; \ddot{y}_G
  \;\longrightarrow\; \mathcal{M}_5.
\end{equation}

\subsubsection*{Runner lateral--torsional subsystem $\mathcal{M}_5$: UKF}

The runner subsystem captures the coupled lateral and torsional dynamics of the turbine runner mass $m_2$,
including the effect of the labyrinth seal.
Because the seal stiffness is treated as an unknown parameter to be
identified, a UKF is used for parameter identification and state estimation.

\paragraph{Augmented state vector.}
The augmented state vector for the runner subsystem is
\begin{equation}
  \mathbf{x}_{\mathcal{M}_5}
  =
  \bigl[
    x_R,\; y_R,\; \dot{x}_R,\; \dot{y}_R,\;
    \alpha_t,\; \beta_t,\;
    \tilde{K}_\mathrm{seal}
  \bigr]^\top
  \in\mathbb{R}^7,
  \label{eq:run_state}
\end{equation}
where $x_R$ and $y_R$ are the runner lateral displacements, $\dot{x}_R$ and $\dot{y}_R$ are the corresponding
velocities, $\alpha_t$ is the torsional twist angle between the generator and runner discs, and
$\beta_t=\dot{\alpha}_t$ is the torsional rate.
The parameter
\begin{equation}
  \tilde{K}_\mathrm{seal}
  =
  \frac{K_\mathrm{seal}}{K_{\mathrm{seal},0}}
\end{equation}
is the scaled seal stiffness to be estimated. The seal damping $D_\mathrm{seal}$ and circumferential velocity ratio
$\tau_\mathrm{seal}$ are held fixed at their nominal values.

\paragraph{Runner lateral process model.}
The runner phase angle relative to the shaft reference is
\begin{equation}
  \phi_2 = \phi_1 - \alpha_t,
\end{equation}
where $\phi_1$ is received from the rotational subsystem $\mathcal{M}_3$.
Following the reduced shafting vibration formulation in~\cite{Shi2022Vibration}, the runner lateral equations are
written as
\begin{align}
  m_2\ddot{x}_R &= F_{x,R}, \label{eq:xRddot}\\
  m_2\ddot{y}_R &= F_{y,R}, \label{eq:yRddot}
\end{align}
where the total lateral forces are
\begin{align}
  F_{x,R}
  &=
  \underbrace{-(k_\mathrm{run}+K_{22})\,x_R - K_{12}\,x_G}_{
      \text{bearing + shaft stiffness}}
  \underbrace{-0.8\,k_\mathrm{cross}\,y_R}_{
      \text{cross-coupling}}
  \nonumber\\
  &\quad
  \underbrace{
    -\tilde{K}_\mathrm{seal}K_{\mathrm{seal},0}\,x_R
    -\tau_\mathrm{seal}\omega D_\mathrm{seal}\,y_R}_{
    \text{seal force}}
  \underbrace{-c_\mathrm{run}\,\dot{x}_R}_{\text{viscous damping}}
  \nonumber\\
  &\quad
  +\underbrace{m_2e_2\omega^2\cos\phi_2}_{
      \text{mass unbalance}}
  +\underbrace{A_\mathrm{vortex}q\sin(f_\mathrm{vortex}t)}_{
      \text{hydraulic excitation}}
  +\underbrace{F_\mathrm{blade}\sin(Z_n\omega t)}_{
      \text{runner--nozzle interaction}},
  \label{eq:FxR}
  \\[6pt]
  F_{y,R}
  &=
  -(k_\mathrm{run}+K_{22})\,y_R - K_{12}\,y_G
  +0.8\,k_\mathrm{cross}\,x_R
  \nonumber\\
  &\quad
  +\tau_\mathrm{seal}\omega D_\mathrm{seal}\,x_R
  -\tilde{K}_\mathrm{seal}K_{\mathrm{seal},0}\,y_R
  -c_\mathrm{run}\,\dot{y}_R
  \nonumber\\
  &\quad
  +m_2e_2\omega^2\sin\phi_2
  +A_\mathrm{vortex}q\cos(f_\mathrm{vortex}t)
  +F_\mathrm{blade}\cos(Z_n\omega t).
  \label{eq:FyR}
\end{align}
Here $x_G$ and $y_G$ are received from the generator lateral subsystem $\mathcal{M}_4$, and $\omega$ is received
from the rotational subsystem $\mathcal{M}_3$. The coefficients $K_{22}$ and $K_{12}$ are equivalent shaft
stiffness terms obtained from the shafting geometry and bearing-support formulation in~\cite{Shi2022Vibration};
their detailed expressions are omitted here for compactness.
The blade-passing force amplitude and hydraulic excitation frequency are defined as
\begin{equation}
  F_\mathrm{blade}=0.1A_\mathrm{vortex},
  \qquad
  f_\mathrm{vortex}=f_{\mathrm{vortex,ratio}}\omega.
\end{equation}
The hydraulic and blade-passing excitations are treated as reduced harmonic disturbance terms representing
low-frequency hydraulic pulsation and runner--nozzle interaction, respectively. Their amplitudes and frequency
ratios are model parameters rather than first-principles quantities.
The seal contribution is represented by a reduced linear direct-stiffness and cross-coupled form. In this
representation, the direct seal stiffness acts as a restoring force, while the cross-coupled term introduces
tangential destabilizing effects. The full nonlinear seal-force formulation and the calculation of equivalent seal
stiffness and damping coefficients are given in~\cite{Shi2022Vibration}.

\paragraph{Torsional process model.}
The torsional dynamics are governed by the reduced two-disc model
\begin{align}
  \dot{\alpha}_t &= \beta_t, \label{eq:alp_dot}\\
  J_\mathrm{eq}\dot{\beta}_t
  &= T_\mathrm{imbalance}
     - c_t\beta_t
     - k_y\alpha_t,
  \label{eq:beta_dot}
\end{align}
where $k_y$ is the shaft torsional stiffness, $c_t$ is the torsional damping coefficient, and $J_\mathrm{eq}$ is the
reduced torsional inertia. The detailed expressions for the equivalent shaft stiffnesses and effective inertias
follow the shafting formulation in~\cite{Shi2022Vibration}.
The coupling torque $T_\mathrm{imbalance}$ arises from the lateral--torsional interaction and is written as
\begin{equation}
  T_\mathrm{imbalance}
  =
  J_\mathrm{eq}
  \left(
    \tau_\mathrm{gen}
    +
    \tau_\mathrm{run}
    -
    \frac{M_g}{J_{1,\mathrm{eff}}}
    -
    \frac{M_t}{J_{2,\mathrm{eff}}}
  \right),
  \label{eq:T_imbalance}
\end{equation}
where $M_g$ and $M_t$ are the generator and turbine torques, respectively. The generator and runner
lateral--torsional coupling contributions are
\begin{align}
  \tau_\mathrm{gen}
  &=
  \frac{
    -m_1e_1
    \left(
      \ddot{y}_G\cos\phi_1
      -
      \ddot{x}_G\sin\phi_1
    \right)
  }{J_{1,\mathrm{eff}}},
  \label{eq:tau_gen}
  \\
  \tau_\mathrm{run}
  &=
  \frac{
    m_2e_2
    \left(
      \ddot{y}_R\cos\phi_2
      -
      \ddot{x}_R\sin\phi_2
    \right)
  }{J_{2,\mathrm{eff}}}.
  \label{eq:tau_run}
\end{align}
These terms represent torque contributions induced by lateral accelerations of the eccentric generator and runner
masses.

\paragraph{Seal parameter dynamics.}
The scaled seal stiffness is modeled as a random walk,
\begin{equation}
  \dot{\tilde{K}}_\mathrm{seal}=0,
  \label{eq:Kseal_rw}
\end{equation}
driven only by process noise. This allows the UKF to track slow changes in the effective seal stiffness caused by
wear, clearance variation, or modeling uncertainty.

\paragraph{Observation model.}
Runner lateral displacements are assumed to be measured by proximity probes:
\begin{equation}
  \mathbf{y}_R
  =
  \begin{bmatrix}
    x_R \\ y_R
  \end{bmatrix}
  +
  \begin{bmatrix}
    v_{x_R} \\ v_{y_R}
  \end{bmatrix},
  \qquad
  v_{x_R},v_{y_R}\sim\mathcal{N}(0,\sigma_R^2),
  \quad
  \sigma_R=5\,\mu\mathrm{m}.
  \label{eq:run_obs}
\end{equation}

\paragraph{Messages produced.}
The runner subsystem sends its lateral displacement estimates to the generator lateral subsystem:
\begin{equation}
  \text{from }\mathcal{M}_5:\qquad
  x_R,\;y_R
  \longrightarrow
  \mathcal{M}_4.
\end{equation}

\subsubsection*{Jacobi message-passing scheme}

At each time step $t_k$, the five estimators are updated in parallel using a Jacobi-style message-passing scheme:
all interface messages are first frozen at their values from the previous estimate $\hat{\mathbf{x}}(t_{k-1})$,
then each estimator performs its full prediction--update cycle independently, and finally the updated states are
stored for the next step.
This avoids constructing a global augmented state while preserving the physically meaningful interface variables.
Table~\ref{tab:messages} summarizes the complete message graph.
Because all messages are built before any estimator advances, the scheme is fully explicit: no iterative inner loop
is required, and the computational cost scales linearly with the number of subsystems.
%
\newpage
\begin{table}[htbp]
\centering
\caption{%
  State vector decomposition for the turbine--generator
  system.
  Rows with a ``$+$'' index offset denote parameters appended to the
  subsystem state during estimation.
  Type: M = measured; L = latent; P = estimated parameter.
}
\label{tab:state_decomp}
\small
\begin{tabular}{clllcl}
\toprule
\textbf{Index} & \textbf{Symbol} & \textbf{Description}
  & \textbf{Units} & \textbf{Type} & \textbf{Module} \\
\midrule
0  & $x_1$          & Penstock wave state 1       & --     & L & $\mathcal{M}_2$ \\
1  & $x_2$          & Penstock wave state 2       & --     & L & $\mathcal{M}_2$ \\
2  & $h$            & Normalized hydraulic head deviation & p.u. & M(NARX) & $\mathcal{M}_2$ \\
3  & $q$            & Normalized flow rate        & p.u.   & M & $\mathcal{M}_2$ \\
4  & $x_n$          & Needle stroke (actuator)    & p.u.   & L & $\mathcal{M}_1$ \\
5  & $x_{\mathrm{int}}$ & Governor integrator     & p.u.   & L & $\mathcal{M}_1$ \\
6  & $n$            & Speed deviation             & p.u.   & L & $\mathcal{M}_3$ \\
7  & $\omega$       & Angular velocity            & rad/s  & M & $\mathcal{M}_3$ \\
8  & $\phi_1$       & Generator rotor angle       & rad    & L & $\mathcal{M}_3$ \\
9  & $x_G$          & Generator $x$-displacement  & m      & M & $\mathcal{M}_4$ \\
10 & $y_G$          & Generator $y$-displacement  & m      & M & $\mathcal{M}_4$ \\
11 & $\dot{x}_G$    & Generator $x$-velocity      & m/s    & L & $\mathcal{M}_4$ \\
12 & $\dot{y}_G$    & Generator $y$-velocity      & m/s    & L & $\mathcal{M}_4$ \\
13 & $x_R$          & Runner $x$-displacement     & m      & M & $\mathcal{M}_5$ \\
14 & $y_R$          & Runner $y$-displacement     & m      & M & $\mathcal{M}_5$ \\
15 & $\dot{x}_R$    & Runner $x$-velocity         & m/s    & L & $\mathcal{M}_5$ \\
16 & $\dot{y}_R$    & Runner $y$-velocity         & m/s    & L & $\mathcal{M}_5$ \\
17 & $\alpha_t$     & Torsional twist angle       & rad    & L & $\mathcal{M}_5$ \\
18 & $\beta_t$ & Torsional rate $(\dot{\alpha}_t)$ & rad/s  & L & $\mathcal{M}_5$ \\
\midrule
\multicolumn{6}{l}{\textit{Augmented estimation states (appended per filter):}} \\
\midrule
$+1$ & $\tilde{e}_x$ & Scaled torque coeff.\ (speed) & -- & P & $\mathcal{M}_3$ EKF \\
$+2$ & $\tilde{e}_y$ & Scaled torque coeff.\ (flow) & -- & P & $\mathcal{M}_3$ EKF \\
$+1$ & $\tilde{K}_\mathrm{seal}$ & Scaled seal stiffness & -- & P & $\mathcal{M}_5$ UKF \\
\bottomrule
\end{tabular}
\end{table}

\begin{table}[htbp]
\centering
\caption{%
  Interface messages exchanged between subsystems at each Jacobi
  step, listed edge by edge (cf.\ Fig.~\ref{fig:sos_turbine_schematic}).
  All messages are computed from the posterior state of the
  sending subsystem at the \emph{previous} time step.
}
\label{tab:messages}
\small
\begin{tabular}{lll}
\toprule
\textbf{Sender} & \textbf{Message variables} & \textbf{Receiver(s)} \\
\midrule
$\mathcal{M}_1$ (Governor)   & $x_n$    & $\mathcal{M}_2$ \\
$\mathcal{M}_2$ (Hydraulics) & $q$, $h$ & $\mathcal{M}_3$ \\
$\mathcal{M}_2$ (Hydraulics) & $q$      & $\mathcal{M}_5$ \\
$\mathcal{M}_3$ (Rotational) & $n$, $\dot{n}$ & $\mathcal{M}_1$ \\
$\mathcal{M}_3$ (Rotational) & $\omega$, $\phi_1$ & $\mathcal{M}_4$ \\
$\mathcal{M}_3$ (Rotational) & $\omega$, $\phi_1$, $M_t$, $M_g$ & $\mathcal{M}_5$ \\
$\mathcal{M}_4$ (Generator)  & $x_G$, $y_G$, $\ddot{x}_G$, $\ddot{y}_G$ & $\mathcal{M}_5$ \\
$\mathcal{M}_5$ (Runner)     & $x_R$, $y_R$ & $\mathcal{M}_4$ \\
\bottomrule
\end{tabular}
\end{table}

\begingroup
\small
\begin{longtable}{llll}
\caption{%
  Physical parameters of the turbine--generator model.
  All values are SI unless noted.
}
\label{tab:physical_params}\\
\toprule
\textbf{Symbol} & \textbf{Description} & \textbf{Value}
  & \textbf{Units} \\
\midrule
\endfirsthead
\caption[]{%
  Physical parameters of the turbine--generator model (continued).
}\\
\toprule
\textbf{Symbol} & \textbf{Description} & \textbf{Value}
  & \textbf{Units} \\
\midrule
\endhead
\midrule
\multicolumn{4}{r}{\textit{Continued on next page}} \\
\endfoot
\bottomrule
\endlastfoot
\multicolumn{4}{l}{\textit{Machine ratings}} \\
$n_\mathrm{rated}$ & Rated speed & 375 & rpm \\
$Q_\mathrm{rated}$ & Rated flow & 27 & m\textsuperscript{3}/s \\
$H_\mathrm{rated}$ & Rated head & 595 & m \\
$\eta$ & Overall efficiency & 0.92 & -- \\
$Z_n$ & Number of nozzles & 6 & -- \\
$\omega_\mathrm{rated}$ & Rated angular velocity & $2\pi \times 375/60$ & rad/s \\
$M_{gB}$ & Rated torque ($P_\mathrm{rated}/\omega_\mathrm{rated}$) & derived & N\,m \\
\midrule
\multicolumn{4}{l}{\textit{Hydraulic / penstock}} \\
$T_e$ & Elastic time constant & 0.5155 & s \\
$T_q$ & Flow inertia time constant & 0.5 & s \\
$h_0$ & Upstream head coefficient & 1.0 & p.u. \\
$y_r$ & Rated guide-vane opening & 0.9 & p.u. \\
$s_e$ & Max.\ needle stroke & 0.102 & m \\
$\alpha_n$ & Nozzle inlet half-angle & 45 & deg \\
$\beta_n$ & Nozzle deflection angle & 62 & deg \\
$C_q$ & Nozzle flow coefficient & derived & -- \\
\midrule
\multicolumn{4}{l}{\textit{Governor PID}} \\
$k_p$ & Proportional gain & 5.0 & -- \\
$k_i$ & Integral gain & 2.12 & -- \\
$k_d$ & Derivative gain & 10.0 & -- \\
$T_y$ & Servo time constant & 0.04 & s \\
\midrule
\multicolumn{4}{l}{\textit{Inertias and masses}} \\
$m_1$ & Generator rotor mass & $6\times10^5$ & kg \\
$m_2$ & Runner mass & $3\times10^5$ & kg \\
$J_1$ & Generator polar inertia & $6.8\times10^6$ & kg\,m\textsuperscript{2} \\
$J_2$ & Runner polar inertia & $3.4\times10^6$ & kg\,m\textsuperscript{2} \\
$J_{1,\mathrm{eff}}$ & Effective generator inertia & derived & kg\,m\textsuperscript{2} \\
$J_{2,\mathrm{eff}}$ & Effective runner inertia & derived & kg\,m\textsuperscript{2} \\
$J_\mathrm{tot}$ & Total inertia & derived & kg\,m\textsuperscript{2} \\
$J_\mathrm{eq}$ & Reduced torsional inertia & derived & kg\,m\textsuperscript{2} \\
\midrule
\multicolumn{4}{l}{\textit{Shaft geometry}} \\
$d_H$ & Shaft outer diameter & 1.15 & m \\
$d_B$ & Shaft bore diameter & 0.3 & m \\
$\ell_\mathrm{shaft}$ & Inter-mass shaft length & 10.3 & m \\
$h_\mathrm{shaft}$ & Overall shaft length & 11.995 & m \\
$a$ & Generator-side overhang & 1.5 & m \\
$b$ & Bearing span & 7.3 & m \\
$c$ & Runner-side overhang & 1.5 & m \\
$E$ & Young's modulus & $200\times10^9$ & Pa \\
$G$ & Shear modulus & $80\times10^9$ & Pa \\
$\phi_m$ & Misalignment angle & 0.01 & rad \\
\midrule
\multicolumn{4}{l}{\textit{Bearing and damping}} \\
$k_\mathrm{gen}$ & Generator bearing stiffness & $2\times10^9$ & N/m \\
$k_\mathrm{run}$ & Runner bearing stiffness & $8\times10^9$ & N/m \\
$k_\mathrm{cross}$ & Lateral cross-coupling stiffness & $2\times10^7$ & N/m \\
$c_\mathrm{gen}$ & Generator bearing damping & $1\times10^6$ & N\,s/m \\
$c_\mathrm{run}$ & Runner bearing damping & $1\times10^6$ & N\,s/m \\
$c_\mathrm{cross}$ & Cross-coupling damping & $2\times10^4$ & N\,s/m \\
$c_t$ & Torsional damping & $1\times10^3$ & N\,m\,s/rad \\
$D_\omega$ & Rotational speed damping & $1\times10^5$ & N\,m\,s/rad \\
\midrule
\multicolumn{4}{l}{\textit{Eccentricities and clearances}} \\
$e_1$ & Generator eccentricity & $0.5\times10^{-3}$ & m \\
$e_2$ & Runner eccentricity & $0.5\times10^{-3}$ & m \\
$\delta_0$ & Generator air-gap clearance & $4\times10^{-3}$ & m \\
$\delta_2$ & Runner seal clearance & $3.5\times10^{-3}$ & m \\
$k_r$ & Rubbing stiffness & $6\times10^9$ & N/m \\
$f_\mathrm{fric}$ & Dry friction coefficient & 0.02 & -- \\
\midrule
\multicolumn{4}{l}{\textit{Seal}} \\
$K_{\mathrm{seal},0}$ & Nominal seal stiffness & $3\times10^7$ & N/m \\
$D_\mathrm{seal}$ & Seal damping & $1\times10^5$ & N\,s/m \\
$\tau_\mathrm{seal}$ & Circumferential velocity ratio & 0.3 & -- \\
\midrule
\multicolumn{4}{l}{\textit{Torque coefficients (Xu model)}} \\
$e_{x,0}$ & Nominal speed torque coefficient & 1.0 & -- \\
$e_{y,0}$ & Nominal flow torque coefficient & 0.5 & -- \\
\midrule
\multicolumn{4}{l}{\textit{Hydraulic excitation}} \\
$A_\mathrm{vortex}$ & Vortex rope force amplitude & $2\times10^4$ & N \\
$f_\mathrm{vortex,ratio}$ & Vortex frequency ratio & 0.25 & -- \\
\end{longtable}
\endgroup

\begin{table}[htbp]
\centering
\caption{%
  Filter hyperparameters for each estimation subsystem.
  $\mathbf{P}_0$ is the initial error covariance,
  $\mathbf{Q}$ the process noise covariance, and
  $\mathbf{R}$ the measurement noise covariance.
  Matrices are diagonal; entries are listed in state order.
  The scaling ratio for initial parameter estimates is
  $\tilde{\theta}_0 = 0.7$ for all estimated parameters,
  corresponding to a 30\,\% deviation from the nominal value.
}
\label{tab:filter_hyperparams}
\small
\begin{tabular}{lllll}
\toprule
\textbf{Module} & \textbf{Filter} & \textbf{Matrix}
  & \textbf{State / obs.} & \textbf{Value (diagonal)} \\
\midrule
\multirow{3}{*}{$\mathcal{M}_3$ Rotational}
  & \multirow{3}{*}{EKF}
  & $\mathbf{P}_0$
  & $[n,\omega,\phi_1,\tilde{e}_x,\tilde{e}_y]$
  & $\mathrm{diag}(10^{-6},\,10^{-6},\,10^{-6},\,10,\,10)$ \\
  & & $\mathbf{Q}$
  & same
  & $\mathrm{diag}(10^{-10},\,10^{-10},\,10^{-10},\,10^{-15},\,10^{-14})$ \\
  & & $\mathbf{R}$
  & $[\omega]$
  & $\sigma_\omega^2 + 10^{-10} = 4\times10^{-4}$ \\
\midrule
\multirow{3}{*}{$\mathcal{M}_4$ Generator}
  & \multirow{3}{*}{Kalman filter}
  & $\mathbf{P}_0$
  & $[x_G,y_G,\dot{x}_G,\dot{y}_G]$
  & $\mathrm{diag}(10^{-8},\,10^{-8},\,10^{-4},\,10^{-4})$ \\
  & & $\mathbf{Q}$
  & same
  & $\mathrm{diag}(10^{-10},\,10^{-10},\,10^{-11},\,10^{-11})$ \\
  & & $\mathbf{R}$
  & $[x_G,\,y_G]$
  & $\mathrm{diag}(\sigma_G^2+10^{-10},\,\sigma_G^2+10^{-10})$,
    $\sigma_G=5\,\mu\mathrm{m}$ \\
\midrule
\multirow{4}{*}{$\mathcal{M}_5$ Runner}
  & \multirow{4}{*}{UKF ($\gamma\!=\!1$)}
  & $\mathbf{P}_0$
  & $[x_R,y_R,\dot{x}_R,\dot{y}_R,\alpha_t,\beta_t,\tilde{K}_{\mathrm{seal}}]$
  & $\mathrm{diag}(10^{-8},\,10^{-8},\,10^{-4},\,10^{-4},
    \,10^{-4},\,10^{-4},\,10^{2})$ \\
  & & $\mathbf{Q}$
  & same
  & $\mathrm{diag}(10^{-10},\,10^{-10},\,10^{-8},\,10^{-8},
    \,10^{-8},\,10^{-8},\,10^{-8})$ \\
  & & $\mathbf{R}$
  & $[x_R,\,y_R]$
  & $\mathrm{diag}(\sigma_R^2+10^{-10},\,\sigma_R^2+10^{-10})$,
    $\sigma_R=5\,\mu\mathrm{m}$ \\
  & & $\tilde{K}_{\mathrm{seal},0}$
  & initial parameter estimate
  & 0.7 \\
\midrule
\multirow{1}{*}{NARX ($\mathcal{M}_2$)}
  & --
  & Training
  & $[L,d_h,\text{dropout},\eta_\mathrm{lr}]$
  & $[200,\,128,\,0.05,\,10^{-4}]$ \\
\midrule
\multicolumn{5}{l}{\textit{Measurement noise standard deviations used
  for synthetic data generation:}} \\
  & & $\sigma_\omega$ & Speed encoder & 0.02\;rad/s \\
  & & $\sigma_q$ & Flow sensor & 0.01\;p.u. \\
  & & $\sigma_G$ & Generator probes & $5\;\mu\mathrm{m}$ \\
  & & $\sigma_R$ & Runner probes & $5\;\mu\mathrm{m}$ \\
\bottomrule
\end{tabular}
\end{table}

\begin{figure}[htbp]
\centering
\includegraphics[width=\linewidth]{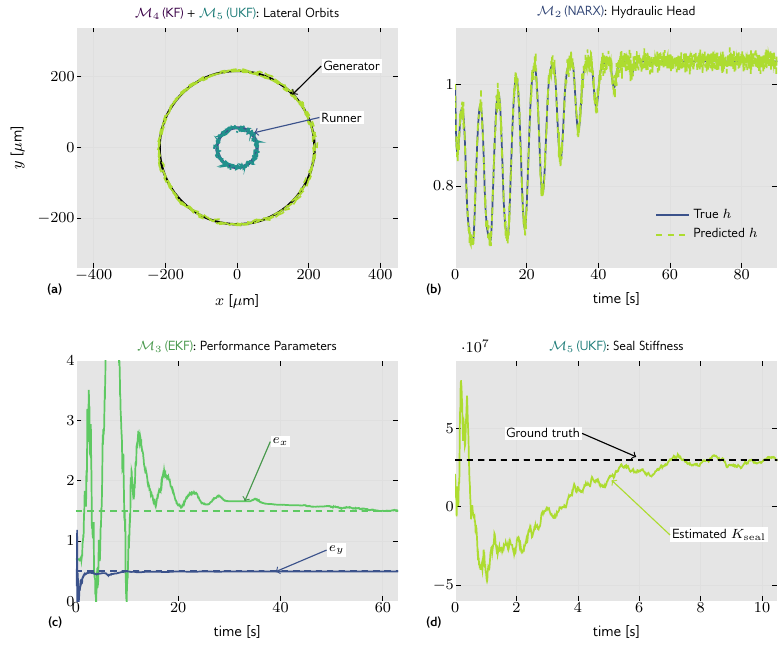}
\caption{%
  Representative subsystem-estimation results for the hydro-turbine case study.
  (a) Lateral orbital motion (in $\mu$m) of the generator (black, ground truth) and runner (dark blue, ground
  truth), recovered by the Kalman filter on $\mathcal{M}_4$ (light-green dashed) and the UKF on $\mathcal{M}_5$
  (teal dash-dotted).
  (b) Normalized hydraulic head $h$ reconstructed online by the NARX surrogate of $\mathcal{M}_2$ (dashed) against
  the ground truth (solid).
  (c) Performance coefficients annotated as $e_x$ and $e_y$ in the rotational subsystem $\mathcal{M}_3$, estimated by
  the EKF (solid) and compared with their dashed reference levels.
  (d) Inferred seal stiffness $K_{\mathrm{seal}}$ [N/m] in the runner subsystem $\mathcal{M}_5$ from the UKF, with the
  estimated curve and dashed ground-truth reference indicated by arrows.
}
\label{fig:sos_turbine_results}
\end{figure}

Figure~\ref{fig:sos_turbine_results} summarizes representative estimation outputs for the decomposed
turbine--generator estimator after the subsystem models and message interfaces have been specified.
Panel~(a) tests the two structural filters that are coupled through the shaft and bearing messages.
The generator orbit is shown together with the light-green dashed Kalman-filter trajectory for $\mathcal{M}_4$, while the runner orbit
is shown together with the teal dash-dotted UKF trajectory for $\mathcal{M}_5$.
This comparison checks that the lateral structural states can be reconstructed separately even though the two
subsystems exchange acceleration and displacement messages at every Jacobi step.
Panel~(b) isolates the hydraulic block $\mathcal{M}_2$.
The NARX surrogate reconstructs the hydraulic-head state $h$ online from the delayed input--output history rather
than by integrating the full penstock equations inside the estimator.

Panel~(c) reports the slowly varying performance coefficients estimated inside the rotational EKF
$\mathcal{M}_3$, with the two annotated traces identifying the plotted $e_x$ and $e_y$ estimates.
These coefficients scale the speed and flow-rate contributions to the turbine torque (cf.~\eqref{eq:Mt}), so errors
in them would propagate directly into the mechanical torque messages sent to the governor, generator, and runner
subsystems.
Panel~(d) shows the runner UKF estimate of the seal stiffness $K_{\mathrm{seal}}$, with direct annotations marking the
estimated curve and the dashed ground-truth reference.
This parameter affects the lateral restoring force in $\mathcal{M}_5$ and is inferred jointly with the runner
displacements, runner velocities, and torsional states.
Taken together, the four result panels show that the decomposition in Fig.~\ref{fig:sos_turbine_schematic} supports
state reconstruction, surrogate-based hydraulic inference, and local parameter estimation within one explicit
message-passing estimator.

Quantitatively, starting from biased initial parameter estimates, the coupled estimator attains the values
reported in the main manuscript: the aggregate estimation RMSE on the four lateral displacements
$(x_G, y_G, x_R, y_R)$ is $9.14\times10^{-6}$\,m, and the trajectory NRMSEs of the seal stiffness
$\tilde{K}_{\mathrm{seal}}$ and the torque coefficients $\tilde{e}_x$ and $\tilde{e}_y$ are
$5.25\times10^{-1}$, $1.19$, and $8.23\times10^{-2}$, respectively.
These trajectory NRMSEs are computed over the estimation window $t \le 80$\,s and normalized by the true value of
the corresponding scaled parameter; because they average over the entire trajectory, they are dominated by the
biased initialization and the convergence transient visible in panels~(c) and~(d) rather than by the converged
estimates.
During coupled online inference, the NARX surrogate maintains a one-step-ahead RMSE of $2.42\times10^{-3}$,
slightly above the held-out test value of $1.94\times10^{-3}$ reported above, because inside the closed
estimation loop the surrogate is driven by estimated rather than measured inputs. The reported turbine results likewise correspond to a single noise realization with a fixed random seed.

\subsection*{Hierarchical composition: IEEE 9-bus network with embedded hydro turbines}
\label{App:turbine_ieee9}

\begin{figure}[htbp]
\centering
\includegraphics[width=\linewidth]{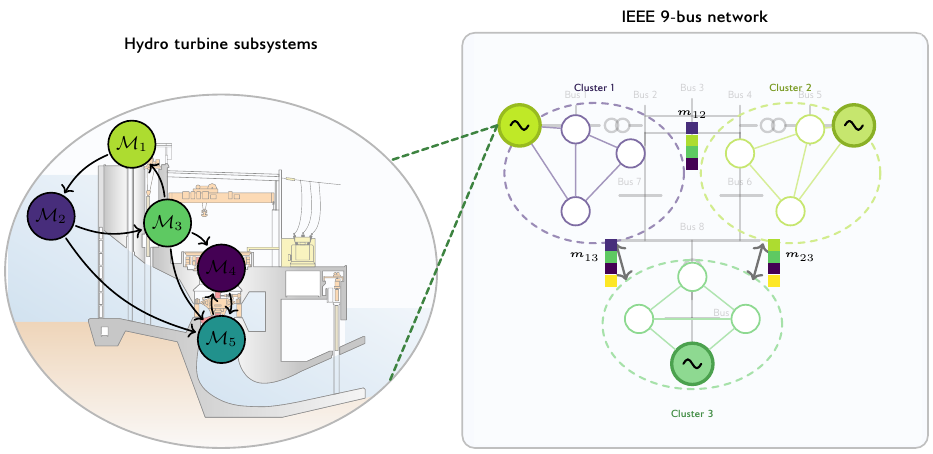}
\caption{%
  Hierarchical composition of the IEEE 9-bus grid
  with embedded hydro turbines.
  Right: the IEEE 9-bus network is partitioned into three clusters
  (Cluster~1--Cluster~3); generator buses are highlighted, and the
  bidirectional arrows between clusters denote inter-cluster
  message-passing $\mathbf{m}_{ij}$ used by the distributed estimator.
  Left: each generator bus is replaced by the full hydro turbine
  model of the \emph{Multi-Physics Turbine--Generator System} section, with the five
  subsystems $\mathcal{M}_1$--$\mathcal{M}_5$ (governor, hydraulic,
  rotational, generator, runner) and their internal interface graph.
  The dashed green lines indicate the embedding of one such turbine
  into a generator bus of Cluster~1; the same construction is
  repeated at every generator bus.
}
\label{fig:ieee_turbine_embed}
\end{figure}

The two previous case studies covered the building blocks separately: the first evaluated the scalability of the
proposed method on several power-grid test cases, and the second used a subsystem decomposition to model a single
hydro turbine as a collection of heterogeneous subsystems.
In this case study we combine both ingredients by embedding the full hydro turbine model inside a power-grid test
case and solving the
\emph{inverse} problem, in which the latent states of every embedded
turbine are reconstructed online from a small set of noisy measurements at each generator while the IEEE 9-bus
network is simultaneously propagated forward in time.
This demonstrates that the proposed method handles
\emph{hierarchical} compositions in which a coarse outer network
and detailed inner subsystem models are simulated and estimated jointly.
Figure~\ref{fig:ieee_turbine_embed} summarizes the construction: the IEEE 9-bus grid (right) is partitioned into
three clusters that exchange inter-cluster messages, and each generator bus in the grid is replaced by the
five-subsystem hydro turbine model (left).

The outer network in the hierarchical case study is the standard IEEE 9-bus system,
retrieved from the PYPOWER library~\cite{Zimmerman2011MATPOWER}.
The network consists of nine buses, three generator buses (buses 1, 2, and 3), three load buses (buses 5, 7, and 9), and the remaining buses carrying no direct
injection.
The electrical parameters are summarized in Table~\ref{tab:ieee9_network}.

In the inner loop of the hierarchical model, each generator bus is replaced by the hydro turbine model
developed in the \emph{Multi-Physics Turbine--Generator System} section, which has 5 subsystems $\mathcal{M}_1$--$\mathcal{M}_5$ and a total of 19
states. However, the non-generator buses are represented by a simple first-order Kuramoto model. 
The three turbines coupled through the IEEE 9-bus network thus form a hierarchical distributed system with $3\times 19 = 57$
subsystem states plus the non-generator bus angles, organized into the three clusters indicated in
Fig.~\ref{fig:ieee_turbine_embed}.

The bus admittance matrix $\mathbf{Y}_\mathrm{bus}$ is constructed from the branch impedance data of the test case
using the standard MATPOWER/PYPOWER admittance-matrix construction, with a system base of $S_\mathrm{base} =
100$\,MVA.
The mechanical coupling strength between nodes $i$ and $j$ used in the dynamical model is taken as the imaginary
part (susceptance) of the corresponding off-diagonal admittance entry,
\begin{equation}
  K_{ij} = \operatorname{Im}(Y_{ij}),
  \quad i \neq j,
  \qquad K_{ii} = 0,
  \label{eq:K_susceptance}
\end{equation}
which enters the power-angle coupling term $K_{ij}\sin(\delta_j - \delta_i)$ in the swing equations.
We note that the hierarchical IEEE-9 case uses the susceptance entry $\operatorname{Im}(Y_{ij})$, whereas the
power-grid scalability study in~\eqref{eq:K_from_Ybus} uses the admittance magnitude $|Y_{ij}|$; the susceptance
choice here is exact for the swing-equation coupling under the standard low-resistance assumption, while the
magnitude choice in the scalability study acts as a sign-invariant proxy across heterogeneous IEEE test cases.
The sign of $K_{ij}$ in~\eqref{eq:K_susceptance} follows the PYPOWER convention for $\mathbf{Y}_\mathrm{bus}$,
which is consistent with the orientation of the coupling term $K_{ij}\sin(\delta_j - \delta_i)$ used
throughout this section.

\subsubsection*{Hybrid electromechanical model}
\label{App:hybrid_model}

The coupled simulation uses a hybrid model in which different bus types are represented at different levels of
fidelity, and within each generator bus the five-subsystem turbine is itself split into deterministically
propagated subsystems and Kalman-filter estimation blocks.
Generator buses (buses 1, 2, 3) are each replaced by the full 19-state hydro turbine model described in
the \emph{Multi-Physics Turbine--Generator System} section, while the non-generator buses are represented by the first-order Kuramoto dynamics
described in the \emph{First-Order Kuramoto Dynamics for Non-Generator Buses} section.
This avoids the need for an algebraic network solve (Newton--Raphson (NR) power-flow) at every time step and keeps the
grid update fully explicit.

Inside each generator bus, the governor ($\mathcal{M}_1$), hydraulic ($\mathcal{M}_2$), and rotational
($\mathcal{M}_3$) subsystems are propagated by the same Heun--Jacobi scheme used in the standalone turbine of
the \emph{Multi-Physics Turbine--Generator System} section.
The only difference is that the generator subsystem $\mathcal{M}_4$ and the runner subsystem
$\mathcal{M}_5$ were estimated with a UKF on the full nonlinear lateral--torsional model; in the hierarchical
co-simulation, assuming known $K_{seal}$, the same subsystem is linearized about the operating trajectory at every step and propagated with
a Kalman filter, which keeps the per-bus update inexpensive enough to be embedded inside the outer-network
Heun loop.
For each generator the available measurements are the lateral displacements $(x_G, y_G)$ of the generator, the
lateral displacements $(x_R, y_R)$ of the runner, the turbine flow rate $q$, and the turbine angular speed
$\omega$.
The forward simulation and synthetic-measurement generation procedure are described below.

\paragraph*{First-order Kuramoto dynamics for non-generator buses.}
\label{App:kur1_nongen}

Unlike the standalone Kuramoto/swing simulations above, each non-generator
bus $i \in \{4, 5, 6, 7, 8, 9\}$ evolves through the following first-order Kuramoto phase dynamics:
\begin{equation}
  \dot{\delta}_i
  =
  \frac{P_{\mathrm{inj},i} + P_{e,i}(\boldsymbol{\delta})}{D_i},
  \label{eq:kur1_nongen}
\end{equation}
where $P_{\mathrm{inj},i}$ is the net power injection (generation minus load) at bus $i$,
\begin{equation}
  P_{e,i}(\boldsymbol{\delta})
  =
  \sum_{j=1}^{9} K_{ij} \sin(\delta_j - \delta_i),
  \label{eq:Pe_kur}
\end{equation}
is the electrical power flowing into bus $i$ from the network, and $D_i > 0$ is the effective damping coefficient
that controls the time scale of the phase response.
All non-generator buses use the same damping value $D_i = D_\mathrm{kur} = 2$ (in units consistent with the
per-unit power base).
The first-order model~\eqref{eq:kur1_nongen} is the overdamped limit of the swing equation, appropriate for load
buses where inertia effects are negligible compared to the network synchronization dynamics.

\paragraph*{Generator bus dynamics: embedded hydro turbine.}
\label{App:turbine_grid_coupling}

Compared to the standalone turbine--generator system, one more modification is needed to couple the turbine dynamics to the grid: in the standalone turbine, the generator was subject to a fixed braking torque $M_g$ that could be tuned to produce a desired operating point; in the hierarchical co-simulation, this fixed torque is replaced by a dynamic electrical torque $T_e$ that depends on the instantaneous electrical power exchanged with the network~\cite{Anderson1994Power,Sauer1998Power}.
Specifically, the electrical power flowing out of generator bus $i$ is
\begin{equation}
  P_{e,i}(\boldsymbol{\delta})
  =
  -\sum_{j=1}^{9} K_{ij} \sin(\delta_j - \delta_i),
  \label{eq:Pe_gen}
\end{equation}
and the corresponding electrical torque applied to the rotational subsystem $\mathcal{M}_3$ is
\begin{equation}
  T_{e,i}
  =
  -\xi \cdot \frac{S_\mathrm{base}}{\omega_{\mathrm{rated},i}}
  \cdot P_{e,i},
  \label{eq:Te_from_Pe}
\end{equation}
where $\xi$ is a dimensionless torque-scaling factor (set to $\xi = 1$ in all simulations) and
$\omega_{\mathrm{rated},i}$ is the rated angular velocity of turbine $i$.
The factor $S_\mathrm{base}/\omega_\mathrm{rated}$ converts the per-unit electrical power into a physical torque in
N\,m.
This torque replaces the generator braking torque $M_g$ in the rotational subsystem dynamics, so that the turbine shaft decelerates when electrical demand increases and accelerates
when demand falls.
The resulting initial torque balance for the three embedded generator buses is verified in
Table~\ref{tab:ieee9_torque}, where the mechanical turbine torque and the network-derived electrical torque agree
to machine precision.
The rotor angle $\delta_i$ of generator bus $i$ is identified with the shaft deviation state
$\delta_{\mathrm{state}}$ in the turbine model, defined as
\begin{equation}
  \delta_{\mathrm{state}}(t)
  =
  \phi_1(t) - \omega_\mathrm{rated}\,t,
  \label{eq:delta_state}
\end{equation}
where $\phi_1$ is the absolute generator rotor angle.
This decomposition avoids catastrophic cancellation that would arise from integrating the large absolute angle
$\phi_1(t) =
\delta_\mathrm{state} + \omega_\mathrm{rated}\,t$ over long
simulation horizons.
The absolute angle is reconstructed when needed for lateral-force computations in subsystems $\mathcal{M}_4$ and
$\mathcal{M}_5$.

\paragraph*{Center-of-inertia reference frame.}
\label{App:COI}

Generator rotor angles are reported relative to the center-of-inertia (COI) reference angle~\cite{Pai1989Energy}, which is a weighted average of the generator angles that tracks the common-mode frequency drift of the system.
\begin{equation}
  \delta_\mathrm{COI}(t)
  =
  \sum_{i \in \mathcal{G}} w_i\, \delta_i(t),
  \qquad
  w_i = \frac{P_{\mathrm{gen},i}}{\sum_{j \in \mathcal{G}} P_{\mathrm{gen},j}},
  \label{eq:COI}
\end{equation}
where $\mathcal{G} = \{1, 2, 3\}$ denotes the set of generator buses and $w_i$ is a power-based weighting
factor that uses the rated generation $P_{\mathrm{gen},i}$ as a proxy for the per-machine inertia.
The COI-relative angle of bus $i$ is then $\tilde{\delta}_i = \delta_i - \delta_\mathrm{COI}$.
This reference removes the common-mode frequency drift from the displayed results and makes inter-machine angle
deviations directly interpretable~\cite{You2021Calculate}.
\subsubsection*{Forward simulation and synthetic measurement data}
\label{App:ieee9_data_generation}

The inverse problem is posed using synthetic measurement data generated from a forward truth simulation of the
hybrid IEEE 9-bus--turbine system.
The IEEE 9-bus operating point and network quantities in Table~\ref{tab:ieee9_network} are used to initialize the
bus injections, generator ratings, steady-state voltage angles, and coupling matrix $\mathbf{K}$.
With these quantities fixed, the coupled system is simulated over a horizon of $T = 60$\,s, using the disturbance
scenario and numerical parameters listed in Table~\ref{tab:ieee9_params}.
The resulting forward trajectory provides the ground-truth turbine and network states.
Synthetic measurements are then produced on the same time grid by adding zero-mean Gaussian noise to the measured
generator displacements $(x_G,y_G)$, runner displacements $(x_R,y_R)$, turbine flow rate $q$, and angular speed
$\omega$, with the noise levels given in Table~\ref{tab:ieee9_params}.
These noisy sensor records are supplied to the inverse Kalman-filter estimator, which reconstructs the latent
turbine states using the linearized Kalman-filter blocks on $\mathcal{M}_4$ and $\mathcal{M}_5$ described in the
\emph{Hybrid Electromechanical Model} section, while the uncorrupted forward trajectory is retained only for
computing reconstruction errors.
\subsubsection*{Time integration and state estimation scheme}
\label{App:ieee9_integration}

The coupled IEEE 9-bus--turbine system is integrated using a predictor--corrector (Heun) scheme at the outer grid
level, with the embedded turbine solved via a sub-stepped Jacobi--Heun scheme at the inner level.

\paragraph*{Outer grid: predictor--corrector.}
\label{App:outer_heun}

At each outer time step $t_k$ with step size $\Delta t = 10^{-3}$\,s, the following operations are performed:
\begin{enumerate}
  \item \textbf{Predictor.}
    For each generator bus $i \in \{1, 2, 3\}$:
    compute the electrical power $P_{e,i}(\boldsymbol{\delta}_k)$
    using the current angles;
    advance the turbine state one step using a single-substep
    Jacobi propagation with 2 Jacobi iterations.
    The implementation-level state vector stores the shaft
    deviation $\delta_{\mathrm{state},i}$ (not the absolute rotor
    angle $\phi_{1,i}$) at index~18, so the predicted grid-side
    angle is obtained as
    $\boldsymbol{\delta}_{k+1}^\mathrm{pred}[i]
     = \hat{x}_{i,k+1}[18]$,
    consistent with~\eqref{eq:delta_state} and the implementation
    layout used in Table~\ref{tab:state_decomp}.
    During the inverse pass, the same predictor step also produces
    a one-step-ahead Kalman prediction of the
    $\mathcal{M}_4$/$\mathcal{M}_5$ states (no measurement
    update applied at the predictor stage).
    For each non-generator bus $b \in \{4,\ldots,9\}$:
    compute the Euler predictor,
    $\delta_{b,k+1}^\mathrm{pred}
     = \delta_{b,k}
       + \Delta t\,
       (P_{\mathrm{inj},b} + P_{e,b}(\boldsymbol{\delta}_k))
       / D_b$.

  \item \textbf{Corrector.}
    Evaluate $\mathbf{P}_e^\mathrm{pred}$ at
    $\boldsymbol{\delta}_{k+1}^\mathrm{pred}$.
    For each generator bus $i$:
    compute the average electrical torque
    $\bar{T}_{e,i} = \tfrac{1}{2}(T_{e,i}^k + T_{e,i}^\mathrm{pred})$
    and advance the turbine state using the full sub-stepped
    Jacobi--Heun scheme with $n_\mathrm{sub} = 2$ substeps and
    2 Jacobi iterations per substep.
    In the inverse simulator the corrector also performs the
    Kalman \emph{measurement update}, fusing the $(x_G, y_G)$ and
    $(x_R, y_R)$ samples available at $t_{k+1}$ into the
    $\mathcal{M}_4$ and $\mathcal{M}_5$ state and covariance
    estimates.
    For each non-generator bus $b$:
    apply the Heun corrector,
    $\delta_{b,k+1}
     = \delta_{b,k}
       + \tfrac{\Delta t}{2}(f_b^k + f_b^\mathrm{pred})$,
    where $f_b = (P_{\mathrm{inj},b} + P_{e,b})/D_b$.
\end{enumerate}

\paragraph*{Inner turbine: sub-stepped Jacobi--Heun.}
\label{App:inner_turbine}

Each turbine time step $\Delta t$ is divided into $n_\mathrm{sub} = 2$ substeps of size $\Delta t_\mathrm{sub} =
\Delta t / n_\mathrm{sub} = 5\times10^{-4}$\,s.
Within each substep, the five turbine subsystems ($\mathcal{M}_1$--$\mathcal{M}_5$) are updated using the Jacobi
message-passing scheme, with the electrical torque $T_e$ treated as a frozen external input for the duration of the
substep.
Each substep performs 2 Jacobi sweeps so that the hydraulics and rotational subsystems exchange interface messages
twice within the same substep before advancing.
This sub-stepping strategy is necessary to resolve the fast structural dynamics of the turbine (characteristic
frequencies of order 1--10\,Hz) within the outer grid time step of $1$\,ms.

\subsubsection*{Disturbance scenario}
\label{App:disturbance}

Two load disturbances are applied sequentially at bus 9 (0-based index 8):
\begin{enumerate}
  \item A demand increase of $+20$\,MW at $t = 5.0$\,s,
    implemented as a step change in the net injection
    $P_{\mathrm{inj},9} \leftarrow P_{\mathrm{inj},9} - 20/S_\mathrm{base}$.

  \item A demand decrease of $-20$\,MW at $t = 30.0$\,s,
    restoring the original injection level.
\end{enumerate}
Both disturbances are applied instantaneously within the time loop without any smoothing or ramp.
Bus 9 is chosen because it carries the largest static load (125\,MW) in the IEEE 9-bus case and is
electrically close to generator bus 3, making the per-turbine response asymmetric across the three generators.

\subsubsection*{Simulation results}
\label{App:ieee9_results}

\begin{figure}[htbp]
\centering
\includegraphics[width=\linewidth]{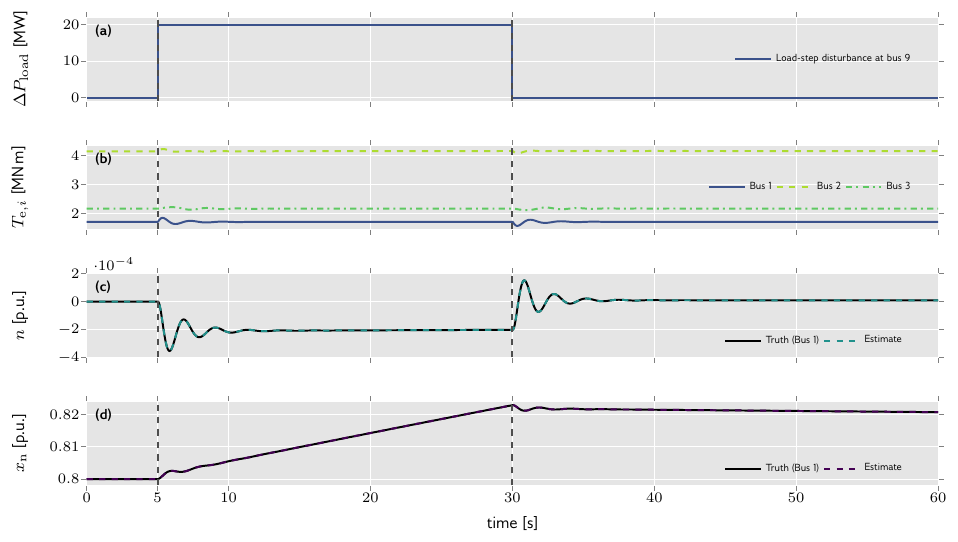}
\caption{%
  Cross-scale propagation of the bus-9 load step into the embedded
  hydro turbines. Panel labels~(a)--(d) are shown in the top-left
  corner of each subplot.
  (a)~Load disturbance $\Delta P_\mathrm{load}$ \emph{applied at bus~9}
  (the network-level event that drives the rest of the figure; the
  same trace is shared by all three generators because it is a
  network-level disturbance).
  (b)~Grid-induced electrical torque $T_{\mathrm{e},i}$ at the three
  generator buses (\emph{bus~1}, \emph{bus~2}, \emph{bus~3}), computed
  from the network power flow via~\eqref{eq:Te_from_Pe}.
  (c)~Rotational-subsystem ($\mathcal{M}_3$) speed deviation $n$
  \emph{at bus~1} (the smallest of the three generators, 67\,MW),
  comparing the forward truth (solid) with the inverse estimate
  (dashed).
  (d)~Governor needle position $x_n$ \emph{at bus~1}, again truth
  versus inverse estimate.
  Vertical dashed lines mark the disturbance times $t = 5$\,s
  and $t = 30$\,s.
}
\label{fig:ieee9_fcross}
\end{figure}

Figure~\ref{fig:ieee9_fcross} shows how the load step at bus~9 propagates from the outer grid into the embedded
turbines.
Panel~(a) is the $\Delta P_\mathrm{load}$ profile \emph{at bus~9} ($+20$\,MW at $t=5$\,s, $-20$\,MW at $t=30$\,s);
this is the only network-level disturbance in the experiment.
Panel~(b) shows the resulting grid-induced electrical torque $T_{\mathrm{e},i}$ delivered to each of the three
turbines (\emph{buses~1, 2, and~3}) via~\eqref{eq:Te_from_Pe}; the three traces differ in DC level because of
the unequal generator sizes (Gen~2 at $163$\,MW carries the largest torque), but they share an almost identical
transient following each step.
Panel~(c) reports the speed deviation $n$ of the rotational subsystem $\mathcal{M}_3$ \emph{at bus~1} (the smallest
unit, 67\,MW), which dips to roughly $-3\times10^{-4}$\,p.u.\ before the governor returns it to a small
steady-state offset of order $-2\times10^{-4}$\,p.u.
Panel~(d) shows the corresponding governor needle position $x_n$ \emph{at bus~1}, slowly opening from $0.80$ to
about $0.82$ over $\sim25$\,s as the integrator clears the wind-up.
Bus~1 is used in panels~(c) and~(d) because it is the smallest of the three generators and therefore the most
sensitive to the load step; the corresponding traces for buses~2 and~3 are qualitatively similar and are omitted
to keep the figure compact.
The estimated traces in panels~(c) and~(d) come from the inverse simulator and overlap the truth lines visually.

\begin{figure}[htbp]
\centering
\includegraphics[width=\linewidth]{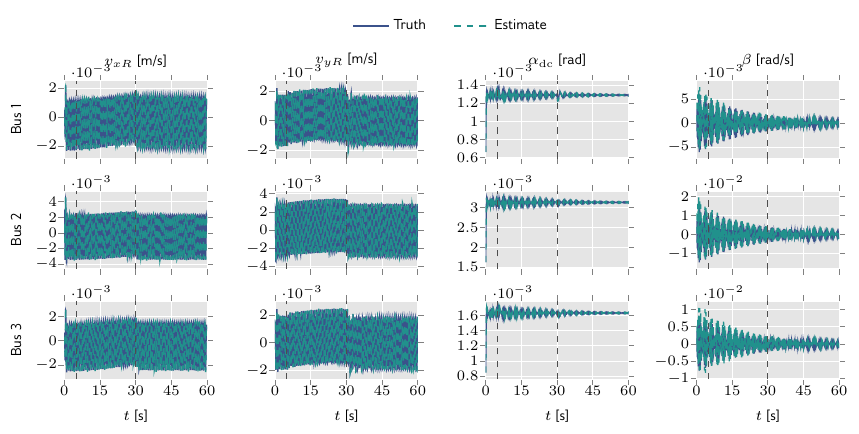}
\caption{%
  Hidden-state recovery for the three embedded turbines over the
  full $T = 60$\,s simulation.
  Rows: generator buses~1, 2, 3.
  Columns: runner $x$- and $y$-velocity $(v_{xR}, v_{yR})$,
  the DC component of the torsional twist $\alpha_{\mathrm{dc}}$,
  and the smoothed torsional rate $\beta$.
  Solid lines: forward truth. Dashed lines: Kalman-filter estimate.
  Vertical dashed lines mark the disturbance times $t = 5$\,s and
  $t = 30$\,s.
  None of these four quantities is directly measured;
  all are reconstructed from the
  $(x_G, y_G, x_R, y_R, q, \omega)$ measurements via the
  dynamic model and inter-subsystem coupling.
}
\label{fig:ieee9_hidden_states}
\end{figure}

\paragraph*{Hidden-state recovery across all three generators.}
Figure~\ref{fig:ieee9_hidden_states} aggregates the four hidden runner-side quantities of greatest engineering
interest across the full $T = 60$\,s simulation horizon: the lateral velocities $(v_{xR}, v_{yR}) \equiv
(\dot{x}_R, \dot{y}_R)$, the (slowly varying) DC component of the torsional twist $\alpha_{\mathrm{dc}}$, and
the smoothed torsional rate $\beta = \dot{\alpha}_t$.
Each row corresponds to one generator bus.
The Kalman-filter estimates track the truth at every bus and across all four hidden quantities; none of these
four channels is directly measured, and each is reconstructed purely from the $(x_G, y_G, x_R, y_R, q, \omega)$
sensor set via the dynamic model and the inter-cluster message passing of
Fig.~\ref{fig:ieee_turbine_embed}.
The corresponding per-bus steady-state RMSE values are tabulated in
Table~\ref{tab:ieee9_results}.

\paragraph*{Embedded versus standalone inference.}
To quantify whether the hierarchical embedding degrades local inference, each embedded turbine estimate is compared with a standalone reference estimator of the same five-subsystem turbine, driven by the grid-induced electrical-torque trajectory $T_{e,i}(t)$ at the corresponding generator bus extracted from a separate grid-only simulation of the same disturbance scenario.
Aggregated across the three generator buses and the full $T=60$\,s horizon, the embedded estimator matches the standalone reference on the speed deviation $n$, the needle position $x_n$, and the flow rate $q$ with an aggregate trajectory RMSE of $1.02\times10^{-3}$, the value reported in the main manuscript.
Composition into the grid hierarchy therefore leaves local turbine inference essentially unchanged: each subsystem retains its own state, covariance, and parameter estimates, and the hierarchy enters only through the exchanged interface messages; all hierarchical results correspond to a single noise realization with a fixed random seed.

\begin{figure}[htbp]
\centering
\includegraphics[width=\linewidth]{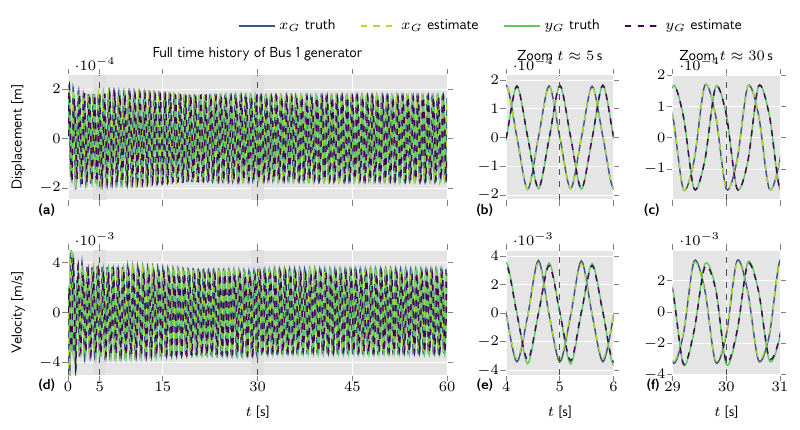}
\caption{%
  Generator subsystem $\mathcal{M}_4$ state estimation at bus~1.
  Top row: lateral displacements $(x_G, y_G)$, with
  (a)~the full $T = 60$\,s history,
  (b)~zoom $t \in [4, 6]$\,s,
  (c)~zoom $t \in [29, 31]$\,s.
  Bottom row: corresponding lateral velocities $(v_{xG}, v_{yG})$
  with the same three views in
  (d)~the full history,
  (e)~zoom $t \in [4, 6]$\,s,
  (f)~zoom $t \in [29, 31]$\,s.
  Truth ($x_G$ blue solid, $y_G$ green solid) and Kalman-filter
  estimate ($x_G$ light-green dashed, $y_G$ purple dashed) overlap to
  graphical accuracy in every panel.
  Shaded bands on the full-history panels mark the two zoom windows;
  vertical dashed lines mark the disturbance times $t = 5$\,s and
  $t = 30$\,s.
}
\label{fig:ieee9_generator_zooms}
\end{figure}

\paragraph*{Generator lateral vibration at bus~1 ($\mathcal{M}_4$).}
Figure~\ref{fig:ieee9_generator_zooms} drills into the generator lateral orbits at the slack bus (bus~1).
The Kalman filter block estimates the four generator states $(x_G, y_G, v_{xG}, v_{yG})$ from the $(x_G, y_G)$ measurements
only; the velocity components $(v_{xG}, v_{yG})$ are therefore \emph{hidden} states recovered purely through the
dynamic model and the message-passing coupling to $\mathcal{M}_5$.
Panel~(a) shows the full-history lateral displacements $(x_G, y_G)$ and panels~(b) and~(c) zoom into the
$t \approx 5$\,s and $t \approx 30$\,s load-step events; panels~(d), (e), and~(f) show the same three views
of the corresponding velocities $(v_{xG}, v_{yG})$.
The shaded bands on the full-history panels mark the zoomed intervals.
At both event windows the estimated displacement and velocity traces overlap the truth to graphical accuracy.

\begin{figure}[htbp]
\centering
\includegraphics[width=\linewidth]{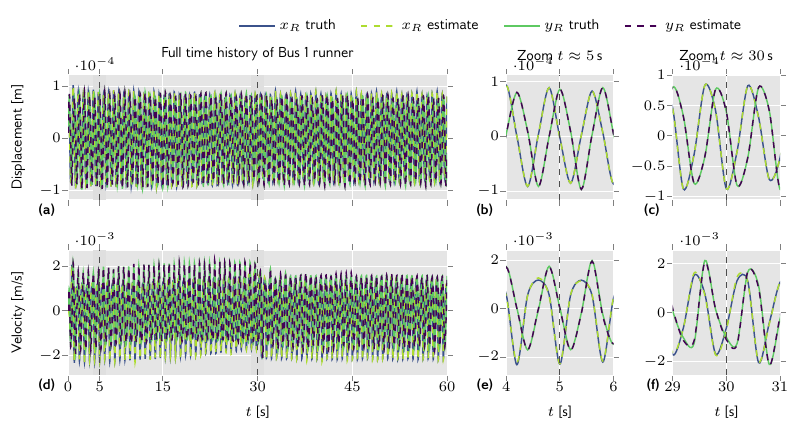}
\caption{%
  Runner subsystem $\mathcal{M}_5$ state estimation at bus~1.
  Layout follows Fig.~\ref{fig:ieee9_generator_zooms}, with the
  runner quantities $(x_R, y_R)$ replacing the generator
  counterparts:
  (a)~full $T = 60$\,s history of $(x_R, y_R)$,
  (b)~zoom $t \in [4, 6]$\,s,
  (c)~zoom $t \in [29, 31]$\,s;
  (d)~full history of $(v_{xR}, v_{yR})$,
  (e)~zoom $t \in [4, 6]$\,s,
  (f)~zoom $t \in [29, 31]$\,s.
  Truth ($x_R$ blue solid, $y_R$ green solid) and Kalman-filter
  estimate ($x_R$ light-green dashed, $y_R$ purple dashed) overlap to
  graphical accuracy.
  Only the two displacement channels $(x_R, y_R)$ are measured;
  the velocity channels are recovered as hidden states.
  Shaded bands on the full-history panels mark the two zoom
  windows centered on $t = 5$\,s and $t = 30$\,s.
}
\label{fig:ieee9_runner_zooms}
\end{figure}

\paragraph*{Runner lateral vibration at bus~1 ($\mathcal{M}_5$).}
Figure~\ref{fig:ieee9_runner_zooms} shows the analogous result for the runner subsystem at bus~1.
Here the Kalman filter estimates six states $(x_R, y_R, v_{xR}, v_{yR}, \alpha_t, \beta_t)$, with
$\beta_t \equiv \dot{\alpha}_t$, from only the two displacement measurements $(x_R, y_R)$, so the four
velocity / torsional components are reconstructed from the model.
Panel~(a) shows the full-history runner displacements $(x_R, y_R)$, with panels~(b) and~(c) magnifying the
response around $t \approx 5$\,s and $t \approx 30$\,s; panels~(d), (e), and~(f) repeat the same three views
for the corresponding velocities $(v_{xR}, v_{yR})$.
Despite the large dimensional gap between observations ($n_z = 2$) and states ($n_x = 6$), the displacement traces
overlap the truth at both event windows and the velocity trajectories match within the displayed amplitude.

Together, Figures~\ref{fig:ieee9_fcross}--\ref{fig:ieee9_runner_zooms} demonstrate that the hierarchical distributed system
approach (i)~propagates disturbances correctly across scales (network $\to$ rotational $\to$ governor $\to$
structural), (ii)~reconstructs all unobserved internal states of every embedded turbine from a small set of
generator-side measurements (including the torsional twist and rate, which are not measured at any generator),
and (iii)~delivers tight truth-versus-estimate agreement at every event window in the lateral displacement and
velocity channels of both the generator and the runner subsystems.
Table~\ref{tab:ieee9_results} summarizes the key system-level observables extracted from the simulation output.
%


\begin{table}[htbp]
\centering
\caption{%
  IEEE 9-bus network parameters for the hierarchical
  case study.
  Bus indices follow 1-based PYPOWER convention throughout this
  table.
  Steady-state angles $\delta^*$ are measured from the slack bus
  (bus 1, $\delta^* = 0$).
  The coupling matrix $\mathbf{K}$ is derived from the susceptance
  (imaginary part) of the admittance matrix.
}
\label{tab:ieee9_network}
\small
\begin{tabular}{llrrl}
\toprule
\textbf{Bus} & \textbf{Type} & \textbf{$P_\mathrm{gen}$ (MW)}
  & \textbf{$P_\mathrm{load}$ (MW)} & \textbf{$\delta^*$ (deg)} \\
\midrule
1 & Generator (slack) &  67 &   0 &   0.00 \\
2 & Generator         & 163 &   0 &  10.12 \\
3 & Generator         &  85 &   0 &   5.34 \\
4 & Load / transit    &   0 &   0 &  $-2.21$ \\
5 & Load              &   0 &  90 &  $-3.79$ \\
6 & Load / transit    &   0 &   0 &   2.49 \\
7 & Load              &   0 & 100 &   1.09 \\
8 & Load / transit    &   0 &   0 &   4.27 \\
9 & Load              &   0 & 125 &  $-4.09$ \\
\midrule
\multicolumn{2}{l}{Total generation} & 315 & -- & -- \\
\multicolumn{2}{l}{Total load} & -- & 315 & -- \\
\multicolumn{2}{l}{System base $S_\mathrm{base}$} & \multicolumn{3}{l}{100 MVA} \\
\multicolumn{2}{l}{NR power-flow residual} & \multicolumn{3}{l}{$1.11\times10^{-16}$\,p.u.} \\
\bottomrule
\end{tabular}
\end{table}

\begin{table}[htbp]
\centering
\caption{%
  Steady-state torque balance verification at initialization.
  $M_t$ is the mechanical (hydraulic) torque computed from the
  hydro turbine model;
  $T_e$ is the electrical torque computed from the network
  power-flow solution via~\eqref{eq:Te_from_Pe}.
  All residuals are below $10^{-15}$\,MN\,m, confirming
  machine-precision consistency between the turbine model and
  the electrical network at $t = 0$.
}
\label{tab:ieee9_torque}
\small
\begin{tabular}{lrrrr}
\toprule
\textbf{Generator} & \textbf{Bus} & \textbf{$M_t$ (MN\,m)}
  & \textbf{$T_e$ (MN\,m)} & \textbf{Residual (MN\,m)} \\
\midrule
Gen 1 & 1 & 1.7061 & 1.7061 & $2.33 \times 10^{-16}$ \\
Gen 2 & 2 & 4.1508 & 4.1508 & $9.31 \times 10^{-16}$ \\
Gen 3 & 3 & 2.1645 & 2.1645 & $4.66 \times 10^{-16}$ \\
\bottomrule
\end{tabular}
\end{table}

\begin{table}[htbp]
\centering
\caption{%
  Key simulation results for the hierarchical co-simulation of the
  IEEE 9-bus network with embedded hydro turbines, including the inverse
  (Kalman-filter) reconstruction of the embedded turbine states.
  Electrical power ranges cover the full $T = 60$\,s simulation
  horizon.
  Speed deviations are peak values relative to rated speed.
  Estimation errors are reported as the steady-state RMSE between
  the inverse Kalman estimate and the synthetic forward truth
  for the four hidden runner-side quantities of
  Fig.~\ref{fig:ieee9_hidden_states}.
  All vibration amplitudes remain below the mechanical
  clearance limits (generator air-gap $\delta_0 = 4$\,mm;
  runner seal clearance $\delta_2 = 3.5$\,mm).
}
\label{tab:ieee9_results}
\small
\begin{tabular}{lrrr}
\toprule
\textbf{Quantity} & \textbf{Gen 1 (Bus 1)} & \textbf{Gen 2 (Bus 2)} & \textbf{Gen 3 (Bus 3)} \\
\midrule
Rated power (MW) & 67 & 163 & 85 \\
DC electrical torque $T_e$ (MN\,m) & $\approx 1.71$ & $\approx 4.15$ & $\approx 2.16$ \\
Peak speed deviation $n$ (p.u.) & $\sim -3\times10^{-4}$ & $\sim -2\times10^{-4}$ & $\sim -2\times10^{-4}$ \\
Steady $x_n$ after $+20$\,MW & $\approx 0.821$ & $\approx 0.821$ & $\approx 0.821$ \\
\midrule
\multicolumn{4}{l}{\textit{Inverse-estimation steady-state RMSE:}} \\
Runner velocity $v_{xR}$ (m/s) & $7.09\times10^{-5}$ & $1.18\times10^{-4}$ & $7.95\times10^{-5}$ \\
Runner velocity $v_{yR}$ (m/s) & $6.09\times10^{-5}$ & $6.35\times10^{-5}$ & $6.02\times10^{-5}$ \\
Torsional twist DC $\alpha_\mathrm{dc}$ (rad) & $1.93\times10^{-6}$ & $4.47\times10^{-6}$ & $4.33\times10^{-6}$ \\
Torsional rate $\beta$ (rad/s) & $3.85\times10^{-4}$ & $8.75\times10^{-4}$ & $6.84\times10^{-4}$ \\
\midrule
\multicolumn{4}{l}{\textit{Disturbance events:}} \\
Event 1 & \multicolumn{3}{l}{$+20$\,MW step at bus 9, $t = 5.0$\,s} \\
Event 2 & \multicolumn{3}{l}{$-20$\,MW step at bus 9, $t = 30.0$\,s} \\
\bottomrule
\end{tabular}
\end{table}

\begin{table}[htbp]
\centering
\caption{%
  Numerical parameters for the hierarchical co-simulation.
  Turbine physical and filter parameters are as reported in
  Tables~\ref{tab:physical_params}--\ref{tab:filter_hyperparams};
  only the co-simulation-specific parameters are listed here.
}
\label{tab:ieee9_params}
\small
\begin{tabular}{lll}
\toprule
\textbf{Parameter} & \textbf{Description} & \textbf{Value} \\
\midrule
\multicolumn{3}{l}{\textit{Outer time integration}} \\
$\Delta t$ & Outer time step & $10^{-3}$\,s \\
$T$ & Simulation horizon & 60.0\,s \\
$n_\mathrm{sub}$ & Turbine substeps per outer step & 2 \\
$\Delta t_\mathrm{sub}$ & Turbine substep size & $5\times10^{-4}$\,s \\
\midrule
\multicolumn{3}{l}{\textit{Jacobi coupling (inner turbine)}} \\
$N_\mathrm{Jacobi}^\mathrm{pred}$ & Jacobi iters (predictor) & 2 \\
$N_\mathrm{Jacobi}^\mathrm{corr}$ & Jacobi iters (corrector) & 2 \\
\midrule
\multicolumn{3}{l}{\textit{Network and coupling}} \\
$S_\mathrm{base}$ & System MVA base & 100\,MVA \\
Coupling type & $K_{ij}$ from $\mathbf{Y}_\mathrm{bus}$ & susceptance $\operatorname{Im}(Y_{ij})$ \\
$\xi$ & Torque scaling factor & 1.0 \\
\midrule
\multicolumn{3}{l}{\textit{Non-generator buses (1st-order Kuramoto)}} \\
$D_\mathrm{kur}$ & Kuramoto damping coefficient & 2.0 \\
Non-gen buses (1-based) & Buses using~\eqref{eq:kur1_nongen} & 4, 5, 6, 7, 8, 9 \\
\midrule
\multicolumn{3}{l}{\textit{Generator turbine sizing}} \\
$\eta$ & Hydraulic efficiency & 0.92 \\
$\rho$ & Water density & 1000\,kg/m\textsuperscript{3} \\
$g$ & Gravitational acceleration & 9.81\,m/s\textsuperscript{2} \\
$H_\mathrm{rated}$ & Rated head (all turbines) & 595\,m \\
$Q_{\mathrm{rated},i}$ & Rated flow (derived per turbine) & $P_{\mathrm{gen},i}/(\rho g H \eta)$ \\
$n_\mathrm{rated}$ & Rated speed & 375\,rpm \\
\midrule
\multicolumn{3}{l}{\textit{Disturbance scenario}} \\
Bus & Disturbance location (1-based) & Bus 9 \\
$\Delta P_1$ & First demand step & $+20$\,MW at $t=5.0$\,s \\
$\Delta P_2$ & Second demand step & $-20$\,MW at $t=30.0$\,s \\
\midrule
\multicolumn{3}{l}{\textit{Governor speed reference}} \\
$n_\mathrm{ref}$ & Speed reference (all turbines) & 0.0\,p.u.\ deviation \\
\midrule
\multicolumn{3}{l}{\textit{Initial turbine conditions}} \\
$x_n(0)$ & Needle position & 0.8 \\
$q(0)$ & Normalized flow & 1.0 \\
$\omega(0)$ & Angular velocity & $\omega_\mathrm{rated}$ \\
$\delta_\mathrm{state}(0)$ & Initial shaft deviation angle & $\delta^*_i$ from NR power flow \\
$x_G(0),\,y_G(0)$ & Generator lateral IC & $10^{-5}$\,m \\
$x_R(0),\,y_R(0)$ & Runner lateral IC & $10^{-5}$\,m \\
\midrule
\multicolumn{3}{l}{\textit{Inverse simulator: sensor noise (Gaussian, zero-mean)}} \\
$\sigma_{xG},\sigma_{yG}$ & Generator displacement noise & $2\times10^{-6}$\,m \\
$\sigma_{xR},\sigma_{yR}$ & Runner displacement noise & $2\times10^{-6}$\,m \\
$\sigma_{q}$ & Flow-rate noise & $2\times10^{-4}$\,p.u. \\
$\sigma_{\omega}$ & Speed noise & $2\times10^{-4}$\,rad/s \\
\midrule
\multicolumn{3}{l}{\textit{Inverse simulator: Kalman filter ($\mathcal{M}_4$, $\mathcal{M}_5$)}} \\
$Q_\mathrm{gen}$ & Process noise (generator block) & $10^{-11}$ \\
$Q_\mathrm{run}$ & Process noise (runner block) & $2\times10^{-12}$ \\
$Q_\mathrm{tor}$ & Process noise (torsional block) & $10^{-8}$ \\
$R_{xy,\mathrm{gen}}$ & Measurement noise (gen.) & $\sigma_{xG}^{2}$ \\
$R_{xy,\mathrm{run}}$ & Measurement noise (runner) & $100\,\sigma_{xR}^{2}$ \\
$P_{0,\mathrm{gen}}$ & Initial covariance (gen.) & $10^{-7}$ \\
$P_{0,\mathrm{run}}$ & Initial covariance (runner) & $10^{-7}$ \\
$P_{0,\mathrm{tor}}$ & Initial covariance (torsional) & $10^{-5}$ \\
\bottomrule
\end{tabular}
\end{table}

\FloatBarrier

\renewcommand{\refname}{Supplementary References}
\putbib[references]
\end{bibunit}


\begin{thebibliography}{39}
\providecommand{\natexlab}[1]{#1}
\providecommand{\url}[1]{\texttt{#1}}
\expandafter\ifx\csname urlstyle\endcsname\relax
  \providecommand{\doi}[1]{doi: #1}\else
  \providecommand{\doi}{doi: \begingroup \urlstyle{rm}\Url}\fi

\bibitem[Kapteyn et~al.(2021)Kapteyn, Pretorius, and
  Willcox]{Kapteyn2021probabilistica}
Michael~G. Kapteyn, Jacob V.~R. Pretorius, and Karen~E. Willcox.
\newblock A probabilistic graphical model foundation for enabling predictive
  digital twins at scale.
\newblock \emph{Nature Computational Science}, 1\penalty0 (5):\penalty0
  337--347, May 2021.
\newblock \doi{10.1038/s43588-021-00069-0}.

\bibitem[Willcox et~al.(2021)Willcox, Ghattas, and
  Heimbach]{Willcox2021imperative}
Karen~E. Willcox, Omar Ghattas, and Patrick Heimbach.
\newblock The imperative of physics-based modeling and inverse theory in
  computational science.
\newblock \emph{Nature Computational Science}, 1\penalty0 (3):\penalty0
  166--168, March 2021.
\newblock \doi{10.1038/s43588-021-00040-z}.

\bibitem[Masison et~al.(2021)Masison, Beezley, Mei, Ribeiro, Knapp,
  Sordo~Vieira, Adhikari, Scindia, Grauer, Helba, Schroeder, Mehrad, and
  Laubenbacher]{Masison2021modular}
J.~Masison, J.~Beezley, Y.~Mei, {\relax HAL}~Ribeiro, A.~C. Knapp,
  L.~Sordo~Vieira, B.~Adhikari, Y.~Scindia, M.~Grauer, B.~Helba, W.~Schroeder,
  B.~Mehrad, and R.~Laubenbacher.
\newblock A modular computational framework for medical digital twins.
\newblock \emph{Proceedings of the National Academy of Sciences}, 118\penalty0
  (20):\penalty0 e2024287118, May 2021.
\newblock \doi{10.1073/pnas.2024287118}.

\bibitem[Makke and Chawla(2024)]{Makke2024Datadriven}
Nour Makke and Sanjay Chawla.
\newblock Data-driven discovery of {{Tsallis-like}} distribution using symbolic
  regression in high-energy physics.
\newblock \emph{PNAS Nexus}, 3\penalty0 (11):\penalty0 pgae467, November 2024.
\newblock \doi{10.1093/pnasnexus/pgae467}.

\bibitem[Champion et~al.(2019)Champion, Lusch, Kutz, and
  Brunton]{Champion2019Datadriven}
Kathleen Champion, Bethany Lusch, J.~Nathan Kutz, and Steven~L. Brunton.
\newblock Data-driven discovery of coordinates and governing equations.
\newblock \emph{Proceedings of the National Academy of Sciences}, 116\penalty0
  (45):\penalty0 22445--22451, November 2019.
\newblock \doi{10.1073/pnas.1906995116}.

\bibitem[Mohammadi and Taylor(2021)]{Mohammadi2021Thinking}
Neda Mohammadi and John~E. Taylor.
\newblock Thinking fast and slow in disaster decision-making with {{Smart City
  Digital Twins}}.
\newblock \emph{Nature Computational Science}, 1\penalty0 (12):\penalty0
  771--773, December 2021.
\newblock \doi{10.1038/s43588-021-00174-0}.

\bibitem[Peherstorfer et~al.(2018)Peherstorfer, Willcox, and
  Gunzburger]{Peherstorfer2018Survey}
Benjamin Peherstorfer, Karen Willcox, and Max Gunzburger.
\newblock Survey of {{Multifidelity Methods}} in {{Uncertainty Propagation}},
  {{Inference}}, and {{Optimization}}.
\newblock \emph{SIAM Review}, 60\penalty0 (3):\penalty0 550--591, January 2018.
\newblock \doi{10.1137/16M1082469}.

\bibitem[Arcucci et~al.(2026)Arcucci, Healy, Dance, Lei, Bach, Weaver, Miyoshi,
  Dillon, Draper, Schneider, Lang, Dueben, Bormann, Lean, Geer, Bonavita, {van
  Leeuwen}, Cheng, Bocquet, Zagar, {de Campos Velho}, Ruiz, Bauer, Boukabara,
  Carrassi, Treadon, Collard, Kleist, Gholoubi, Wang, Samrat, Ralton, Moore,
  Lamer, and Caltabiano]{Arcucci2026convergence}
Rossella Arcucci, Sean Healy, Sarah Dance, Lili Lei, Eviatar Bach, Anthony~T.
  Weaver, Takemasa Miyoshi, Maria~Eugenia Dillon, Clara Draper, Rochelle
  Schneider, Simon Lang, Peter Dueben, Niels Bormann, Peter Lean, Alan Geer,
  Massimo Bonavita, Peter~Jan {van Leeuwen}, Sibo Cheng, Marc Bocquet,
  Nedjeljka Zagar, Haroldo~Fraga {de Campos Velho}, Juan~Jose Ruiz, Peter
  Bauer, Sid~Ahmed Boukabara, Alberto Carrassi, Russ Treadon, Andrew Collard,
  Daryl Kleist, Azadeh Gholoubi, Xuguang Wang, Nahidul Samrat, Gemma Ralton,
  Andrew~M. Moore, Katia Lamer, and Nico Caltabiano.
\newblock The convergence of machine learning and data assimilation in
  {{Earth}} system science.
\newblock \emph{npj Artificial Intelligence}, 2\penalty0 (1):\penalty0 48,
  April 2026.
\newblock \doi{10.1038/s44387-026-00107-0}.

\bibitem[Leach et~al.(2021)Leach, Weisheimer, Allen, and
  Palmer]{Leach2021Forecastbased}
Nicholas~J. Leach, Antje Weisheimer, Myles~R. Allen, and Tim Palmer.
\newblock Forecast-based attribution of a winter heatwave within the limit of
  predictability.
\newblock \emph{Proceedings of the National Academy of Sciences}, 118\penalty0
  (49):\penalty0 e2112087118, December 2021.
\newblock \doi{10.1073/pnas.2112087118}.

\bibitem[Kalnay et~al.(2023)Kalnay, Sluka, Yoshida, Da, and
  Mote]{Kalnay2023Review}
Eugenia Kalnay, Travis Sluka, Takuma Yoshida, Cheng Da, and Safa Mote.
\newblock Review article: {{Towards}} strongly coupled ensemble data
  assimilation with additional improvements from machine learning.
\newblock https://repository.library.noaa.gov, 2023.

\bibitem[Tokuda and Dirmeyer(2026)]{Tokuda2026Selective}
Daisuke Tokuda and Paul~A. Dirmeyer.
\newblock Selective reuse of prior ensemble data improves the latest air
  temperature forecast over {{North America}}.
\newblock \emph{Proceedings of the National Academy of Sciences}, 123\penalty0
  (15):\penalty0 e2524516123, April 2026.
\newblock \doi{10.1073/pnas.2524516123}.

\bibitem[Karniadakis et~al.(2021)Karniadakis, Kevrekidis, Lu, Perdikaris, Wang,
  and Yang]{Karniadakis2021Physicsinformeda}
George~Em Karniadakis, Ioannis~G. Kevrekidis, Lu~Lu, Paris Perdikaris, Sifan
  Wang, and Liu Yang.
\newblock Physics-informed machine learning.
\newblock \emph{Nature Reviews Physics}, 3\penalty0 (6):\penalty0 422--440,
  June 2021.
\newblock \doi{10.1038/s42254-021-00314-5}.

\bibitem[Carrassi et~al.(2018)Carrassi, Bocquet, Bertino, and
  Evensen]{Carrassi2018Data}
Alberto Carrassi, Marc Bocquet, Laurent Bertino, and Geir Evensen.
\newblock Data assimilation in the geosciences: {{An}} overview of methods,
  issues, and perspectives.
\newblock \emph{Wiley Interdisciplinary Reviews: Climate Change}, 9\penalty0
  (5), 2018.

\bibitem[Gomes et~al.(2018)Gomes, Thule, Broman, Larsen, and
  Vangheluwe]{Gomes2018CoSimulation}
Cl{\'a}udio Gomes, Casper Thule, David Broman, Peter~Gorm Larsen, and Hans
  Vangheluwe.
\newblock Co-{{Simulation}}: {{A Survey}}.
\newblock \emph{ACM Computing Surveys (CSUR)}, 51\penalty0 (3):\penalty0
  49:1--49:33, May 2018.
\newblock \doi{10.1145/3179993}.

\bibitem[Tencer et~al.(2023)Tencer, Rojas, and Schroeder]{Tencer2023Network}
John Tencer, Edward Rojas, and Benjamin~B. Schroeder.
\newblock Network {{Uncertainty Quantification}} for {{Analysis}} of
  {{Multi-Component Systems}}.
\newblock \emph{ASCE-ASME Journal of Risk and Uncertainty in Engineering
  Systems, Part B: Mechanical Engineering}, 9\penalty0 (2):\penalty0 021203,
  June 2023.
\newblock \doi{10.1115/1.4055688}.

\bibitem[Chavali and Nehorai(2015)]{Chavali2015Distributed}
Phani Chavali and Arye Nehorai.
\newblock Distributed {{Power System State Estimation Using Factor Graphs}}.
\newblock \emph{IEEE Transactions on Signal Processing}, 63\penalty0
  (11):\penalty0 2864--2876, June 2015.
\newblock \doi{10.1109/TSP.2015.2413297}.

\bibitem[{Sanchez-Gonzalez} et~al.(2020){Sanchez-Gonzalez}, Godwin, Pfaff,
  Ying, Leskovec, and Battaglia]{Sanchez-Gonzalez2020Learning}
Alvaro {Sanchez-Gonzalez}, Jonathan Godwin, Tobias Pfaff, Rex Ying, Jure
  Leskovec, and Peter~W. Battaglia.
\newblock Learning to simulate complex physics with graph networks.
\newblock In \emph{Proceedings of the 37th {{International Conference}} on
  {{Machine Learning}}}, volume 119 of \emph{{{ICML}}'20}, pages 8459--8468.
  JMLR.org, July 2020.

\bibitem[Sharma and Fink(2026)]{Sharma2026physicsinformed}
Vinay Sharma and Olga Fink.
\newblock A physics-informed graph neural network conserving linear and angular
  momentum for dynamical systems.
\newblock \emph{Nature Communications}, 17\penalty0 (1):\penalty0 1045, January
  2026.
\newblock \doi{10.1038/s41467-025-67802-5}.

\bibitem[Yu and Wang(2024)]{Yu2024Learning}
Rose Yu and Rui Wang.
\newblock Learning dynamical systems from data: {{An}} introduction to
  physics-guided deep learning.
\newblock \emph{Proceedings of the National Academy of Sciences}, 121\penalty0
  (27):\penalty0 e2311808121, July 2024.
\newblock \doi{10.1073/pnas.2311808121}.

\bibitem[Cantwell and Newman(2019)]{Cantwell2019Message}
George~T. Cantwell and M.~E.~J. Newman.
\newblock Message passing on networks with loops.
\newblock \emph{Proceedings of the National Academy of Sciences}, 116\penalty0
  (47):\penalty0 23398--23403, November 2019.
\newblock \doi{10.1073/pnas.1914893116}.

\bibitem[Julier and Uhlmann(2004)]{Julier2004Unscented}
S.J. Julier and J.K. Uhlmann.
\newblock Unscented filtering and nonlinear estimation.
\newblock \emph{Proceedings of the IEEE}, 92\penalty0 (3):\penalty0 401--422,
  March 2004.
\newblock \doi{10.1109/JPROC.2003.823141}.

\bibitem[Ruozzi and Tatikonda(2013)]{Ruozzi2013Messagepassing}
Nicholas Ruozzi and Sekhar Tatikonda.
\newblock Message-passing algorithms for quadratic minimization.
\newblock \emph{J. Mach. Learn. Res.}, 14\penalty0 (1):\penalty0 2287--2314,
  January 2013.

\bibitem[Brunton et~al.(2016)Brunton, Proctor, and
  Kutz]{Brunton2016Discovering}
Steven~L. Brunton, Joshua~L. Proctor, and J.~Nathan Kutz.
\newblock Discovering governing equations from data by sparse identification of
  nonlinear dynamical systems.
\newblock \emph{Proceedings of the National Academy of Sciences}, 113\penalty0
  (15):\penalty0 3932--3937, April 2016.
\newblock \doi{10.1073/pnas.1517384113}.

\bibitem[Thurner et~al.(2018)Thurner, Scheidler, Sch{\"a}fer, Menke, Dollichon,
  Meier, Meinecke, and Braun]{Thurner2018Pandapower}
Leon Thurner, Alexander Scheidler, Florian Sch{\"a}fer, Jan-Hendrik Menke,
  Julian Dollichon, Friederike Meier, Steffen Meinecke, and Martin Braun.
\newblock Pandapower---{{An Open-Source Python Tool}} for {{Convenient
  Modeling}}, {{Analysis}}, and {{Optimization}} of {{Electric Power Systems}}.
\newblock \emph{IEEE Transactions on Power Systems}, 33\penalty0 (6):\penalty0
  6510--6521, November 2018.
\newblock \doi{10.1109/TPWRS.2018.2829021}.

\bibitem[Filatrella et~al.(2008)Filatrella, Nielsen, and
  Pedersen]{Filatrella2008Analysis}
G.~Filatrella, A.~H. Nielsen, and N.~F. Pedersen.
\newblock Analysis of a power grid using a {{Kuramoto-like}} model.
\newblock \emph{The European Physical Journal B}, 61\penalty0 (4):\penalty0
  485--491, February 2008.
\newblock \doi{10.1140/epjb/e2008-00098-8}.

\bibitem[Cohen and Migliorati(2017)]{Cohen2017Optimal}
Albert Cohen and Giovanni Migliorati.
\newblock Optimal weighted least-squares methods.
\newblock \emph{The SMAI Journal of computational mathematics}, 3:\penalty0
  181--203, October 2017.
\newblock \doi{10.5802/smai-jcm.24}.

\bibitem[Xu et~al.(2017)Xu, Yan, Chen, Gao, and Wu]{Xu2017Sensitivity}
Beibei Xu, Donglin Yan, Diyi Chen, Xiang Gao, and Changzhi Wu.
\newblock Sensitivity analysis of a {{Pelton}} hydropower station based on a
  novel approach of turbine torque.
\newblock \emph{Energy Conversion and Management}, 148:\penalty0 785--800,
  September 2017.
\newblock \doi{10.1016/j.enconman.2017.06.019}.

\bibitem[Sch{\"a}r et~al.(2025)Sch{\"a}r, Marelli, and
  Sudret]{Schar2025Surrogate}
Styfen Sch{\"a}r, Stefano Marelli, and Bruno Sudret.
\newblock Surrogate modeling with functional nonlinear autoregressive models
  ({{F}}{$<$}math{$><$}mi mathvariant="script"
  is="true"{$>$}{{F}}{$<$}/mi{$><$}/math{$>$}-{{NARX}}).
\newblock \emph{Reliability Engineering \& System Safety}, 264:\penalty0
  111276, December 2025.
\newblock \doi{10.1016/j.ress.2025.111276}.

\bibitem[Tcheumchoua et~al.(2022)Tcheumchoua, Nam, and
  Chung]{Tcheumchoua2022Torque}
Kamgang~Blaise Tcheumchoua, Seokho Nam, and Wan~Kyun Chung.
\newblock Torque {{Control}} of {{Hydraulic Pressure Servo Valve Driven
  Actuator}} with {{Deep Neural Network}}.
\newblock In \emph{2022 {{IEEE}}/{{RSJ International Conference}} on
  {{Intelligent Robots}} and {{Systems}} ({{IROS}})}, pages 12512--12519,
  October 2022.
\newblock \doi{10.1109/IROS47612.2022.9981211}.

\bibitem[Benner et~al.(2015)Benner, Gugercin, and Willcox]{Benner2015Survey}
Peter Benner, Serkan Gugercin, and Karen Willcox.
\newblock A {{Survey}} of {{Projection-Based Model Reduction Methods}} for
  {{Parametric Dynamical Systems}}.
\newblock \emph{SIAM Review}, 57\penalty0 (4):\penalty0 483--531, January 2015.
\newblock \doi{10.1137/130932715}.

\bibitem[Yang et~al.(2020)Yang, Meng, and Karniadakis]{Yang2020BPINNs}
Liu Yang, Xuhui Meng, and George~Em Karniadakis.
\newblock B-{{PINNs}}: {{Bayesian}} physics-informed neural networks for
  forward and inverse {{PDE}} problems with noisy data.
\newblock \emph{Journal of Computational Physics}, 425, October 2020.
\newblock \doi{10.1016/j.jcp.2020.109913}.

\bibitem[Carter et~al.(2023)Carter, Feddema, Kothe, Neely, Pruet, Stevens,
  Balaprakash, Beckman, Foster, Iskra, Ramanathan, Taylor, Thakur, Agarwal,
  Crivelli, {de Jong}, Rouson, Sohn, Wetter, Wild, Bremer, Goldman, Kupresanin,
  Peterson, Spears, Stevens, Van~Essen, Bent, Grosskopf, Lawrence, Shipman,
  Rose, Grout, Kouakpaizan, Omitaomu, Peles, Ramuhalli, Shankar, Womble, Zhang,
  Catanach, Oldfield, Rajamanickam, Ray, Leung, Catlett, and
  Dietrich]{Carter2023Advanced}
Jonathan Carter, John Feddema, Doug Kothe, Rob Neely, Jason Pruet, Rick
  Stevens, Prasanna Balaprakash, Pete Beckman, Ian Foster, Kamil Iskra, Arvind
  Ramanathan, Valerie Taylor, Rajeev Thakur, Deb Agarwal, Silvia Crivelli, Bert
  {de Jong}, Damian Rouson, Mike Sohn, Michael Wetter, Stefan Wild, Timo
  Bremer, Michael Goldman, Ana Kupresanin, Luc Peterson, Brian Spears, Dave
  Stevens, Brian Van~Essen, Russell Bent, Mike Grosskopf, Earl Lawrence, Galen
  Shipman, Kelly Rose, Ray Grout, Nicholson Kouakpaizan, Femi Omitaomu, Slaven
  Peles, Pradeep Ramuhalli, Arjun Shankar, David Womble, Guannan Zhang, Tommie
  Catanach, Ron Oldfield, Sivasankaran Rajamanickam, Jaideep Ray, Mary~Ann
  Leung, Charles Catlett, and Emily~M. Dietrich.
\newblock Advanced {{Research Directions}} on {{AI}} for {{Science}},
  {{Energy}}, and {{Security}}: {{Report}} on {{Summer}} 2022 {{Workshops}}.
\newblock Technical Report ANL-22/91, Argonne National Laboratory (ANL), May
  2023.

\bibitem[Zechner et~al.(2016)Zechner, Seelig, Rullan, and
  Khammash]{Zechner2016Molecular}
Christoph Zechner, Georg Seelig, Marc Rullan, and Mustafa Khammash.
\newblock Molecular circuits for dynamic noise filtering.
\newblock \emph{Proceedings of the National Academy of Sciences}, 113\penalty0
  (17):\penalty0 4729--4734, April 2016.
\newblock \doi{10.1073/pnas.1517109113}.

\bibitem[Singh et~al.(2020)Singh, Wang, Braver, and Ching]{Singh2020Scalable}
Matthew~F. Singh, Anxu Wang, Todd~S. Braver, and ShiNung Ching.
\newblock Scalable surrogate deconvolution for identification of
  partially-observable systems and brain modeling.
\newblock \emph{Journal of Neural Engineering}, 17\penalty0 (4):\penalty0
  046025, August 2020.
\newblock \doi{10.1088/1741-2552/aba07d}.

\bibitem[ASTROM(1995)]{ASTROM1995PID}
K.~ASTROM.
\newblock {{PID Controllers-Theory}}.
\newblock \emph{Design and Tuning}, 1995.

\bibitem[Yu et~al.(2019)Yu, Kim, Cho, and Mago]{Yu2019Nonlinear}
Byeongho Yu, Dongsu Kim, Heejin Cho, and Pedro Mago.
\newblock A {{Nonlinear Autoregressive With Exogenous Inputs Artificial Neural
  Network Model}} for {{Building Thermal Load Prediction}}.
\newblock \emph{Journal of Energy Resources Technology}, 142\penalty0 (050902),
  December 2019.
\newblock \doi{10.1115/1.4045543}.

\bibitem[Wan and Nelson(2001)]{Wan2001Dual}
Eric~A. Wan and Alex~T. Nelson.
\newblock Dual {{Extended Kalman Filter Methods}}.
\newblock In \emph{Kalman {{Filtering}} and {{Neural Networks}}}, chapter~5,
  pages 123--173. John Wiley \& Sons, Ltd, 2001.
\newblock \doi{10.1002/0471221546.ch5}.

\bibitem[Kalman(1960)]{Kalman1960New}
R.~E. Kalman.
\newblock A {{New Approach}} to {{Linear Filtering}} and {{Prediction
  Problems}}.
\newblock \emph{Journal of Basic Engineering}, 82\penalty0 (1):\penalty0
  35--45, March 1960.
\newblock \doi{10.1115/1.3662552}.

\bibitem[Yedidia(2011)]{Yedidia2011MessagePassing}
Jonathan~S. Yedidia.
\newblock Message-{{Passing Algorithms}} for {{Inference}} and
  {{Optimization}}: ``{{Belief Propagation}}'' and ``{{Divide}} and
  {{Concur}}''.
\newblock \emph{Journal of Statistical Physics}, 145\penalty0 (4):\penalty0
  860--890, November 2011.
\newblock \doi{10.1007/s10955-011-0384-7}.

\end{thebibliography}


\begin{thebibliography}{49}
\providecommand{\natexlab}[1]{#1}
\providecommand{\url}[1]{\texttt{#1}}
\expandafter\ifx\csname urlstyle\endcsname\relax
  \providecommand{\doi}[1]{doi: #1}\else
  \providecommand{\doi}{doi: \begingroup \urlstyle{rm}\Url}\fi

\bibitem[Ascher et~al.(1995)Ascher, Ruuth, and
  Wetton]{Ascher1995ImplicitExplicit}
Uri~M. Ascher, Steven~J. Ruuth, and Brian T.~R. Wetton.
\newblock Implicit-{{Explicit Methods}} for {{Time-Dependent Partial
  Differential Equations}}.
\newblock \emph{SIAM Journal on Numerical Analysis}, 32\penalty0 (3):\penalty0
  797--823, June 1995.
\newblock \doi{10.1137/0732037}.

\bibitem[Yedidia(2011)]{Yedidia2011MessagePassing}
Jonathan~S. Yedidia.
\newblock Message-{{Passing Algorithms}} for {{Inference}} and
  {{Optimization}}: ``{{Belief Propagation}}'' and ``{{Divide}} and
  {{Concur}}''.
\newblock \emph{Journal of Statistical Physics}, 145\penalty0 (4):\penalty0
  860--890, November 2011.
\newblock \doi{10.1007/s10955-011-0384-7}.

\bibitem[Gilmer et~al.(2017)Gilmer, Schoenholz, Riley, Vinyals, and
  Dahl]{Gilmer2017Neural}
Justin Gilmer, Samuel~S. Schoenholz, Patrick~F. Riley, Oriol Vinyals, and
  George~E. Dahl.
\newblock Neural {{Message Passing}} for {{Quantum Chemistry}}.
\newblock In \emph{Proceedings of the 34th {{International Conference}} on
  {{Machine Learning}}}, pages 1263--1272. PMLR, July 2017.

\bibitem[Ding et~al.(2025)Ding, Maros, and Scutari]{Ding2025New}
Kuangyu Ding, Marie Maros, and Gesualdo Scutari.
\newblock A {{New Decomposition Paradigm}} for {{Graph-structured Nonlinear
  Programs}} via {{Message Passing}}, 2025.

\bibitem[Shang(2009)]{Shang2009distributed}
Yueqiang Shang.
\newblock A distributed memory parallel {{Gauss}}--{{Seidel}} algorithm for
  linear algebraic systems.
\newblock \emph{Computers \& Mathematics with Applications}, 57\penalty0
  (8):\penalty0 1369--1376, April 2009.
\newblock \doi{10.1016/j.camwa.2009.01.034}.

\bibitem[Hai(1993)]{Hairer1993RungeKutta}
Runge-{{Kutta}} and {{Extrapolation Methods}}.
\newblock In Ernst Hairer, Gerhard Wanner, and Syvert~P. N{\o}rsett, editors,
  \emph{Solving {{Ordinary Differential Equations I}}: {{Nonstiff Problems}}},
  pages 129--353. Springer, Berlin, Heidelberg, 1993.
\newblock \doi{10.1007/978-3-540-78862-1_2}.

\bibitem[Borodulin(2025)]{Borodulin2025Accuracy}
Mikhail~Y. Borodulin.
\newblock Accuracy and {{Stability}} of the {{Second-Order Adams-Bashforth
  Method}} and the {{Trapezoidal Method}} in {{Power System Dynamic
  Simulations}}.
\newblock In \emph{2025 {{IEEE Power}} \& {{Energy Society General Meeting}}
  ({{PESGM}})}, pages 1--5, July 2025.
\newblock \doi{10.1109/PESGM52009.2025.11225339}.

\bibitem[Kalman(1960)]{Kalman1960New}
R.~E. Kalman.
\newblock A {{New Approach}} to {{Linear Filtering}} and {{Prediction
  Problems}}.
\newblock \emph{Journal of Basic Engineering}, 82\penalty0 (1):\penalty0
  35--45, March 1960.
\newblock \doi{10.1115/1.3662552}.

\bibitem[Ghorbani(2021)]{Ghorbani2021Nonlinear}
Esmaeil Ghorbani.
\newblock \emph{Nonlinear {{Kalman}} Filtering Based Damage Quantification for
  Civil Infrastructure}.
\newblock PhD thesis, University of Manitoba, 2021.

\bibitem[Khodarahmi and Maihami(2023)]{Khodarahmi2023Review}
Masoud Khodarahmi and Vafa Maihami.
\newblock A {{Review}} on {{Kalman Filter Models}}.
\newblock \emph{Archives of Computational Methods in Engineering}, 30\penalty0
  (1):\penalty0 727--747, January 2023.
\newblock \doi{10.1007/s11831-022-09815-7}.

\bibitem[Jazwinski(2007)]{Jazwinski2007Stochastic}
Andrew~H. Jazwinski.
\newblock \emph{Stochastic {{Processes}} and {{Filtering Theory}}}.
\newblock Courier Corporation, January 2007.

\bibitem[Wan and Nelson(2001)]{Wan2001Dual}
Eric~A. Wan and Alex~T. Nelson.
\newblock Dual {{Extended Kalman Filter Methods}}.
\newblock In \emph{Kalman {{Filtering}} and {{Neural Networks}}}, chapter~5,
  pages 123--173. John Wiley \& Sons, Ltd, 2001.
\newblock \doi{10.1002/0471221546.ch5}.

\bibitem[Wan and Van Der~Merwe(2000)]{Wan2000unscented}
E.A. Wan and R.~Van Der~Merwe.
\newblock The unscented {{Kalman}} filter for nonlinear estimation.
\newblock In \emph{Proceedings of the {{IEEE}} 2000 {{Adaptive Systems}} for
  {{Signal Processing}}, {{Communications}}, and {{Control Symposium}}
  ({{Cat}}. {{No}}.{{00EX373}})}, pages 153--158, October 2000.
\newblock \doi{10.1109/ASSPCC.2000.882463}.

\bibitem[Julier and Uhlmann(2004)]{Julier2004Unscented}
S.J. Julier and J.K. Uhlmann.
\newblock Unscented filtering and nonlinear estimation.
\newblock \emph{Proceedings of the IEEE}, 92\penalty0 (3):\penalty0 401--422,
  March 2004.
\newblock \doi{10.1109/JPROC.2003.823141}.

\bibitem[Ghorbani and Cha(2018)]{Ghorbani2018iterated}
Esmaeil Ghorbani and Young-Jin Cha.
\newblock An iterated cubature unscented {{Kalman}} filter for large-{{DoF}}
  systems identification with noisy data.
\newblock \emph{Journal of Sound and Vibration}, 420:\penalty0 21--34, April
  2018.
\newblock \doi{10.1016/j.jsv.2018.01.035}.

\bibitem[Leontaritis and Billings(1985)]{Leontaritis1985Inputoutput}
I.~J. Leontaritis and S.~A. Billings.
\newblock Input-output parametric models for non-linear systems {{Part II}}:
  Stochastic non-linear systems.
\newblock \emph{International Journal of Control}, 41\penalty0 (2):\penalty0
  329--344, February 1985.
\newblock \doi{10.1080/0020718508961130}.

\bibitem[Yu et~al.(2019)Yu, Kim, Cho, and Mago]{Yu2019Nonlinear}
Byeongho Yu, Dongsu Kim, Heejin Cho, and Pedro Mago.
\newblock A {{Nonlinear Autoregressive With Exogenous Inputs Artificial Neural
  Network Model}} for {{Building Thermal Load Prediction}}.
\newblock \emph{Journal of Energy Resources Technology}, 142\penalty0 (050902),
  December 2019.
\newblock \doi{10.1115/1.4045543}.

\bibitem[Biswas et~al.(2024)Biswas, Abbasi, and
  Chakrabortty]{Biswas2024Generalized}
Tarun~Kumer Biswas, Alireza Abbasi, and Ripon~Kumar Chakrabortty.
\newblock Generalized hop-based approaches for identifying influential nodes in
  social networks.
\newblock \emph{Expert Systems}, 41\penalty0 (10):\penalty0 e13649, 2024.
\newblock \doi{10.1111/exsy.13649}.

\bibitem[Lafferty and Lebanon(2005)]{Lafferty2005Diffusion}
John Lafferty and Guy Lebanon.
\newblock Diffusion {{Kernels}} on {{Statistical Manifolds}}.
\newblock \emph{Journal of Machine Learning Research}, 6\penalty0 (5):\penalty0
  129--163, 2005.

\bibitem[Patel and Habeeb(2025)]{Patel2025HKDMGIN}
B.~Patel and Z.~Habeeb.
\newblock {{HKD-MGIN}}: A physics informed graph neural network using heat
  kernel diffusion for mapping adolescent functional brain connectivity.
\newblock \emph{Machine Learning for Computational Science and Engineering},
  1\penalty0 (2):\penalty0 37, October 2025.
\newblock \doi{10.1007/s44379-025-00038-8}.

\bibitem[Brunton et~al.(2016)Brunton, Proctor, and
  Kutz]{Brunton2016Discovering}
Steven~L. Brunton, Joshua~L. Proctor, and J.~Nathan Kutz.
\newblock Discovering governing equations from data by sparse identification of
  nonlinear dynamical systems.
\newblock \emph{Proceedings of the National Academy of Sciences}, 113\penalty0
  (15):\penalty0 3932--3937, April 2016.
\newblock \doi{10.1073/pnas.1517384113}.

\bibitem[Kaiser et~al.(2018)Kaiser, Kutz, and Brunton]{Kaiser2018Sparse}
E.~Kaiser, J.~N. Kutz, and S.~L. Brunton.
\newblock Sparse identification of nonlinear dynamics for model predictive
  control in the low-data limit.
\newblock \emph{Proceedings of the Royal Society A: Mathematical, Physical and
  Engineering Sciences}, 474\penalty0 (2219):\penalty0 20180335, November 2018.
\newblock \doi{10.1098/rspa.2018.0335}.

\bibitem[Forootani et~al.(2025)Forootani, Goyal, and
  Benner]{Forootani2025robust}
Ali Forootani, Pawan Goyal, and Peter Benner.
\newblock A robust sparse identification of nonlinear dynamics approach by
  combining neural networks and an integral form.
\newblock \emph{Engineering Applications of Artificial Intelligence},
  149:\penalty0 110360, June 2025.
\newblock \doi{10.1016/j.engappai.2025.110360}.

\bibitem[Kaptanoglu et~al.(2022)Kaptanoglu, de~Silva, Fasel, Kaheman,
  Goldschmidt, Callaham, Delahunt, Nicolaou, Champion, Loiseau, Kutz, and
  Brunton]{Kaptanoglu2022PySINDy}
Alan~A. Kaptanoglu, Brian~M. de~Silva, Urban Fasel, Kadierdan Kaheman, Andy~J.
  Goldschmidt, Jared Callaham, Charles~B. Delahunt, Zachary~G. Nicolaou,
  Kathleen Champion, Jean-Christophe Loiseau, J.~Nathan Kutz, and Steven~L.
  Brunton.
\newblock {{PySINDy}}: {{A}} comprehensive {{Python}} package for robust sparse
  system identification.
\newblock \emph{Journal of Open Source Software}, 7\penalty0 (69):\penalty0
  3994, January 2022.
\newblock \doi{10.21105/joss.03994}.

\bibitem[Zhu et~al.(2019)Zhu, Fei, Jiang, and Cao]{Zhu2019Maintaining}
Rui Zhu, Qingguo Fei, Dong Jiang, and Zhifu Cao.
\newblock Maintaining {{Specific Natural Frequency}} of {{Damped System}}
  despite {{Mass Modification}}.
\newblock \emph{International Journal of Aerospace Engineering}, 2019\penalty0
  (1):\penalty0 1947506, 2019.
\newblock \doi{10.1155/2019/1947506}.

\bibitem[Thurner et~al.(2018)Thurner, Scheidler, Sch{\"a}fer, Menke, Dollichon,
  Meier, Meinecke, and Braun]{Thurner2018Pandapower}
Leon Thurner, Alexander Scheidler, Florian Sch{\"a}fer, Jan-Hendrik Menke,
  Julian Dollichon, Friederike Meier, Steffen Meinecke, and Martin Braun.
\newblock Pandapower---{{An Open-Source Python Tool}} for {{Convenient
  Modeling}}, {{Analysis}}, and {{Optimization}} of {{Electric Power Systems}}.
\newblock \emph{IEEE Transactions on Power Systems}, 33\penalty0 (6):\penalty0
  6510--6521, November 2018.
\newblock \doi{10.1109/TPWRS.2018.2829021}.

\bibitem[Zimmerman et~al.(2011)Zimmerman, {Murillo-S{\'a}nchez}, and
  Thomas]{Zimmerman2011MATPOWER}
Ray~Daniel Zimmerman, Carlos~Edmundo {Murillo-S{\'a}nchez}, and Robert~John
  Thomas.
\newblock {{MATPOWER}}: {{Steady-State Operations}}, {{Planning}}, and
  {{Analysis Tools}} for {{Power Systems Research}} and {{Education}}.
\newblock \emph{IEEE Transactions on Power Systems}, 26\penalty0 (1):\penalty0
  12--19, February 2011.
\newblock \doi{10.1109/TPWRS.2010.2051168}.

\bibitem[Kundur(2012)]{Kundur2012Power}
Prabha~S. Kundur.
\newblock Power {{System Stability}}.
\newblock In \emph{Power {{System Stability}} and {{Control}}}. CRC Press, 3
  edition, 2012.

\bibitem[D{\"o}rfler et~al.(2013)D{\"o}rfler, Chertkov, and
  Bullo]{Dorfler2013Synchronization}
Florian D{\"o}rfler, Michael Chertkov, and Francesco Bullo.
\newblock Synchronization in complex oscillator networks and smart grids.
\newblock \emph{Proceedings of the National Academy of Sciences}, 110\penalty0
  (6):\penalty0 2005--2010, February 2013.
\newblock \doi{10.1073/pnas.1212134110}.

\bibitem[Kuramoto(1984)]{Kuramoto1984Chemical}
Y~Kuramoto.
\newblock Chemical oscillations, waves, and turbulence.
\newblock \emph{(No Title)}, 8:\penalty0 156, 1984.

\bibitem[Kuramoto(2005)]{Kuramoto2005Selfentrainment}
Yoshiki Kuramoto.
\newblock Self-entrainment of a population of coupled non-linear oscillators.
\newblock In \emph{International Symposium on Mathematical Problems in
  Theoretical Physics: {{January}} 23--29, 1975, Kyoto University,
  Kyoto/{{Japan}}}, pages 420--422. Springer, 2005.

\bibitem[Acebr{\'o}n et~al.(2005)Acebr{\'o}n, Bonilla, P{\'e}rez~Vicente,
  Ritort, and Spigler]{Acebron2005Kuramoto}
Juan~A. Acebr{\'o}n, L.~L. Bonilla, Conrad~J. P{\'e}rez~Vicente, F{\'e}lix
  Ritort, and Renato Spigler.
\newblock The {{Kuramoto}} model: {{A}} simple paradigm for synchronization
  phenomena.
\newblock \emph{Reviews of Modern Physics}, 77\penalty0 (1):\penalty0 137--185,
  April 2005.
\newblock \doi{10.1103/RevModPhys.77.137}.

\bibitem[Filatrella et~al.(2008)Filatrella, Nielsen, and
  Pedersen]{Filatrella2008Analysis}
G.~Filatrella, A.~H. Nielsen, and N.~F. Pedersen.
\newblock Analysis of a power grid using a {{Kuramoto-like}} model.
\newblock \emph{The European Physical Journal B}, 61\penalty0 (4):\penalty0
  485--491, February 2008.
\newblock \doi{10.1140/epjb/e2008-00098-8}.

\bibitem[D{\"o}rfler and Bullo(2014)]{Dorfler2014Synchronization}
Florian D{\"o}rfler and Francesco Bullo.
\newblock Synchronization in complex networks of phase oscillators: {{A}}
  survey.
\newblock \emph{Automatica}, 50\penalty0 (6):\penalty0 1539--1564, June 2014.
\newblock \doi{10.1016/j.automatica.2014.04.012}.

\bibitem[Golub and Loan(2013)]{Golub2013Matrixa}
Gene~H. Golub and Charles F.~Van Loan.
\newblock \emph{Matrix {{Computations}}}.
\newblock Johns Hopkins University Press, 2013.
\newblock \doi{10.56021/9781421407944}.

\bibitem[Cohen and Migliorati(2017)]{Cohen2017Optimal}
Albert Cohen and Giovanni Migliorati.
\newblock Optimal weighted least-squares methods.
\newblock \emph{The SMAI Journal of computational mathematics}, 3:\penalty0
  181--203, October 2017.
\newblock \doi{10.5802/smai-jcm.24}.

\bibitem[Yang(2019)]{Yang2019Hydropower}
Weijia Yang.
\newblock \emph{Hydropower {{Plants}} and {{Power Systems}}: {{Dynamic
  Processes}} and {{Control}} for {{Stable}} and {{Efficient Operation}}}.
\newblock Springer, April 2019.

\bibitem[Chaudhry(2014)]{Chaudhry2014Applieda}
M.~Hanif Chaudhry.
\newblock \emph{Applied {{Hydraulic Transients}}}.
\newblock Springer, New York, NY, 2014.
\newblock \doi{10.1007/978-1-4614-8538-4}.

\bibitem[Gurecky et~al.(2023)Gurecky, Wang, and Ou]{Gurecky2023Elastic}
William Gurecky, Hong Wang, and Shawn Ou.
\newblock Elastic {{Flow Modeling}} for {{Hydropower Digital Twins}}.
\newblock Technical Report ORNL/SPR--2023/2889, Oak Ridge National Laboratory
  (ORNL), Oak Ridge, TN (United States), May 2023.

\bibitem[Xu et~al.(2017)Xu, Yan, Chen, Gao, and Wu]{Xu2017Sensitivity}
Beibei Xu, Donglin Yan, Diyi Chen, Xiang Gao, and Changzhi Wu.
\newblock Sensitivity analysis of a {{Pelton}} hydropower station based on a
  novel approach of turbine torque.
\newblock \emph{Energy Conversion and Management}, 148:\penalty0 785--800,
  September 2017.
\newblock \doi{10.1016/j.enconman.2017.06.019}.

\bibitem[Loshchilov and Hutter(2018)]{Loshchilov2018Decoupled}
Ilya Loshchilov and Frank Hutter.
\newblock Decoupled {{Weight Decay Regularization}}.
\newblock In \emph{International {{Conference}} on {{Learning
  Representations}}}, September 2018.

\bibitem[Zhou et~al.(2024)Zhou, Xie, Lin, and Yan]{Zhou2024Understanding}
Pan Zhou, Xingyu Xie, Zhouchen Lin, and Shuicheng Yan.
\newblock Towards {{Understanding Convergence}} and {{Generalization}} of
  {{AdamW}}.
\newblock \emph{IEEE Transactions on Pattern Analysis and Machine
  Intelligence}, 46\penalty0 (9):\penalty0 6486--6493, September 2024.
\newblock \doi{10.1109/TPAMI.2024.3382294}.

\bibitem[Genta(2005)]{Genta2005Dynamics}
Giancarlo Genta.
\newblock \emph{Dynamics of {{Rotating Systems}}}.
\newblock Springer Science \& Business Media, April 2005.

\bibitem[Goldman and Muszynska(1995)]{Goldman1995Rotortostator}
Paul Goldman and Agnes Muszynska.
\newblock Rotor-to-stator, rub-related, thermal/mechanical effects in rotating
  machinery.
\newblock \emph{Chaos, Solitons \& Fractals}, 5\penalty0 (9):\penalty0
  1579--1601, September 1995.
\newblock \doi{10.1016/0960-0779(94)00165-M}.

\bibitem[Shi et~al.(2022)Shi, Zhou, Huang, Xu, and Liu]{Shi2022Vibration}
Yousong Shi, Jianzhong Zhou, Jie Huang, Yanhe Xu, and Baonan Liu.
\newblock A {{Vibration Fault Identification Framework}} for {{Shafting
  Systems}} of {{Hydropower Units}}: {{Nonlinear Modeling}}, {{Signal
  Processing}}, and {{Holographic Identification}}.
\newblock \emph{Sensors}, 22\penalty0 (11):\penalty0 4266, January 2022.
\newblock \doi{10.3390/s22114266}.

\bibitem[Anderson and Fouad(1994)]{Anderson1994Power}
Paul~M. Anderson and A.~A. Fouad.
\newblock \emph{Power {{System Control}} and {{Stability}}}.
\newblock Wiley, 1994.

\bibitem[Sauer and Pai(1998)]{Sauer1998Power}
Peter~W. Sauer and M.~A. Pai.
\newblock \emph{Power {{System Dynamics}} and {{Stability}}}.
\newblock Prentice Hall, 1998.

\bibitem[Pai(1989)]{Pai1989Energy}
M.~A. Pai.
\newblock \emph{Energy {{Function Analysis}} for {{Power System Stability}}}.
\newblock Kluwer Academic Publishers, Norwell, MA, 1989.

\bibitem[You et~al.(2021)You, Li, Liu, Sun, Wang, Qiu, and
  Liu]{You2021Calculate}
Shutang You, Hongyu Li, Shengyuan Liu, Kaiqi Sun, Weikang Wang, Wei Qiu, and
  Yilu Liu.
\newblock Calculate {{Center-of-Inertia Frequency}} and {{System RoCoF Using
  PMU Data}}, April 2021.

\end{thebibliography}
\end{document}